\documentclass[pra,nofootinbib]{revtex4}

\usepackage{amsmath,amssymb}
\usepackage{mathrsfs}
\usepackage{mathbbol}
\usepackage{pst-func}
\usepackage{epsfig}
\usepackage{boldline}
\usepackage{cancel}
\usepackage[math]{cellspace}
\usepackage{listings}

\usepackage{chngcntr}
\counterwithout{equation}{section}

\allowdisplaybreaks

\newtheorem{theorem}{\bf Theorem}[section]
\newtheorem{corollary}{\bf Corollary}[section]

\def\dbl{\hbox{${1\hskip -2.4pt{\rm l}}$}}

\begin{document}

\title{Quantum Correlations are Weaved by the Spinors of the Euclidean Primitives}

\author{Joy Christian}

\email{jjc@alum.bu.edu}

\affiliation{Einstein Centre for Local-Realistic Physics, 15 Thackley End, Oxford OX2 6LB, United Kingdom}

\begin{abstract}
The exceptional Lie group ${E_8}$ plays a prominent role in both mathematics and theoretical physics. It is the largest symmetry group associated with the most general possible normed division algebra, namely, that of the non-associative real octonions, which --- thanks to their non-associativity --- form the only possible closed set of spinors (or rotors) that can parallelize the 7-sphere. By contrast, here we show how a similar 7-sphere also arises naturally from the algebraic interplay of the graded Euclidean primitives, such as points, lines, planes, and volumes, which characterize the three-dimensional conformal geometry of the ambient physical space, set within its eight-dimensional Clifford-algebraic representation. Remarkably, the resulting algebra remains associative, and allows us to understand the origins and strengths of all quantum correlations locally, in terms of the geometry of the compactified physical space, namely, that of a quaternionic 3-sphere, ${S^3}$, with ${S^7}$ being its algebraic representation space. Every quantum correlation can thus be understood as a correlation among a set of points of this ${S^7}$, computed using manifestly local spinors within ${S^3}$, thereby extending the stringent bounds of ${\pm2}$ set by Bell inequalities to the bounds of ${\pm2\sqrt{2}}$ on the strengths of all possible strong correlations, in the same quantitatively precise manner as that predicted within quantum mechanics. The resulting geometrical framework thus overcomes Bell's theorem by producing a strictly deterministic and realistic framework that allows a locally causal understanding of all quantum correlations, without requiring either remote contextuality or backward causation. We demonstrate this by first proving a general theorem concerning the geometrical origins of the correlations predicted by arbitrarily entangled quantum states, and then reproducing the correlations predicted by the EPR-Bohm and the GHZ states. The {\it raison d'\^etre} of strong correlations turns out to be the M\"obius-like twists in the Hopf bundles of ${S^3}$ and ${S^7}$.
\end{abstract}

\maketitle

\section{Introduction}\label{intro}

The central source of intrinsic coherence, geometrical elegance, and empirical success of Einstein's theory of gravity is undoubtedly its strict adherence to local causality \cite{MTW,Wald}. Indeed, despite the phenomenal empirical success of Newton's theory of gravity for over two centuries \cite{Cartan}, its founding on the unexplained ``action at a distance" was a reason enough for Einstein to search for its locally causal generalization. Today we face a similar challenge in search of a theory that may unify quantum theory with Einstein's theory of gravity. But in sharp contrast to Einstein's theory, quantum theory seems to harbor a peculiar form of non-signalling non-locality, as noticed long ago by Einstein, Podolsky, and Rosen (EPR) \cite{EPR}. They hoped, however, that quantum theory can be completed into a locally causal theory with addition of ``hidden" parameters or supplementary variables. Today such a hope of completing quantum theory into a realistic and local theory envisaged by Einstein is widely believed to have been dashed by Bell's theorem \cite{Bell-1964}, its variants \cite{GHZ}, and the related experimental investigations \cite{Aspect}. Indeed, the claim of Bell's theorem is remarkably comprehensive in scope: {\it no physical theory which is local and realistic as hoped for by Einstein can reproduce all of the strong correlations predicted by quantum mechanics} \cite{Stanford}.

By contrast, our primary concern in this paper is not Bell's theorem but understanding the origins and strengths of all quantum correlations in terms of the algebraic, geometrical, and topological properties of the physical space in which we are confined to perform our experiments. In our view Bell's theorem is a distraction that prevents us from understanding the true origins of quantum correlations, especially because it is neither a theorem in the strict mathematical sense, nor a result within quantum theory itself. Indeed, not a single concept from quantum theory is used in the derivation of the Bell-CHSH inequalities \cite{Bell-1964,CHSH}. It is, in fact, an argument that depends on a number of physical assumptions about what is and what is not possible within any locally causal theory, and these assumptions can be, and have been questioned before \cite{disproof,IJTP}. Consequently, by circumventing Bell's argument, in this paper we set out to explain the origins and strengths of all quantum correlations within a locally causal framework of octonion-like spinors, which are constructed using a geometric algebra \cite{Clifford,Dorst} of rudimentary Euclidean primitives, such as points, lines, planes, and volumes. This is accomplished by recognizing and overcoming two neglected shortcomings of Bell's argument \cite{Bell-1964,Stanford,CHSH,Clauser}. The first of these shortcomings, which is discussed in greater detail in subsection \ref{Bell-flaw} below, amounts to averaging over measurement events in the derivation of the experimentally violated absolute bound of ${2}$ on the CHSH string of expectation values that are {\it impossible} to occur in {\it any} possible world, classical or quantum, stemming from a mistaken application of the criterion of reality propounded by EPR \cite{EPR,GHZ}. The second shortcoming of Bell's argument stems from the unjustified identification of the {\it image} ${\{+1,\,-1\}}$ of the measurement functions, which represents the actual measurement results in Bell's prescription \cite{Bell-1964}, with the {\it co-domain} of these functions, which is neither specified by Bell explicitly nor observable directly in the so-called Bell-test experiments \cite{Aspect,Stanford}. An explicit specification of the latter, however, is a prerequisite for the very definition of a mathematical function \cite{disproof}. By contrast, in our prescription (\ref{theoequa}) of measurement results discussed in subsection \ref{Mfunc}, the locally unobservable co-domain ${S^7}$ of the measurement functions is explicitly specified with considerable detail, without compromising Bell's pristine bivalued prescription, ${\pm1}$, for the actually observed measurement results. It incorporates the Clifford algebraic properties of the physical space in which all such experiments are necessarily situated and performed \cite{Aspect,Clauser}.

As noted above, however, our primary focus in this paper is not on Bell's theorem but on understanding the origins and strengths of all quantum correlations as a consequence of the geometry and topology of the physical space (or more generally of spacetime). Since quantum correlations are necessarily observed within the confines of spacetime, it is natural to view them as correlations among measurement events in spacetime --- {\it i.e.}, among the ``clicks" of a set of detectors configured within spacetime. On the other hand, what is actually recorded in the Bell-test experiments are coincidence counts among bivalued measurement results, observed simultaneously within space at a given time \cite{Aspect}. Therefore, without loss of generality, we will restrict our analysis to the physical space. With that in mind, in the next section we extensively review the algebraic properties of the compactified physical space, captured in the definition (\ref{3-sphere}), which is a quaternionic 3-sphere, and construct its algebraic representation space (\ref{spring}), which is an octonion-like 7-sphere. Since such parallelizable 3- and 7-spheres play a vital role\textsuperscript{\ref{Hardy}} in our local-realistic framework, we have devoted a brief appendix (appendix \ref{ApA}) to discuss their wider significance in physics and mathematics at a pedagogical level \cite{disproof}. Our central theorem concerning the origins of all quantum correlations is then stated and proved in subsection \ref{theo}.

The proof presented in subsection \ref{theo} includes a local-realistic derivation of the simplest yet emblematic quantum correlations --- namely, those predicted by the rotationally invariant singlet or EPR-Bohm state --- the strengths of which are well known to violate the theoretical bounds of ${\pm2}$ set by the Bell-CHSH inequalities, in the Bell-test experiments \cite{Aspect}. Then, in subsection \ref{T-bound}, we derive the closely related Tsirel'son's bounds of ${\pm2\sqrt{2}}$ on the strengths of {\it all} quantum correlations within our framework. In the subsequent subsection, \ref{Frag}, we then explain the geometrical reasons for the fragility of quantum correlations as a counterpart of that of quantum entanglement. This brings us to subsection \ref{SecGHZ} in which we derive the strong correlations predicted by the rotationally non-invariant 4-particle GHSZ state \cite{GHZ}, together with the proof of Bell's condition of factorizability within ${S^7}$ in appendix \ref{ApB} ${\big(\text{a similar proof also goes through within}\; S^3\big)}$. 

In subsection \ref{Bell-test} we then point out that the predictions of our local-realistic ${S^7}$ framework is not in conflict with what is actually observed in the Bell-test experiments \cite{Aspect}, since they simply reproduce the predictions of quantum mechanics \cite{Bell-1964,GHZ}. In the subsequent subsection, \ref{Bell-flaw}, in the light of the widespread belief in Bell's theorem, we reveal a serious oversight in Bell's argument in some detail, independently of the constructive counterexamples provided by our ${S^7}$ model for the strong correlations. This brings us to subsection \ref{GHZ-flaw} in which we present an analytical disproof of the GHZ variant of Bell's theorem, which does not involve Boole-type mathematical inequalities\textsuperscript{\ref{BooleIn}} used by Bell in his original argument. In section \ref{Numerical} we then present event-by-event numerical simulations of the 2-particle EPR-Bohm and 4-particle GHSZ correlations predicted by our local-realistic framework based on ${S^7}$. Finally, in section \ref{conc}, we summarize our findings.

\section{Modern Perspective on the Euclidean Primitives}\label{222}

In physical experiments --- which are usually confined to the three-dimensional physical space by necessity --- we often measure relevant quantities by setting up a Cartesian coordinate system ${\left\{x,y,z\right\}}$ in that space. Mathematically this is equivalent to identifying the Euclidean space ${{\mathbb E}^3}$ with a three-fold product of the real line, ${{\rm I\!R}^3}$. In practice we sometimes even think of ${{\rm I\!R}^3}$ as {\it the} Euclidean space. Euclid himself, however, did not think of ${{\mathbb E}^3}$ in terms of such a Cartesian triple of real numbers. He defined a representation of ${{\mathbb E}^3}$ axiomatically, in terms of primitive geometric objects such as points and lines, together with a list of their properties, from which his theorems of geometry follow.

It is, however, not always convenient to model the physical space in the spirit of Euclid. Therefore in practice we tend to identify ${{\mathbb E}^3}$ with ${{\rm I\!R}^3}$ whenever possible. But there is no intrinsic way of identifying the two spaces in this manner without introducing an {\it unphysical} element of arbitrarily chosen coordinate system. This difficulty is relevant for understanding the origins of quantum correlations, for time and again we have learned that careless introduction of unphysical ideas in physics could lead to distorted views of the physical reality \cite{disproof,Dechant}. An intrinsic, coordinate-free representation of the Euclidean space is surely preferable, if what is at stake is the very nature of the physical reality [cf. subsection \ref{Bell-flaw}]. 

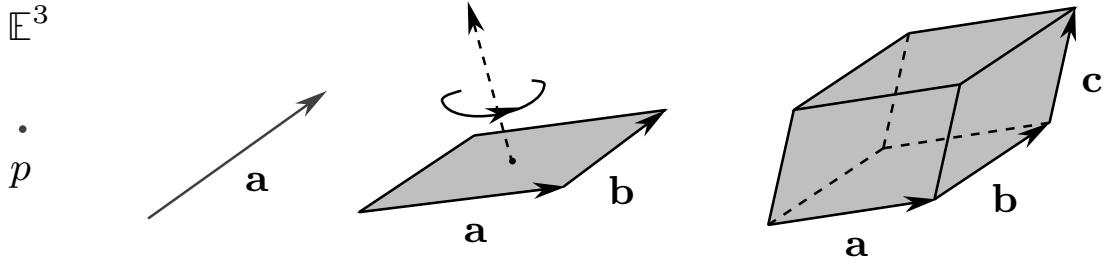
\begin{figure}
\hrule
\scalebox{1.7}{
\begin{pspicture}(7.2,2.2)(0.5,5.0)

\psline[linewidth=0.00001mm,linecolor=darkgray,dotsize=1.7pt 4]{*-}(-0.38,3.55)(-0.38,3.55)

\put(-0.5,4.25){{${{\mathbb E}^3}$}}

\put(-0.475,3.15){{${p}$}}

\psline[linewidth=0.2mm,linecolor=darkgray,arrowinset=0.3,arrowsize=2pt 3,arrowlength=2]{->}(0.6,2.85)(2.0,3.85)

\put(1.36,3.07){{${\bf a}$}}

\pspolygon[linewidth=0.1mm,fillstyle=solid,fillcolor=lightgray](2.25,2.9)(3.15,3.5)(4.65,3.7)(3.85,3.1)

\pspolygon[linewidth=0.1mm,fillstyle=solid,fillcolor=lightgray](5.65,3.7)(6.55,4.3)(7.85,4.5)(6.95,3.9)

\pspolygon[linewidth=0.1mm,fillstyle=solid,fillcolor=lightgray](5.45,2.8)(5.65,3.7)(6.95,3.9)(6.75,3.0)

\pspolygon[linewidth=0.1mm,fillstyle=solid,fillcolor=lightgray](6.95,3.9)(7.85,4.5)(7.65,3.6)(6.75,3.0)

\psline[linewidth=0.2mm,arrowinset=0.3,arrowsize=2pt 3,arrowlength=2]{-}(2.25,2.9)(3.15,3.5)

\psline[linewidth=0.2mm,arrowinset=0.3,arrowsize=2pt 3,arrowlength=2]{->}(2.25,2.9)(3.87,3.1)

\psline[linewidth=0.2mm,arrowinset=0.3,arrowsize=2pt 3,arrowlength=2]{-}(3.15,3.5)(4.65,3.7)

\psline[linewidth=0.2mm,arrowinset=0.3,arrowsize=2pt 3,arrowlength=2]{->}(3.85,3.1)(4.65,3.7)

\put(3.07,2.67){{${\bf a}$}}

\psline[linewidth=0.2mm,linestyle=dashed,dash=2pt 2pt,arrowinset=0.3,arrowsize=2pt 3,arrowlength=2]{->}(3.45,3.28)(3.1,4.55)

\pscurve[linewidth=0.2mm]{-}(3.05,3.85)(2.9,3.77)(3.7,3.9)(3.6,3.93)

\psline[linewidth=0.2mm,arrowinset=0.3,arrowsize=2pt 3,arrowlength=2]{->}(3.41,3.695)(3.5,3.72)

\psline[linewidth=0.00001mm,dotsize=1.5pt 4]{*-}(3.45,3.3)(3.45,3.3)

\put(4.2,2.97){{${\bf b}$}}

\psline[linewidth=0.2mm,linestyle=dashed,dash=2pt 2pt,arrowinset=0.3,arrowsize=2pt 3,arrowlength=2]{-}(5.45,2.8)(6.35,3.4)

\psline[linewidth=0.2mm,arrowinset=0.3,arrowsize=2pt 3,arrowlength=2]{->}(5.45,2.8)(6.77,3.0)

\psline[linewidth=0.2mm,linestyle=dashed,dash=2pt 2pt,arrowinset=0.3,arrowsize=2pt 3,arrowlength=2]{-}(6.35,3.4)(7.65,3.6)

\psline[linewidth=0.2mm,arrowinset=0.3,arrowsize=2pt 3,arrowlength=2]{->}(6.75,3.0)(7.65,3.6)

\put(6.05,2.56){{${\bf a}$}}

\put(7.2,2.9){{${\bf b}$}}

\psline[linewidth=0.2mm,arrowinset=0.3,arrowsize=2pt 3,arrowlength=2]{-}(5.45,2.8)(5.65,3.7)

\psline[linewidth=0.2mm,arrowinset=0.3,arrowsize=2pt 3,arrowlength=2]{->}(7.65,3.6)(7.85,4.5)

\psline[linewidth=0.2mm,linestyle=dashed,dash=2pt 2pt,arrowinset=0.3,arrowsize=2pt 3,arrowlength=2]{-}(6.35,3.4)(6.55,4.3)

\psline[linewidth=0.2mm,arrowinset=0.3,arrowsize=2pt 3,arrowlength=2]{-}(6.75,3.0)(6.95,3.9)

\put(7.9,3.84){{${\bf c}$}}

\psline[linewidth=0.2mm,arrowinset=0.3,arrowsize=2pt 3,arrowlength=2]{-}(5.65,3.7)(6.55,4.3)

\psline[linewidth=0.2mm,arrowinset=0.3,arrowsize=2pt 3,arrowlength=2]{-}(5.65,3.7)(6.95,3.9)

\psline[linewidth=0.2mm,arrowinset=0.3,arrowsize=2pt 3,arrowlength=2]{-}(6.55,4.3)(7.85,4.5)

\psline[linewidth=0.2mm,arrowinset=0.3,arrowsize=2pt 3,arrowlength=2]{-}(6.95,3.9)(7.85,4.5)

\end{pspicture}}
\hrule
\caption{Euclidean subspaces such as points (${p}$), lines (${\bf a}$), areas (${{\bf a}\wedge{\bf b}}$), and volumes (${{\bf a}\wedge{\bf b}\wedge{\bf c}}$) are taken in Clifford algebra ${Cl_{3,0}}$ as \underbar{primitive entities} constituting the Euclidean space ${{\mathbb E}^3}$, with each subspace specified only by its magnitude, direction, and orientation (or handedness), providing a unified and complete algebraic framework of directed numbers across dimensions, spanned by the geometric product ${{\bf a}\,{\bf b}={\bf a}\cdot{\bf b}+{\bf a}\wedge{\bf b}}$, reminiscent of a complex number.}
\vspace{5pt}
\hrule
\label{fig0}
\end{figure}

Fortunately, precisely such a representation of ${{\mathbb E}^3}$ was proposed by Grassmann in 1844 \cite{Dorst}. In the Euclidean spirit, the basic elements of this powerful algebraic representation of ${{\mathbb E}^3}$ are not coordinate systems, but points, lines, planes, and volumes, {\it all treated on equal footing}. Given a set ${\{\,{\bf e}_x,\,{\bf e}_y,\,{\bf e}_z\}}$ of basis vectors representing lines in ${{\mathbb E}^3}$, the algebra of corresponding vector space is constructed as follows. One begins by defining a volume element (or a trivector) in ${{\mathbb E}^3}$:
\begin{equation}
I_3:=\,{{\bf e}_x}{{\bf e}_y}{{\bf e}_z}\,,\label{1-1}
\end{equation}
with ${\{{\bf e}_i\}}$ being a set of anti-commuting orthonormal vectors in ${{\rm I\!R}^3}$ such that ${{\bf e}_j{\bf e}_i=-\,{\bf e}_i{\bf e}_j}$ for any ${i,j=x,\,y}$, or ${z}$. More generally the unit vectors ${{\bf e}_i}$ satisfy the fundamental geometric or Clifford product in this (by definition) {\it associative} algebra,
\begin{equation}
{\bf e}_i\,{\bf e}_j\,=\,{\bf e}_i\cdot{\bf e}_j+\,{\bf e}_i\wedge\,{\bf e}_j\,, \label{gp}
\end{equation}
with
\begin{equation}
{\bf e}_i\cdot{\bf e}_j:=\,\frac{1}{2}\left\{{\bf e}_i{\bf e}_j+{\bf e}_j{\bf e}_i\right\}
\end{equation}
being the symmetric inner product and
\begin{equation}
{\bf e}_i\wedge{\bf e}_j:=\,\frac{1}{2}\left\{{\bf e}_i{\bf e}_j-{\bf e}_j{\bf e}_i\right\}\,
\end{equation}
being the anti-symmetric outer product, implying ${({\bf e}_i\wedge{\bf e}_j)^2=-1}$. Any vector ${{\bf x}\in{\mathbb E}^3}$ is then a solution of the equation 
\begin{equation}
I_3\wedge{\bf x}=0\,.
\end{equation}
The normalized volume element ${I_3}$ thus represents an element of the highest grade in the corresponding algebra, namely grade-3. It is also referred to as a pseudo-scalar, dual to the scalar, which in turn is the lowest possible grade in the algebra:
\begin{equation}
1=I_3\,I^{\dagger}_3\,,
\end{equation}   
where the conjugate ${I^{\dagger}_3:=\,{{\bf e}_z}{{\bf e}_y}{{\bf e}_x}=-I_3}$ is the ``reverse" of ${I_3}$ implying ${(I_3)^2=-1}$, and the duality relation between the elements ${\widetilde{\Omega}}$ and ${\Omega}$ of arbitrary grades is defined as
\begin{equation}
\widetilde{\Omega}:=\Omega\,I^{\dagger}_3\,,
\end{equation}
with the norm ${||\;\,||}$ of ${\Omega}$ and scalar part ${\langle\;\;\rangle_s}$ of the product of mixed-grade vectors ${\bf X}$ and ${\bf Y}$ of ${n}$-components defined${\;}$as
\begin{equation}
\left|\left|\,\Omega\,\right|\right|\,:=\,\sqrt{\,\Omega\cdot\Omega^{\dagger}\,}\,\equiv\,\sqrt{\,\langle\,\Omega\,\Omega^{\dagger}\rangle_s}\;\;\;\;\;\text{and}\;\;\;\,\langle\,{\bf X}\,{\bf Y}^{\dagger}\rangle_s =  \sum_{l\,=\,0}^{n}\langle\,{\bf X}_l{\bf Y}_{\!l}^{\dagger}\,\rangle_s\,. 
\label{s-norm}
\end{equation}
Thus, for example, the orthonormal vectors ${{\bf e}_k}$ of grade-1 can be easily recovered from the unit bivectors ${{\bf e}_i\wedge\,{\bf e}_j}$ of grade-2 using the above duality relation:
\begin{equation}
{\bf e}_k=\,({\bf e}_i\wedge\,{\bf e}_j)\,I^{\dagger}_3\,=\,({\bf e}_i{\bf e}_j)\,I^{\dagger}_3\,.
\end{equation}
In three-dimensional Euclidean space there are thus basis elements of four different grades: An identity element ${{\bf e}^2_i=1}$ of grade-0, three orthonormal vectors ${{\bf e}_i}$ of grade-1, three orthonormal bivectors ${{\bf e}_j{\bf e}_k}$ of grade-2, and a trivector ${{\bf e}_i{\bf e}_j{\bf e}_k}$ of grade-3. Respectively, they represent points, lines, planes, and volumes in ${{\mathbb E}^3}$, as shown in Fig.${\,}$\ref{fig0}. Since in ${{\rm I\!R}^3}$ there are ${2^3=8}$ ways to combine the vectors ${{\bf e}_i}$ using the geometric product (\ref{gp}) such that no two products are linearly dependent, the resulting algebra, ${Cl_{3,0}\,}$, is a linear vector space of ${2^3=8}$ dimensions, spanned by these graded bases:
\begin{equation}
Cl_{3,0}={\rm span}\!\left\{\,1,\;{\bf e}_x,\,{\bf e}_y,\,{\bf e}_z,\;{\bf e}_x{\bf e}_y,\,
{\bf e}_z{\bf e}_x,\,{\bf e}_y{\bf e}_z,\;{\bf e}_x{\bf e}_y{\bf e}_z\,\right\}\!. \label{cl}
\end{equation}
This algebra intrinsically characterizes the Euclidean space ${{\mathbb E}^3}$ without requiring a coordinate system, by the bijection 
\begin{equation}
{\cal F}: {\rm I\!R}^3:={\rm span}\!\left\{\,{\bf e}_x,\,{\bf e}_y,\,{\bf e}_z\right\}\longrightarrow\, {\rm I\!R}^8:={\rm span}\!\left\{\,1,\;{\bf e}_x,\,{\bf e}_y,\,{\bf e}_z,\;{\bf e}_x{\bf e}_y,\,
{\bf e}_z{\bf e}_x,\,{\bf e}_y{\bf e}_z,\;{\bf e}_x{\bf e}_y{\bf e}_z\,\right\}=Cl_{3,0}.
\end{equation}

\subsection{One-point Compactification of the 3-Dimensional Euclidean Space}

\begin{figure}
\hrule
\scalebox{0.9}{
\begin{pspicture}(0.5,-3.5)(4.2,3.5)

\psline[linewidth=0.4mm,arrowinset=0.3,arrowsize=3pt 3,arrowlength=2]{<->}(-5.47,-2.3)(10.2,-2.3)

\psline[linewidth=0.3mm,arrowinset=0.3,arrowsize=3pt 3,arrowlength=2]{->}(-5.0,0.0)(-4.0,0.0)

\psline[linewidth=0.3mm,arrowinset=0.3,arrowsize=3pt 3,arrowlength=2]{->}(-5.0,0.0)(-5.0,1.0)

\psline[linewidth=0.2mm,arrowinset=0.3,arrowsize=3pt 3,arrowlength=2]{->}(2.35,-2.3)(8.35,-2.3)

\psline[linewidth=0.2mm,arrowinset=0.3,arrowsize=3pt 3,arrowlength=2]{->}(2.35,-2.3)(4.55,0.63)

\put(2.25,2.55){\large {${{\bf e}_{\infty}}$}}

\put(2.89,2.55){${\sim(0,2)}$}

\put(-2.9,-2.85){\large {${{\mathbb E}^3}$}}

\put(2.3,-2.7){\large {${\rm o}$}}

\put(-4.0,0.8){\large {${{\rm I\!R}^4}$}}

\put(-5.1,1.1){\large {${\hat{x}_4}$}}

\put(-3.85,-0.1){\large {${\bf x}$}}

\put(8.45,-2.15){\large {${\bf x}$}}

\put(4.7,0.75){\large {${\vec{\phi}({\bf x})}$}}

\put(-1.79,1.25){\large {${S^3\sim\vec{\phi}({\mathbb E}^3)}$}}

\put(-5.97,-2.4){{${\infty}$}}

\put(-5.97,2.25){{${\infty}$}}

\put(10.37,2.25){{${\infty}$}}

\put(10.37,-2.4){{${\infty}$}}

\psline[linewidth=0.2mm,linestyle=dashed,arrowinset=0.3,arrowsize=3pt 3,arrowlength=2]{<->}(-5.47,2.3)(10.2,2.3)

\psline[linewidth=0.2mm,linestyle=dashed](2.35,2.3)(8.35,-2.3)

\psline[linewidth=0.2mm,linestyle=dashed](2.35,2.3)(6.35,-2.3)

\psline[linewidth=0.2mm,linestyle=dashed](2.35,2.3)(4.35,-2.3)

\psline[linewidth=0.3mm,linestyle=dashed](2.35,-2.3)(2.35,2.2)

\psline[linewidth=0.1mm,dotsize=2pt 3]{*-}(2.35,2.27)(2.4,2.27)

\psline[linewidth=0.1mm,dotsize=2pt 2]{*-}(2.35,0)(2.35,0)

\put(2.45,-0.1){{${(0,1)}$}}

\psline[linewidth=0.2mm,linestyle=dashed](2.35,2.3)(0.35,-2.3)

\psline[linewidth=0.2mm,linestyle=dashed](2.35,2.3)(-1.65,-2.3)

\psline[linewidth=0.2mm,linestyle=dashed](2.35,2.3)(-3.65,-2.3)

\pscircle[linewidth=0.4mm](2.35,0){2.3}

\psline[linewidth=0.1mm,dotsize=2pt 3]{*-}(2.35,2.3)(2.4,2.3)

\psline[linewidth=0.1mm,dotsize=2.5pt 3]{*-}(2.35,-2.29)(2.4,-2.31)

\psline[linewidth=0.1mm,dotsize=2pt 3]{*-}(8.35,-2.3)(8.4,-2.3)

\psline[linewidth=0.1mm,dotsize=2pt 3]{*-}(6.35,-2.3)(6.4,-2.3)

\psline[linewidth=0.1mm,dotsize=2pt 3]{*-}(4.35,-2.3)(4.4,-2.3)

\psline[linewidth=0.1mm,dotsize=2pt 3]{*-}(0.35,-2.3)(0.34,-2.3)

\psline[linewidth=0.1mm,dotsize=2pt 3]{*-}(-1.65,-2.3)(-1.64,-2.3)

\psline[linewidth=0.1mm,dotsize=2pt 3]{*-}(-3.65,-2.3)(-3.64,-2.3)

\psline[linewidth=0.1mm,dotsize=2pt 3]{*-}(4.55,0.62)(4.55,0.64)

\psline[linewidth=0.1mm,dotsize=2pt 3]{*-}(0.15,0.61)(0.15,0.63)

\psline[linewidth=0.1mm,dotsize=2pt 3]{*-}(4.61,-0.3)(4.615,-0.3)

\psline[linewidth=0.1mm,dotsize=2pt 3]{*-}(0.09,-0.3)(0.095,-0.3)

\psline[linewidth=0.1mm,dotsize=2pt 3]{*-}(4.02,-1.55)(4.07,-1.55)

\psline[linewidth=0.1mm,dotsize=2pt 3]{*-}(0.68,-1.55)(0.7,-1.55)

\end{pspicture}}
\hrule
\caption{One-point compactification of the Euclidean space ${{\mathbb E}^3}$ by means of a stereographic projection onto ${S^3\in{\rm I\!R}^4}$.}
\vspace{5pt}
\hrule
\label{fig2}
\end{figure}
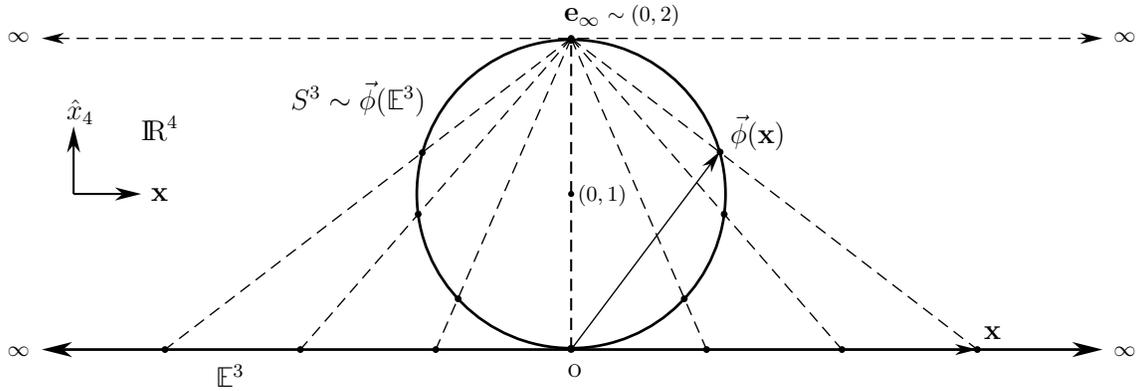

The physical space represented by the above algebraic model is, however, not quite satisfactory. Stemming from an arbitrarily chosen origin, its points run off to infinity along every radial direction \cite{Dorst}. Moreover, there is no reason for these infinitely many infinities --- which can be approached from infinitely many possible different directions --- to be distinct from one another. It is therefore natural to assume that one and the same infinity is encountered along any radial direction, and identify it with a single point. One way to achieve this is by compactifying the space ${{\mathbb E}^3}$ by adding a single point to it at infinity. This well known procedure of one-point compactification is illustrated in Fig.${\,}$\ref{fig2}.

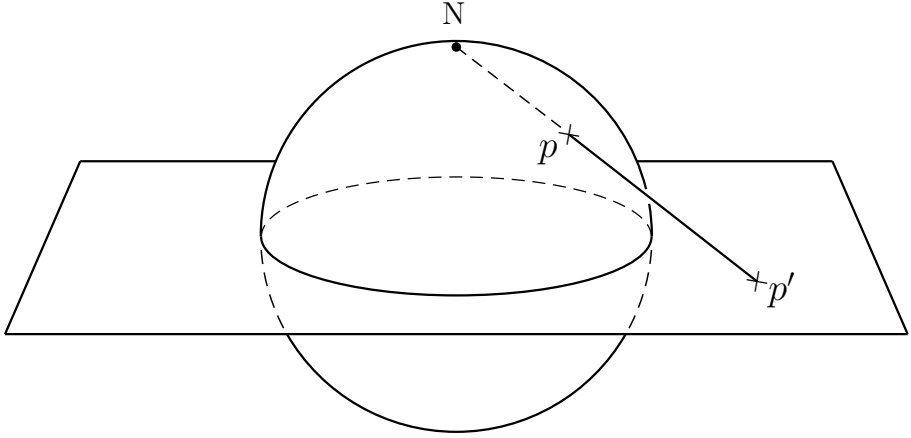
\begin{figure}
\hrule
\scalebox{1}{
\begin{pspicture}(-5.2,-3.7)(3.5,3.2)

\psarc[linewidth=0.2mm,linestyle=dashed](-1.0,-0.45){2.6}{184}{208}

\psarc[linewidth=0.2mm,linestyle=dashed](-1.0,-0.45){2.6}{332}{356}

\psarc[linewidth=0.3mm](-1.0,-0.45){2.6}{0}{9.6}

\psarc[linewidth=0.3mm](-1.0,-0.45){2.6}{14}{180}

\psarc[linewidth=0.3mm](-1.0,-0.45){2.6}{210}{330}

\psellipticarc[linewidth=0.17mm,linestyle=dashed](-1.0,-0.45)(2.6,0.8){0}{180}

\psellipticarc[linewidth=0.3mm](-1.0,-0.45)(2.6105,0.8){180}{360}

\psline[linewidth=0.3mm](-6.0,0.55)(-3.4,0.55)

\psline[linewidth=0.3mm](1.393,0.55)(4.0,0.55)

\psline[linewidth=0.3mm](-7.0,-1.75)(5.0,-1.75)

\psline[linewidth=0.3mm](-7.0,-1.75)(-6.0,0.55)

\psline[linewidth=0.3mm](4.0,0.55)(5.0,-1.75)

\psdot*(-1.0,2.07)

\rput{35}(3.0,-1.05){\large ${\times}$}

\rput{35}(0.5,0.904){\large ${\times}$}

\put(0.1,0.6){{\Large ${p}$}}

\put(-1.19,2.37){{\large ${\rm N}$}}

\put(3.15,-1.3){{\Large ${p'}$}}

\psline[linewidth=0.2mm,linestyle=dashed](-1.0,2.07)(0.39,1.0)

\psline[linewidth=0.3mm](0.5,0.904)(3.0,-1.05)

\end{pspicture}}
\hrule
\caption{Stereographic projection of ${S^2}$ onto the plane of ${{\rm I\!R}^2}$. Both ${S^2}$ and ${{\rm I\!R}^2}$ contain infinite number of points. Each point ${p}$ of ${S^2}$ is mapped to a point ${p'}$ of ${{\rm I\!R}^2}$, except North pole, which has no meaningful finite image under this projection.}
\vspace{5pt}
\label{fig7}
\hrule
\end{figure}

Intuitively this procedure is not difficult to understand with a two-dimensional analogue of ${{\mathbb E}^3}$. Imagine a stretchable balloon, which is topologically a two-dimensional surface, ${S^2}$ [cf. Fig.${\,}$\ref{fig7}]. If we surgically remove a single point from this surface and stretch the remainder out to infinity in every radial direction (like an infinite bed-sheet), then it provides an intuitive model for the two-dimensional Euclidean space, ${{\mathbb E}^2}$. The one-point, or Alexandroff compactification of ${{\mathbb E}^2}$ is an inverse of this process whereby all points at infinity from all possible radial directions in ${{\mathbb E}^2}$ are brought together again and identified with the previously removed point, thereby reconstructing the ${S^2}$-balloon from an ${{\mathbb E}^2}$-bed-sheet.

Similarly, Fig.${\,}$\ref{fig2} depicts an inverse stereographic projection of ${{\mathbb E}^3}$ onto a unit 3-sphere, ${S^3}$, by the embedding map ${\vec{\phi}:{\mathbb E}^3\rightarrow S^3}$, which is given by
\begin{equation}
\vec{\phi}\left({\bf x}\in{\mathbb E}^3\right) = \left(\frac{2}{||{\bf x}||^2+1}\right){\bf x}+\left(\frac{2\,||{\bf x}||^2}{||{\bf x}||^2+1}\right){{\hat x}_4}\,,
\end{equation}
where two of the dimensions of ${{\mathbb E}^3}$ are suppressed in the figure and ${{\hat x}_4}$ represents the fourth dimension in the embedding space ${{\rm I\!R}^4}$. The crucial observation here is that, as an arbitrary vector ${{\bf x}\in {\mathbb E}^3}$ from the origin approaches infinity, it is mapped to the same point ${{\bf e}_{\infty}}$ located at ${(0,2)}$, thereby closing the non-compact space ${{\mathbb E}^3}$ into the compact 3-sphere. By shifting the origin to ${(0,1)}$ the above set of points can be inscribed by a radial 4-vector originating from ${(0,1)}$ as
\begin{equation}
\vec{\psi}\left({\bf x}\in{\mathbb E}^3\right) = \left(\frac{2}{||{\bf x}||^2+1}\right){\bf x}+\left(\frac{2\,||{\bf x}||^2}{||{\bf x}||^2+1}-1\right){{\hat x}_4}\,.
\end{equation}
The magnitude of this vector then confirms the unity of the radius of our conformally embedded 3-sphere within ${{\rm I\!R}^4}$:
\begin{equation}
1=\left|\left|\,\vec{\psi}\left({\bf x}\in{\mathbb E}^3\right)\right|\right| = \text{radius of}\;S^3 \hookrightarrow {\rm I\!R}^4.
\end{equation}
The embedding operator ${\vec{\phi}({\bf x})}$ [or ${\vec{\psi}({\bf x})}$] thus transforms the entire space ${{\mathbb E}^3}$ into a unit 3-sphere within ${{\rm I\!R}^4}$,
thereby accomplishing a one-point compactification of ${{\mathbb E}^3}$:
\begin{equation}
S^3 = \,{\mathbb E}^3 \cup \left\{{\bf e}_{\infty}\right\}.
\end{equation}
Such a conformal mapping is angle-preserving in the sense that a small angle between two curves on ${S^3}$ projects to the same angle between the projected curves on ${{\mathbb E}^3}$, with a circle of any size on ${S^3}$ projecting to an exact circle on ${{\mathbb E}^3}$.

Now the tangent bundle of ${S^3}$ happens to be trivial: ${{\rm T}S^3 = S^3 \times{\rm I\!R}^3}$. This renders the tangent space at each point of ${S^3}$ to be isomorphic to ${{\rm I\!R}^3}$. Consequently, local experiences of the experimenters within ${S^3}$ are no different from those of their counterparts within ${{\mathbb E}^3}$. The global topology of ${S^3}$, however, is clearly different from that of ${{\rm I\!R}^3}$ \cite{disproof,IJTP}. In particular, the triviality of the bundle ${{\rm T}S^3}$ means that ${S^3}$ is parallelizable. As a result, a global {\it anholonomic} frame can be defined on ${S^3}$ that fixes each of its points uniquely. Such a frame renders ${S^3}$ diffeomorphic to the group SU(2) --- {\it i.e.}, to the set of all unit quaternions:
\begin{equation}
S^3:=\left\{\,{\bf q}(\theta,\,{\bf r}):=\cos\frac{\theta}{2}\,+\,{\boldsymbol\xi}({\bf r})\,\sin\frac{\theta}{2}\;\,
\bigg|\;\left|\left|\,{\bf q}(\theta,\,{\bf r})\,\right|\right| = 1\right\}\!, \label{3-sphere}
\end{equation}
where ${{\boldsymbol\xi}({\bf r})}$ is a bivector rotating about ${{\bf r}\in{\rm I\!R}^3}$
with the rotation angle ${\theta}$ in the range ${0\leq\theta < 4\pi}$. In terms of the even sub-algebra of (\ref{cl}), the bivector ${{\boldsymbol\xi}({\bf r})\in S^3}$ can be parameterized by the dual vector ${{\bf r}=r_x\,{\bf e}_x+r_y\,{\bf e}_y+r_z\,{\bf e}_z\in{\rm I\!R}^3}$ as
\begin{align}
{\boldsymbol\xi}({\bf r})\,:=\,(\,I_3\cdot{\bf r}\,)\,&=\,r_x\,(\,I_3\cdot{\bf e}_x\,)
\,+\,r_y\,(\,I_3\cdot{\bf e}_y\,)\,+\,r_z\,(\,I_3\cdot{\bf e}_z\,)\,=\,r_x\;{{\bf e}_y}{{\bf e}_z}\,+\,r_y\;{{\bf e}_z}{{\bf e}_x}\,+\,r_z\;{{\bf e}_x}{{\bf e}_y}\,,
\end{align}
with ${{\boldsymbol\xi}^2({\bf r})=-1}$. Each configuration of any rotating rigid body can thus be represented by a quaternion ${{\bf q}(\theta,\,{\bf r})}$, which in turn can always be decomposed into a product of two bivectors, say ${{\boldsymbol\xi}({\bf u})}$ and ${{\boldsymbol\xi}({\bf v})}$, belonging to an ${S^2\subset S^3}$,
\begin{equation}
{\boldsymbol\xi}({\bf u})\,{\boldsymbol\xi}({\bf v})\,
=\,\cos\frac{\theta}{2}\,+\,{\boldsymbol\xi}({\bf r})\,\sin\frac{\theta}{2}\,, \label{new3}
\end{equation}
in accordance with the bivector subalgebra \cite{Clifford}
\begin{equation}
{\boldsymbol \xi}_{a}\,{\boldsymbol \xi}_{b} \,=\,-\,\delta_{ab}\,-\sum^{3}_{c\,=\,1}\epsilon_{abc}\;{\boldsymbol \xi}_{c}\,,
\end{equation} 
with ${\theta}$ being its rotation angle from ${{\bf q}(0,\,{\bf r})=1}$. Note also that ${{\bf q}(\theta,\,{\bf r})}$ reduces to ${\pm\,1}$ as ${\theta\rightarrow\,2\kappa\pi}$ for ${\kappa\,=\,0,\,1,\;\text{or}\,\;2}$.

\subsection{Conformal Completion of the Euclidean Primitives}

Our interest now lies in the point ${{\bf e}_{\infty}}$ which represents the multitude of infinities of ${{\mathbb E}^3}$. Within three dimensions we continue to view it as a dimensionless point and take its algebraic counterpart to be a {\it non-zero} vector of zero norm:
\begin{equation}
{\bf e}_{\infty}\not=\,\bf{0},\;\,\text{but}\;\,\left|\left|{\bf e}_{\infty}\right|\right|^{2}=\,{\bf e}_{\infty}\cdot{\bf e}_{\infty}=\,0\iff\,{\bf e}^2_{\infty}=\,0\,.
\end{equation}
Such a vector that is orthogonal to itself is called a {\it null vector} in Conformal Geometric Algebra\footnote{\label{Conformal}The conformal space we are considering is an {\it in}-homogeneous version of the space usually studied in Conformal Geometric Algebra \cite{Dorst}. It can be viewed as an ${8}$-dimensional subspace of the ${32}$-dimensional representation space postulated in Conformal Geometric Algebra. The larger representation space results from a homogeneous freedom of the origin within ${{\mathbb E}^3}$, which is neither necessary nor useful for our purposes here.} \cite{Clifford}. It is introduced to represent both finite points in space as well as points at infinity \cite{Dorst}. Since points thus defined are null-dimensional or dimensionless, addition of ${{\bf e}_{\infty}}$ into the algebraic structure of ${{\mathbb E}^3}$ does not alter the latter's dimensions but only its point-set topology, rendering it diffeomorphic to a closed, compact, simply-connected 3-sphere, as we discussed above.

Equipped with ${{\bf e}_{\infty}}$, we are now ready to rebuild the compactified Euclidean space and its algebraic representation as follows. We begin by identifying the set ${\left\{\,{\bf e}_x{\bf e}_y,\;{\bf e}_z{\bf e}_x,\;{\bf e}_y{\bf e}_z\right\}}$ of bivectors as the orthonormal basis of the space ${{\mathbb E}^3}$:
\begin{equation}
{\mathbb E}^3=\,{\rm span}\!\left\{\,{\bf e}_x{\bf e}_y,\,{\bf e}_z{\bf e}_x,\,{\bf e}_y{\bf e}_z\right\}\!.\label{espa}
\end{equation}
Using the orthonormality and anti-commutativity of the vectors ${{\bf e}_i}$ the product of the basis bivectors works out to be
\begin{equation}
({\bf e}_x{\bf e}_y)({\bf e}_z{\bf e}_x)({\bf e}_y{\bf e}_z)={\bf e}_x{\bf e}_y{\bf e}_z{\bf e}_x{\bf e}_y{\bf e}_z=-1.
\end{equation}
The associativity of geometric product then allows us to rediscover the volume form ${I_3}$ for the Euclidean space (\ref{espa}):
\begin{equation}
({\bf e}_x{\bf e}_y)({\bf e}_z{\bf e}_x)({\bf e}_y{\bf e}_z)=\,({\bf e}_x{\bf e}_y{\bf e}_z)({\bf e}_x{\bf e}_y{\bf e}_z)=\,({\bf e}_x{\bf e}_y{\bf e}_z)^2=:\,(I_3)^2=-1.
\end{equation}
As it stands, this volume form is open and has the topology of ${{\rm I\!R}^3}$. But we can now close it with the null vector ${{\bf e}_{\infty}}$:
\begin{equation}
I_c:=\,I_3\,{\bf e}_{\infty}=\,{\bf e}_x{\bf e}_y{\bf e}_z{\bf e}_{\infty}\,,
\end{equation}
where we have used the subscript ${c}$ on ${I_c}$ to indicate that it is a volume element of the compact 3-sphere, ${S^3}$. As we noted earlier, in the Euclidean space the reverse of ${I_3}$ is ${I_3^{\dagger}=-I_3}$. Likewise in the conformal space the reverse of ${I_c}$ is
\begin{equation}
I^{\dagger}_c=\,I_3^{\dagger}{\bf e}_{\infty}=\,-I_3\,{\bf e}_{\infty}=\,-I_c\,.
\end{equation}
As a result, in the conformal space the general duality operation between elements ${\widetilde{\Omega}}$ and ${\Omega}$ of any grade is given by
\begin{equation}
\widetilde{\Omega}_c:=\Omega\,I^{\dagger}_c=\,\Omega\,I^{\dagger}_3\,{\bf e}_{\infty}\,. \label{duality}
\end{equation}
This allows us, in particular, to work out the dual elements of all of the basis bivectors in (\ref{espa}) in the conformal space:
\begin{align}
{\bf e}_x{\bf e}_yI^{\dagger}_3{\bf e}_{\infty}=\,{\bf e}_x{\bf e}_y{\bf e}_z{\bf e}_y{\bf e}_x{\bf e}_{\infty}&=\,{\bf e}_z{\bf e}_{\infty}\,, \\
{\bf e}_z{\bf e}_xI^{\dagger}_3{\bf e}_{\infty}=\,{\bf e}_z{\bf e}_x{\bf e}_z{\bf e}_y{\bf e}_x{\bf e}_{\infty}&=\,{\bf e}_y{\bf e}_{\infty}\,, \\
\text{and}\;\;{\bf e}_y{\bf e}_zI^{\dagger}_3{\bf e}_{\infty}=\,{\bf e}_y{\bf e}_z{\bf e}_z{\bf e}_y{\bf e}_x{\bf e}_{\infty}&=\,{\bf e}_x{\bf e}_{\infty}\,.
\end{align}
Moreover, analogous to how the dual of ${+1}$ in ${{\mathbb E}^3}$ is ${-I_3\,}$, the dual of ${+1}$ in the conformal space also works out to be 
\begin{equation}
(+1)\,I^{\dagger}_c = -I_3\,{\bf e}_{\infty} = -I_c\,.
\end{equation}
We have thus worked out the conformal counterparts of all of the basis elements appearing in the algebraic vector space (\ref{cl}). Putting them together we can now formalize the desired algebraic representation of our conformal space as 
\begin{equation}
{\cal K}^-=\,{\rm span}\!\left\{\,1,\,{\bf e}_x{\bf e}_y,\,{\bf e}_z{\bf e}_x,\,{\bf e}_y{\bf e}_z,\,{\bf e}_x{\bf e}_{\infty},\,
{\bf e}_y{\bf e}_{\infty},\,{\bf e}_z{\bf e}_{\infty},\,-I_3{\bf e}_{\infty}\,\right\}\!. \label{RepR}
\end{equation}
Evidently, not unlike (\ref{cl}), this vector space too is eight-dimensional. Unlike (\ref{cl}), however, it is closed and compact. The three-dimensional physical space --- {\it i.e.}, the compact 3-sphere we discussed above --- can now be viewed as embedded in the four-dimensional ambient space, ${{\rm I\!R}^4}$, as depicted in Fig.${\,}$\ref{fig2}. In this higher dimensional space ${{\bf e}_{\infty}}$ is then a {\it unit} vector,
\begin{equation}
\left|\left|{\bf e}_{\infty}\right|\right|^{2}=\,{\bf e}_{\infty}\cdot{\bf e}_{\infty}=\,1\iff\,{\bf e}^2_{\infty}=\,1\,,
\end{equation}
and the corresponding algebraic representation space (\ref{RepR}) is nothing but the eight-dimensional {\it even} sub-algebra of the ${2^4=16}$-dimensional Clifford algebra ${Cl_{4,0}}$. Thus a one-dimensional subspace --- represented by the unit vector ${{\bf e}_{\infty}}$ in the ambient space ${{\rm I\!R}^4}$ --- represents a {\it null}-dimensional space --- {\it i.e.}, the infinite point of ${{\mathbb E}^3}$ --- in the physical space ${S^3}$.

\subsection{Orientation of Representation Space as a Binary Degree of Freedom}

Before we explore the properties of the above vector space, let us endow it with one more degree of freedom without which it is unjustifiably restrictive. To that end, we first recall what is meant by an orientation of a vector space \cite{Milnor}:

{\bf Definition of Orientation}: An orientation of a finite dimensional vector space ${{\cal V}_n}$ is an equivalence class of ordered basis, say ${\left\{b_1,\,\dots,\,b_n\right\}}$, which determines the same orientation of ${\,{\cal V}_n}$ as the basis ${\left\{b'_1,\,\dots,\,b'_n\right\}}$ if ${b'_i =  \omega_{ij}\, b_j}$ holds with ${{\rm det}(\omega_{ij})>0}$, and the opposite orientation of ${{\cal V}_n}$ as the basis ${\left\{b'_1,\,\dots,\,b'_n\right\}}$ if ${b'_i = \omega_{ij}\, b_j}$ holds with ${{\rm det}(\omega_{ij}) < 0}$.

Thus each positive dimensional real vector space has precisely two possible orientations, which we will denote as ${\lambda=+1}$ or ${\lambda=-1}$. More generally an oriented smooth manifold consists of that manifold together with a choice of orientation for each of its tangent spaces. It is worth noting that orientation is a {\it relative} concept. The orientation of a tangent space ${{\cal V}_n}$ of a manifold defined by the equivalence class of ordered basis such as ${\{b_1,\,\dots,\,b_n\}}$ is meaningful only with respect to that defined by the equivalence class of ordered basis ${\{b'_1,\,\dots,\,b'_n\}}$, and vice versa.

Now in geometric algebra the choice of the sign of the unit pseudoscalar amounts to choosing an orientation of the space \cite{Clifford,Dorst}. In our three-dimensional Euclidean space defined in (\ref{espa}) with an orthonormal set of unit bivector basis, ${I_3=\,{{\bf e}_x}{{\bf e}_y}{{\bf e}_z}}$ picks out the right-handed orientation for ${{\mathbb E}^3}$. The convention usually is to assume such a right-handed set of basis bivectors (or vectors) {\it ab initio}. But the algebra itself does not fix the handedness of the basis. In our presentation above we could have equally well started out with a left-handed set of bivectors in (\ref{espa}) by letting ${-I_3}$ instead of ${+I_3}$ select the basis. Instead of the representation space (\ref{RepR}) we would have then ended up with the space
\begin{equation}
{\cal K}^+=\,{\rm span}\!\left\{\,1,\,+{\bf e}_x{\bf e}_y,\,+{\bf e}_z{\bf e}_x,\,+{\bf e}_y{\bf e}_z,\,+{\bf e}_x{\bf e}_{\infty},\,
+{\bf e}_y{\bf e}_{\infty},\,+{\bf e}_z{\bf e}_{\infty},\,+I_3{\bf e}_{\infty}\,\right\}\!.
\end{equation}
On the other hand, in the light of the above definition of orientation, the representation space (\ref{RepR}) can be written as 
\begin{equation}
{\cal K}^-=\,{\rm span}\!\left\{\,1,\,-{\bf e}_x{\bf e}_y,\,-{\bf e}_z{\bf e}_x,\,-{\bf e}_y{\bf e}_z,\,-{\bf e}_x{\bf e}_{\infty},\,
-{\bf e}_y{\bf e}_{\infty},\,-{\bf e}_z{\bf e}_{\infty},\,-I_3{\bf e}_{\infty}\,\right\}\!.
\end{equation}
It is easy to verify that the bases of ${{\cal K}^+}$ and ${{\cal K}^-}$ are indeed related by an ${8\times 8}$ diagonal matrix whose determinant is ${(-1)^7 < 0}$. Consequently, ${{\cal K}^+}$ and ${{\cal K}^-}$ indeed represent right-oriented and left-oriented vector spaces, respectively, in accordance with our definition of orientation. We can therefore leave the orientation unspecified and write ${{\cal K}^{\pm}}$ as 
\begin{equation}
{\cal K}^{\lambda}=\,{\rm span}\!\left\{\,1,\,\lambda{\bf e}_x{\bf e}_y,\,\lambda{\bf e}_z{\bf e}_x,\,\lambda{\bf e}_y{\bf e}_z,\,\lambda{\bf e}_x{\bf e}_{\infty},\,\lambda{\bf e}_y{\bf e}_{\infty},\,\lambda{\bf e}_z{\bf e}_{\infty},\,\lambda I_3{\bf e}_{\infty}\,\right\}\!,\,\;\lambda^2=1\iff\lambda=\pm1. \label{clind}
\end{equation}

\begin{table}[t]
\hrule
\vspace{18pt}
\begin{center}
{\def\arraystretch{1.7}\tabcolsep=4.9pt
 \begin{tabular}[b]{@{} | Sc!{\vrule width 1.32pt} Sc | Sc | Sc | Sc | Sc | Sc | Sc | Sc | @{}} \hline
${*}$ &${1}$ &${\lambda\,{\bf e}_x{\bf e}_y}$ &${\lambda\,{\bf e}_z{\bf e}_x}$ &${\lambda\,{\bf e}_y{\bf e}_z}$ &${\lambda\,{\bf e}_x{\bf e}_{\infty}}$ &${\lambda\,{\bf e}_y{\bf e}_{\infty}}$ &${\lambda\,{\bf e}_z{\bf e}_{\infty}}$ &${\lambda\,I_3{\bf e}_{\infty}}$ \\ \hlineB{3.5}
${1}$ &${1}$ &${\lambda\,{\bf e}_x{\bf e}_y}$ &${\lambda\,{\bf e}_z{\bf e}_x}$ &${\lambda\,{\bf e}_y{\bf e}_z}$ &${\lambda\,{\bf e}_x{\bf e}_{\infty}}$ &${\lambda\,{\bf e}_y{\bf e}_{\infty}}$ &${\lambda\,{\bf e}_z{\bf e}_{\infty}}$ &${\lambda\,I_3{\bf e}_{\infty}}$ \\ \hline
${\lambda\,{\bf e}_x{\bf e}_y}$ &${\lambda\,{\bf e}_x{\bf e}_y}$ &${-1}$ &${{\bf e}_y{\bf e}_z}$ &${-{\bf e}_z{\bf e}_x}$ &${-{\bf e}_y{\bf e}_{\infty}}$ &${{\bf e}_x{\bf e}_{\infty}}$ &${I_3{\bf e}_{\infty}}$ &${-{\bf e}_z{\bf e}_{\infty}}$ \\ \hline
${\lambda\,{\bf e}_z{\bf e}_x}$ &${\lambda\,{\bf e}_z{\bf e}_x}$ &${-{\bf e}_y{\bf e}_z}$ &${-1}$ &${{\bf e}_x{\bf e}_y}$ &${{\bf e}_z{\bf e}_{\infty}}$ &${I_3{\bf e}_{\infty}}$ &${-{\bf e}_x{\bf e}_{\infty}}$ &${-{\bf e}_y{\bf e}_{\infty}}$ \\ \hline
${\lambda\,{\bf e}_y{\bf e}_z}$ &${\lambda\,{\bf e}_y{\bf e}_z}$ &${{\bf e}_z{\bf e}_x}$ &${-{\bf e}_x{\bf e}_y}$ &${-1}$ &${I_3{\bf e}_{\infty}}$ &${-{\bf e}_z{\bf e}_{\infty}}$ &${{\bf e}_y{\bf e}_{\infty}}$ &${-{\bf e}_x{\bf e}_{\infty}}$ \\ \hline
${\lambda\,{\bf e}_x{\bf e}_{\infty}}$ &${\lambda\,{\bf e}_x{\bf e}_{\infty}}$ &${{\bf e}_y{\bf e}_{\infty}}$ &${-{\bf e}_z{\bf e}_{\infty}}$ &${I_3{\bf e}_{\infty}}$ &${-1}$ &${-{\bf e}_x{\bf e}_y}$ &${{\bf e}_z{\bf e}_x}$ &${-{\bf e}_y{\bf e}_z}$ \\ \hline
${\lambda\,{\bf e}_y{\bf e}_{\infty}}$ &${\lambda\,{\bf e}_y{\bf e}_{\infty}}$ &${-{\bf e}_x{\bf e}_{\infty}}$ &${I_3{\bf e}_{\infty}}$ &${{\bf e}_z{\bf e}_{\infty}}$ &${{\bf e}_x{\bf e}_y}$ &${-1}$ &${-{\bf e}_y{\bf e}_z}$ &${-{\bf e}_z{\bf e}_x}$ \\ \hline
${\lambda\,{\bf e}_z{\bf e}_{\infty}}$ &${\lambda\,{\bf e}_z{\bf e}_{\infty}}$ &${I_3{\bf e}_{\infty}}$ &${{\bf e}_x{\bf e}_{\infty}}$ &${-{\bf e}_y{\bf e}_{\infty}}$ &${-{\bf e}_z{\bf e}_x}$ &${{\bf e}_y{\bf e}_z}$ &${-1}$ &${-{\bf e}_x{\bf e}_y}$ \\ \hline
${\lambda\,I_3{\bf e}_{\infty}}$ &${\lambda\,I_3{\bf e}_{\infty}}$ &${-{\bf e}_z{\bf e}_{\infty}}$ &${-{\bf e}_y{\bf e}_{\infty}}$ &${-{\bf e}_x{\bf e}_{\infty}}$ &${-{\bf e}_y{\bf e}_z}$ &${-{\bf e}_z{\bf e}_x}$ &${-{\bf e}_x{\bf e}_y}$ &${1}$ \\
\hline
 \end{tabular}%
}
\end{center}
\vspace{5pt}
\hrule
\caption{Multiplication Table for a ``Conformal Geometric Algebra\textsuperscript{\ref{Conformal}}" of ${{\mathbb E}^3}$. Here ${I_3={\bf e}_x{\bf e}_y{\bf e}_z}$, ${{\bf e}_{\infty}^2=+1}$, and ${\lambda =\pm 1}$.\break}
\vspace{5pt}
\hrule
\label{T+1}
\end{table}

\subsection{Representation Space ${{\cal K}^{\lambda}}$ Remains Closed Under Multiplication}\label{1D}

As an eight-dimensional linear vector space, ${{\cal K}^{\lambda}}$ has some remarkable properties. To begin with, ${{\cal K}^{\lambda}}$ is {\it closed} under multiplication. Suppose ${\bf X}$ and ${\bf Y}$ are two unit vectors in ${{\cal K}^{\lambda}}$. Then ${\bf X}$ and ${\bf Y}$ can be expanded in the basis of ${{\cal K}^{\lambda}}$ as
\begin{equation}
{\bf X}=\,X_0+X_1\,\lambda{\bf e}_x{\bf e}_y+X_2\,\lambda{\bf e}_z{\bf e}_x+X_3\,\lambda{\bf e}_y{\bf e}_z+X_4\,\lambda{\bf e}_x{\bf e}_{\infty}+X_5\,\lambda{\bf e}_y{\bf e}_{\infty}+X_6\,\lambda{\bf e}_z{\bf e}_{\infty}+X_7\,\lambda I_3{\bf e}_{\infty}
\end{equation}
and
\begin{equation}
{\bf Y}=\,Y_0+Y_1\,\lambda{\bf e}_x{\bf e}_y+Y_2\,\lambda{\bf e}_z{\bf e}_x+Y_3\,\lambda{\bf e}_y{\bf e}_z+Y_4\,\lambda{\bf e}_x{\bf e}_{\infty}+Y_5\,\lambda{\bf e}_y{\bf e}_{\infty}+Y_6\,\lambda{\bf e}_z{\bf e}_{\infty}+Y_7\,\lambda I_3{\bf e}_{\infty}\,,
\end{equation}
and using (\ref{s-norm}) they can be normalized as
\begin{equation}
||{\bf X}||^2=\sum_{\mu\,=\,0}^{7} \;X_{\mu}^2\,=\,1\;\;\;\text{and}\;\;\;||{\bf Y}||^2=\sum_{\nu\,=\,0}^{7} \;Y_{\nu}^2\,=\,1\,.
\end{equation}
Now it is evident from the multiplication table above (Table \ref{T+1}) that if ${{\bf X},{\bf Y}\in{\cal K}^{\lambda}}$, then so is their product ${{\bf Z}={\bf X}{\bf Y}}$:
\begin{equation}
{\bf Z}=\,Z_0+Z_1\,\lambda{\bf e}_x{\bf e}_y+Z_2\,\lambda{\bf e}_z{\bf e}_x+Z_3\,\lambda{\bf e}_y{\bf e}_z+Z_4\,\lambda{\bf e}_x{\bf e}_{\infty}+Z_5\,\lambda{\bf e}_y{\bf e}_{\infty}+Z_6\,\lambda{\bf e}_z{\bf e}_{\infty}+Z_7\,\lambda I_3{\bf e}_{\infty}={\bf X}{\bf Y}.
\end{equation}
More importantly, we shall soon see that for vectors ${\bf X}$ and ${\bf Y}$ in ${{\cal K}^{\lambda}}$ (not necessarily unit) the following relation holds:
\begin{equation}
||{\bf X}{\bf Y}|| \,=\, ||{\bf X}||\;||{\bf Y}||\,. \label{norms}
\end{equation}
In particular, this means that for any two unit vectors ${\bf X}$ and ${\bf Y}$ in ${{\cal K}^{\lambda}}$ with the geometric product ${{\bf Z}={\bf X}{\bf Y}}$ we have
\begin{equation}
||\,{\bf Z}\,||^2=\sum_{\rho\,=\,0}^{7} \;Z_{\rho}^2\,=\,1\,.
\end{equation}
One of the important observations here is that, without loss of generality, we can restrict our representation space to a set of {\it unit} vectors in ${{\cal K}^{\lambda}}$. We are then dealing with a unit 7-sphere as an algebraic representation of the compactified physical space (\ref{3-sphere}). If, for convenience, we now identify the basis elements of ${{\cal K}^{\lambda}}$ (in order) with the ordered elements of the following set 
\begin{equation}
\left\{\,{\boldsymbol \zeta_0},\,{\boldsymbol \zeta_1},\,{\boldsymbol \zeta_2},\,{\boldsymbol \zeta_3},\,{\boldsymbol \zeta_4},\,{\boldsymbol \zeta_5},\,{\boldsymbol \zeta_6},\,{\boldsymbol \zeta_7}\,\right\}, \label{newbasis}
\end{equation}
then the algebra generated by them --- which has been explicitly displayed in Table \ref{T+1} --- can be succinctly rewritten as
\begin{equation}
{\boldsymbol \zeta}_{\mu}\,{\boldsymbol \zeta}_{\nu} \,=\,\left\{-1\right\}^{\delta_{\mu 7}}\left\{\,-\,\delta_{\mu\nu}\right\}\,+\,\lambda\sum_{\rho\,=\,1}^7\left[\,f_{\mu\nu\rho}\,+\,\left\{-1\right\}^{\delta_{\rho 7}}\,l_{\mu\nu\rho}\,\right]{\boldsymbol \zeta}_{\rho}\,,\;\;\;\;\mu,\,\nu\,=\,1,\,2,\,\dots,\,7, \label{neoprod}
\end{equation}
where ${f_{\mu\nu\rho}}$ is a totally anti-symmetric permutation tensor with only non-vanishing independent components being
\begin{equation}
f_{123}\,=\,f_{246}\,=\,f_{365}\,=\,f_{415}\,=\,+1\,,
\end{equation}
and similarly ${l_{\mu\nu\rho}}$ is a totally {\it symmetric} permutation tensor with only non-vanishing independent components being
\begin{equation}
l_{176}\,=\,l_{257}\,=\,l_{347}\,=\,-1\,.
\end{equation}
The 8-dimensional multi-vectors ${\bf X}$ and ${\bf Y}$ within ${{\cal K}^{\lambda}}$ can now be expanded more conveniently in the basis (\ref{newbasis}) as
\begin{equation}
{\bf X}\,=\,\sum_{\mu\,=\,0}^{7} \;X_{\mu}\,{\boldsymbol \zeta}_{\mu}\;\;\;\;\;\text{and}\;\;\;\;\;{\bf Y}\,=\,\sum_{\nu\,=\,0}^{7} \;Y_{\nu}\,{\boldsymbol \zeta}_{\nu}\,.
\end{equation}

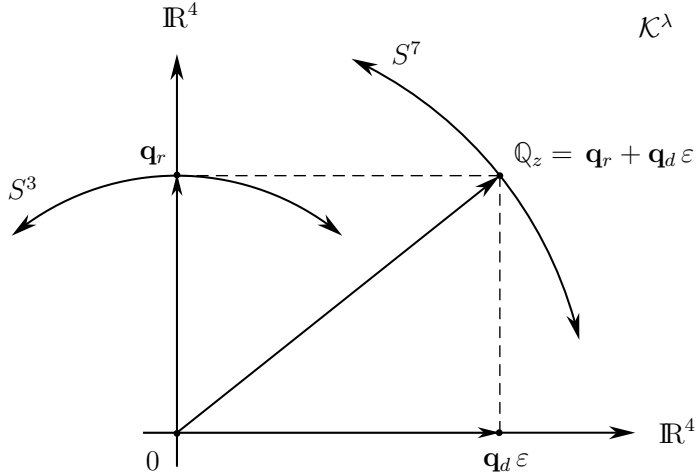
\begin{figure}
\hrule
\scalebox{0.9}{
\begin{pspicture}(0.5,-3.5)(4.2,4.7)

\psline[linewidth=0.3mm,arrowinset=0.3,arrowsize=3pt 3,arrowlength=2]{->}(-1.27,-2.3)(6.0,-2.3)

\psline[linewidth=0.3mm,arrowinset=0.3,arrowsize=3pt 3,arrowlength=2]{->}(-0.77,-2.8)(-0.77,3.3)

\psline[linewidth=0.2mm,arrowinset=0.3,arrowsize=3pt 3,arrowlength=2]{->}(-1.27,-2.3)(4.0,-2.3)

\psline[linewidth=0.1mm,dotsize=2pt 3]{*-}(3.99,-2.3)(4.01,-2.3)

\psline[linewidth=0.2mm,arrowinset=0.3,arrowsize=3pt 3,arrowlength=2]{->}(-0.77,-2.8)(-0.77,1.53)

\psline[linewidth=0.1mm,dotsize=2pt 3]{*-}(-0.77,1.51)(-0.77,1.52)

\psline[linewidth=0.3mm,arrowinset=0.3,arrowsize=3pt 3,arrowlength=2]{->}(-0.77,-2.3)(4.0,1.5)

\psarc[linewidth=0.3mm,arrowinset=0.3,arrowsize=3pt 3,arrowlength=2]{<->}(-0.77,-2.3){6.1}{13}{65}

\put(2.4,3.15){\large ${S^7}$}

\psarc[linewidth=0.3mm,arrowinset=0.3,arrowsize=3pt 3,arrowlength=2]{<->}(-0.77,-2.3){3.8}{50}{130}

\put(-3.27,1.15){\large ${S^3}$}

\psline[linewidth=0.1mm,dotsize=2pt 3]{*-}(4.0,1.5)(4.0,1.55)

\psline[linewidth=0.1mm,dotsize=2pt 3]{*-}(-0.77,-2.31)(-0.77,-2.27)

\psline[linewidth=0.2mm,linestyle=dashed,arrowinset=0.3,arrowsize=3pt 3,arrowlength=2]{-}(-0.77,1.5)(4.0,1.5)

\psline[linewidth=0.2mm,linestyle=dashed,arrowinset=0.3,arrowsize=3pt 3,arrowlength=2]{-}(4.0,-2.3)(4.0,1.5)

\put(6.05,3.55){\large ${{\cal K}^{\lambda}}$}

\put(4.2,1.7){\large ${{\mathbb Q}_z =\, {\bf q}_r + {\bf q}_d\,\varepsilon}$}

\put(3.75,-2.80){\large ${{\bf q}_d\,\varepsilon}$}

\put(-1.32,1.75){\large ${{\bf q}_r}$}

\put(6.35,-2.45){\large ${{\rm I\!R}^4}$}

\put(-0.99,3.7){\large ${{\rm I\!R}^4}$}

\put(-1.23,-2.85){\large ${0}$}

\end{pspicture}}
\hrule
\caption{An illustration of the 8D plane of ${{\cal K}^{\lambda}}$, which may be interpreted as an Argand diagram for a pair of quaternions.}
\vspace{5pt}
\hrule
\label{fig3333}
\end{figure}

\subsection{Representation Space as a Set of Orthogonal Pairs of Quaternions}

In his seminal works Clifford introduced the concept of dual numbers, ${z}$, analogous to complex numbers, as follows:
\begin{equation}
z = r + d\,\varepsilon, \;\;\text{where}\;\,\varepsilon\not=0 \;\,\text{but}\;\, \varepsilon^2=0\,.
\end{equation}
Here ${\varepsilon}$ is the dual operator, ${r}$ is the real part, and ${d}$ is the dual part \cite{Dorst}. Similar to how the ``imaginary" operator ${i}$ is introduced in the complex number theory to distinguish the ``real" and ``imaginary" parts of a complex number, Clifford introduced the dual operator ${\varepsilon}$ to distinguish the ``real" and ``dual" parts of a dual number. The dual number theory can be extended to numbers of higher grades, including to numbers of composite grades, such as quaternions:
\begin{equation}
{\mathbb Q}_z =\, {\bf q}_r + {\bf q}_d\,\varepsilon\,,
\end{equation}
where ${{\bf q}_r}$ and ${{\bf q}_d}$ are quaternions and ${{\mathbb Q}_z}$ is a dual-quaternion (or in Clifford's terminology, ${{\mathbb Q}_z}$ is a bi-quaternion). Recall that, as defined in (\ref{3-sphere}), the set of all quaternions is a 3-sphere, which can be normalized to radius ${\varrho}$ and rewritten as
\begin{equation}
S^3=\left\{\,{\bf q}_r:=\,q_0 + q_1\,\lambda\,{\bf e}_x{\bf e}_y+q_2\,\lambda\,{\bf e}_z{\bf e}_x+q_3\,\lambda\,{\bf e}_y{\bf e}_z \;\Big|\;||{\bf q}_r|| = \varrho \, \right\}.
\end{equation}
Consider now a second, dual copy of the set of quaternions within ${{\cal K}^{\lambda}}$, corresponding to the fixed orientation ${\lambda=+1}$:
\begin{equation}
S^3=\left\{\,{\bf q}_d:=-q_7 + q_6\,{\bf e}_x{\bf e}_y+q_5\,{\bf e}_z{\bf e}_x+q_4\,{\bf e}_y{\bf e}_z \;\Big|\;||{\bf q}_d|| = \varrho \, \right\}. \label{s-r}
\end{equation}
If we now identify ${\varepsilon}$ with the duality operator ${I^{\dagger}_c=-\lambda\,I_3{\bf e}_{\infty}}$ used in (\ref{duality}), then (in the reverse additive order) we have
\begin{align}
\varepsilon &\equiv\,-\lambda\,I_3{\bf e}_{\infty}\;\;\text{with}\;\;\varepsilon^2=+1\;\;(\text{since ${{\bf e}_{\infty}}$ is a unit vector within ${{\cal K}^{\lambda}}$}) \\
\text{and}\;\;\;{\bf q}_d\,\varepsilon &\equiv\, -{\bf q}_d\,\lambda\,I_3{\bf e}_{\infty}=\,q_4\,\lambda\,{\bf e}_x{\bf e}_{\infty}+\,q_5\,\lambda\,{\bf e}_y{\bf e}_{\infty}+\,q_6\,\lambda\,{\bf e}_z{\bf e}_{\infty}+\,q_7\,\lambda\,I_3{\bf e}_{\infty}\,, \label{s-d}
\end{align}
which is a multi-vector ``dual" to the quaternion ${{\bf q}_d}$ at infinity. Note that we continue to write ${\varepsilon}$ as if it were a scalar because it commutes with ${{\bf q}_d}$. Comparing (\ref{s-r}) and (\ref{s-d}) with (\ref{clind}) we can now rewrite ${{\cal K}^{\lambda}}$ as a set of paired quaternions:
\begin{equation}
{\cal K}^{\lambda}=\,\left\{\,{\mathbb Q}_z :=\, {\bf q}_r + {\bf q}_d\,\varepsilon\;\Big|\; ||{\mathbb Q}_z|| = \sqrt{2}\,\varrho \,\right\}.
\end{equation}
Now the normalization of ${{\mathbb Q}_z}$ in fact necessitates that every ${{\bf q}_r}$ be orthogonal to its dual ${{\bf q}_d}$:
\begin{equation}
||{\mathbb Q}_z|| = \sqrt{2}\,\varrho\;\;\Longrightarrow\;\;{\bf q}_r\,{\bf q}^{\dagger}_d+{\bf q}_d\,{\bf q}^{\dagger}_r = 0\,,
\end{equation}
or equivalently, ${\langle\,{\bf q}_r\,{\bf q}^{\dagger}_d\,\rangle_s = 0}$ ({\it i.e.}, ${{\bf q}_r\,{\bf q}^{\dagger}_d}$ is a pure quaternion). We can see this by working out the product of ${{\mathbb Q}_{z}}$ with ${{\mathbb Q}^{\dagger}_{z}}$ while using ${\varepsilon^2=+1}$, which gives
\begin{equation}
{\mathbb Q}_{z}\,{\mathbb Q}^{\dagger}_{z}\,=\;\left({\bf q}_{r}\,{\bf q}^{\dagger}_{r}\,+\,{\bf q}_{d}\,{\bf q}^{\dagger}_{d}\right)\,+\,\left({\bf q}_{r}\,{\bf q}^{\dagger}_{d}\,+\,{\bf q}_{d}\,{\bf q}^{\dagger}_{r}\right)\,\varepsilon\,.
\end{equation}
Now, using the definition of ${\bf q}$ in (\ref{3-sphere}), it is not difficult to see that 
${{\bf q}_{r}\,{\bf q}^{\dagger}_{r}={\bf q}_{d}\,{\bf q}^{\dagger}_{d}=\varrho^2}$, reducing the above product to
\begin{equation}
{\mathbb Q}_{z}\,{\mathbb Q}^{\dagger}_{z}\,=\;2\,\varrho^2\,+\,\left({\bf q}_{r}\,{\bf q}^{\dagger}_{d}\,+\,{\bf q}_{d}\,{\bf q}^{\dagger}_{r}\right)\,\varepsilon\,.
\end{equation}
It is thus clear that for ${{\mathbb Q}_{z}\,{\mathbb Q}^{\dagger}_{z}}$ to be a scalar ${{\bf q}_{r}\,{\bf q}^{\dagger}_{d}+{\bf q}_{d}\,{\bf q}^{\dagger}_{r}}$ must vanish, or equivalently ${{\bf q}_r}$ must be orthogonal to ${{\bf q}_d}$. 

But there is more to the normalization condition ${{\bf q}_{r}\,{\bf q}^{\dagger}_{d}+{\bf q}_{d}\,{\bf q}^{\dagger}_{r}=0}$ then meets the eye. It also leads to the crucial norm relation (\ref{norms}), which is at the very heart of the only possible four normed division algebras associated with the four parallelizable spheres ${S^0}$, ${S^1}$, ${S^3}$, and ${S^7}$ (cf. appendix \ref{ApA}). To verify it, consider a product of two different members of the set ${{\cal K}^{\lambda}}$,
\begin{equation}
{\mathbb Q}_{z1}\,{\mathbb Q}_{z2}\,=\;\left({\bf q}_{r1}\,{\bf q}_{r2}\,+\,{\bf q}_{d1}\,{\bf q}_{d2}\right)\,+\,\left({\bf q}_{r1}\,{\bf q}_{d2}\,+\,{\bf q}_{d1}\,{\bf q}_{r2}\right)\,\varepsilon\,,
\end{equation}
together with their individual definitions
\begin{equation}
{\mathbb Q}_{z1} =\, {\bf q}_{r1} + {\bf q}_{d1}\,\varepsilon\;\;\;\text{and}\;\;\;{\mathbb Q}_{z2} =\, {\bf q}_{r2} + {\bf q}_{d2}\,\varepsilon\,.
\end{equation}
If we now work out the products ${{\mathbb Q}_{z1}{\mathbb Q}^{\dagger}_{z1}}$,
${\,{\mathbb Q}_{z2}{\mathbb Q}^{\dagger}_{z2}}$ and ${({\mathbb Q}_{z1}\,{\mathbb Q}_{z2})({\mathbb Q}_{z1}\,{\mathbb Q}_{z2})^{\dagger}}$, then, thanks to the orthogonality condition ${{\bf q}_{r}\,{\bf q}^{\dagger}_{d}+{\bf q}_{d}\,{\bf q}^{\dagger}_{r}=0}$, the norm relation is not difficult to verify:
\begin{equation}
||{\mathbb Q}_{z1}\,{\mathbb Q}_{z2}||\,=\;||{\mathbb Q}_{z1}||\;||{\mathbb Q}_{z2}||\,.
\end{equation}
Without loss of generality we can now restrict our algebraic representation space ${{\cal K}^{\lambda}}$ to a unit 7-sphere by setting the radius ${\varrho}$ of ${S^3}$ to ${\frac{1}{\sqrt{2}}}$. In what follows ${S^7}$ will provide the conformal\textsuperscript{\ref{Conformal}} counterpart of the algebra ${Cl_{3,0}}$ given in (\ref{cl}):
\begin{equation}
{\cal K}^{\lambda}\supset\,S^7:=\,\left\{\,{\mathbb Q}_z :=\, {\bf q}_r + {\bf q}_d\,\varepsilon\;\Big|\; ||{\mathbb Q}_z||=1\;\,\text{and}\,\;{\bf q}_{r}\,{\bf q}^{\dagger}_{d}+{\bf q}_{d}\,{\bf q}^{\dagger}_{r}=0\,\right\}, \label{spring}
\end{equation}
where ${\varepsilon =-\lambda\,I_3{\bf e}_{\infty}\,}$, ${\,\varepsilon^2=\,{\bf e}^2_{\infty}=\,+1\,}$,
\begin{equation}
{\bf q}_r =\,q_0 + q_1\,\lambda\,{\bf e}_x{\bf e}_y+q_2\,\lambda\,{\bf e}_z{\bf e}_x+q_3\,\lambda\,{\bf e}_y{\bf e}_z\,,\;\;\,
\text{and}\;\;\,{\bf q}_d =\,-q_7 + q_6\,{\bf e}_x{\bf e}_y+q_5\,{\bf e}_z{\bf e}_x+q_4\,{\bf e}_y{\bf e}_z\,,
\end{equation}
so that
\begin{equation}
{\mathbb Q}_z=\,q_0+q_1\,\lambda{\bf e}_x{\bf e}_y+q_2\,\lambda{\bf e}_z{\bf e}_x+q_3\,\lambda{\bf e}_y{\bf e}_z+q_4\,\lambda{\bf e}_x{\bf e}_{\infty}+q_5\,\lambda{\bf e}_y{\bf e}_{\infty}+q_6\,\lambda{\bf e}_z{\bf e}_{\infty}+q_7\,\lambda I_3{\bf e}_{\infty}\,. \label{snons}
\end{equation}
Needless to say, since all Clifford algebras are associative division algebras by definition, unlike the non-associative octonionic algebra the 7-sphere we have constructed here corresponds to an {\it associative} (but of course non-commutative) algebra.

Thus, to summarize this section, we started out with the observation that the correct model of the physical space is provided by the algebra of Euclidean primitives, such as points, lines, planes and volumes, as discovered by Grassmann and Clifford in the ${19^{th}}$ century. We then recognized the need to ``close'' the Euclidean space with a non-zero null vector ${{\bf e}_{\infty}}$ representing its infinities, thereby compactifying ${{\mathbb E}^3}$ to a 3-sphere, ${S^3}$. The corresponding algebraic representation space of ${{\mathbb E}^3}$ then turned out to be a unit 7-sphere, ${S^7}$. It is quite remarkable that ${S^3}$ and ${S^7}$, which are the two spheres associated with the only two non-trivially possible normed division algebras, namely the quaternionic and octonionic algebras \cite{disproof,related}, emerge in this manner from the elementary algebraic properties of the Euclidean primitives (cf. appendix \ref{ApA}). Unlike the non-associative octonionic algebra and the exceptional Lie groups such as ${E_8}$ it gives rise to, however, the compact 7-sphere we have arrived at corresponds to an {\it associative} Clifford (or geometric) algebra \cite{Dechant,Dorst}, as noted above. And yet, as we shall soon see, it is sufficient to explain the origins of {\it all} quantum correlations. It remains to be seen what role, if any, the exceptional groups such as ${G_2}$ and ${E_8}$ may eventually play when the current framework is developed further.

\section{Derivation of Quantum Correlations from Euclidean Primitives}

\subsection{Constructing Measurement Functions in the Manner of Bell}\label{Mfunc}

In order to derive quantum correlations predicted by arbitrary quantum states, our first task is to construct a set of measurement functions of the form:  
\begin{equation}
\pm\,1\,=\,{\mathscr N}
({\bf n},\,\lambda): {\rm I\!R}^3\!\times\Lambda\longrightarrow S^7 \hookrightarrow{\rm I\!R}^8.\label{theoequa}
\end{equation}
These functions describe {\it local} detections of binary measurement results, ${{\mathscr N}({\bf n},\,\lambda)=\pm1}$, by some analyzers fixed along freely chosen directions represented by the vectors ${\bf n}$. They are of the same realistic and deterministic form as that considered by Bell\footnote{\label{Bellfoot}Readers not familiar with Bell's locally causal (or local-realistic) framework \cite{Bell-1964,GHZ} may benefit from reviewing it from the appendix of Ref.${\,}$\cite{local} before proceeding further.} \cite{Bell-1964,GHZ}, except for their locally unobservable co-domain, which we have taken to be the algebraic representation space ${S^7}$ constructed above, embedded in ${{\rm I\!R}^8}$. For an explicit construction of the functions ${{\mathscr N}({\bf n},\,\lambda)}$, let us consider the following multi-vector in ${{\rm I\!R}^8}$ analogous to (\ref{snons}):
\begin{align}
{\mathbb N}_z&=\,n_0+\{n_1\,\lambda{\bf e}_x{\bf e}_y+n_2\,\lambda{\bf e}_z{\bf e}_x+n_3\,\lambda{\bf e}_y{\bf e}_z\}+\{n_4\,\lambda{\bf e}_x{\bf e}_{\infty}+n_5\,\lambda{\bf e}_y{\bf e}_{\infty}+n_6\,\lambda{\bf e}_z{\bf e}_{\infty}\}+n_7\,\lambda I_3{\bf e}_{\infty} \label{64444} \\
&\equiv\,n_0+\lambda\,{\boldsymbol \xi}({\bf n}_r)\,+\,\lambda\,{\boldsymbol \xi}({\bf n}_d)\,\varepsilon_+\,-\,\lambda\,n_7\,\varepsilon_+ \\
&\equiv\,n_0+\lambda\,{\bf D}({\bf n}_r,\,{\bf n}_d,\,-n_7)\,, \label{mandef}
\end{align}
where
\begin{align}
{\bf D}({\bf n}_r,\,{\bf n}_d,\,-n_7)\,&:=\,{\boldsymbol \xi}({\bf n}_r)\,+\,{\boldsymbol \xi}({\bf n}_d)\,\varepsilon_+\,-\,n_7\,\varepsilon_+\,, \\
S^3\ni\text{bivector}\;{\boldsymbol \xi}({\bf n}_r)\,&:=\,n_1\,{\bf e}_x{\bf e}_y+\,n_2\,{\bf e}_z{\bf e}_x+\,n_3\,{\bf e}_y{\bf e}_z\,\equiv\,I_3\cdot{\bf n}_r\,, \\
S^3\ni\text{bivector}\;{\boldsymbol \xi}({\bf n}_d)\,&:=\,n_6\,{\bf e}_x{\bf e}_y+\,n_5\,{\bf e}_z{\bf e}_x+\,n_4\,{\bf e}_y{\bf e}_z\,\equiv\,I_3\cdot{\bf n}_d\,, \\
\text{pseudoscalar}\;\;\varepsilon_+\,&:=\,-\,I_3{\bf e}_{\infty}\,,
\end{align}
\begin{equation}
\text{3D vector}\;\;{\bf n}_r\,:=\,n_3\,{\bf e}_x\,+\,n_2\,{\bf e}_y\,+\,n_1\,{\bf e}_z\,,\;\;\;\text{with}\;\;\;||{\bf n}_r||=\sqrt{n^2_1+n^2_2+n^2_3}\,=\frac{1}{\sqrt{2}}\,, \label{71-root}
\end{equation}
and
\begin{equation}
\text{3D vector}\;\;{\bf n}_d\,:=\,n_4\,{\bf e}_x\,+\,n_5\,{\bf e}_y\,+\,n_6\,{\bf e}_z\,, \;\;\;\text{with}\;\;\;||{\bf n}_d||=\sqrt{n^2_4+n^2_5+n^2_6}\,=\frac{1}{\sqrt{2}}\,. 
\end{equation}
Next, consider the non-scalar part ${{\bf N}({\bf n}_r,\,{\bf n}_d,\,-n_7,\,\lambda)}$ of the above ${S^7}$-vector,
\begin{equation}\tag{\ref{mandef}}
{\mathbb N}_z\,=\,n_0\,+\,{\bf N}({\bf n}_r,\,{\bf n}_d,\,-n_7,\,\lambda),
\end{equation}
so that
\begin{equation}
{\bf N}({\bf n}_r,\,{\bf n}_d,\,-n_7,\,\lambda)\,=\,\lambda\,{\bf D}({\bf n}_r,\,{\bf n}_d,\,-n_7)\;\;\Longleftrightarrow\;\,{\bf D}({\bf n}_r,\,{\bf n}_d,\,-n_7)\,=\,\lambda\,{\bf N}({\bf n}_r,\,{\bf n}_d,\,-n_7,\,\lambda)\,, \label{ori-bela}
\end{equation}
since ${\lambda^2=1}$. For our purposes it is suffice to represent the detectors with the special case of this non-scalar part for which ${n_7 \equiv 0\,}$:
\begin{equation}
{\bf N}({\bf n}_r,\,{\bf n}_d,\,0,\,\lambda)\,=\,\lambda\,{\bf D}({\bf n}_r,\,{\bf n}_d,\,0)\;\;\;\,\Longleftrightarrow\;\;\;{\bf D}({\bf n}_r,\,{\bf n}_d,\,0)\,=\,\lambda\,{\bf N}({\bf n}_r,\,{\bf n}_d,\,0,\,\lambda)\,, \label{ori-rela}
\end{equation}
where
\begin{equation}
{\bf D}({\bf n}_r,\,{\bf n}_d,\,0)\,:=\,{\boldsymbol \xi}({\bf n}_r)\,+\,{\boldsymbol \xi}({\bf n}_d)\,\varepsilon_+\,=\,(I_3\cdot{\bf n}_r)\,+\,(I_3\cdot{\bf n}_d)\,\varepsilon_+\,=\,I_3\cdot\left\{\,{\bf n}_r+\,{\bf n}_d\,\varepsilon_+\right\}. \label{det}
\end{equation}

Next recall that, although global topology of ${S^3}$ is different from that of ${{\rm I\!R}^3}$, local experiences of experimenters within ${S^3}$ are no different from those of their counterparts within ${{\rm I\!R}^3}$, not the least because the tangent space at any point of ${S^3}$ is isomorphic to ${{\rm I\!R}^3}$. With this in mind, we identify the counterparts of measurement directions ${\bf n}$ within ${{\mathbb E}^3}$ with the dual vectors ${{\bf n}_r+\,{\bf n}_d\,\varepsilon_+}$ within its algebraic representation space ${S^7}$. Then ${{\bf n}}$ relates to ${{\bf D}({\bf n}_r,\,{\bf n}_d,\,0)}$ as 
\begin{equation}
S^3\supset S^2\ni\,{\boldsymbol \xi}({\bf n})\,=\,I_3\cdot{\bf n}\,\longleftrightarrow\,{\bf D}({\bf n}_r,\,{\bf n}_d,\,0)\, \in S^5 \subset S^7.
\end{equation}
This allows us to identity the anti-symmetric part ${{\bf D}({\bf n}_r,\,{\bf n}_d,\,0)}$ in (\ref{64444}) as a detector of the physical system represented by ${{\bf N}({\bf n}_r,\,{\bf n}_d,\,0,\,\lambda)}$, originating in the initial state ${\lambda}$ and producing the measurement results ${{\mathscr N}({\bf n},\,\lambda)=\pm1}$ along freely chosen unit directions ${{\bf n}\longleftrightarrow{\bf n}_r+{\bf n}_d\,\varepsilon_+}$ within ${{\rm I\!R}^3}$. Indeed, using the definitions
(\ref{mandef}) to (\ref{det}) it is easy to verify that
\begin{equation}
{\bf N}^2({\bf n}_r,\,{\bf n}_d,\,0,\,\lambda)\,=\,\lambda^2\,{\bf D}^2({\bf n}_r,\,{\bf n}_d,\,0)\,=\,{\bf D}^2({\bf n}_r,\,{\bf n}_d,\,0)\,=\,-1\,. \label{sitam}
\end{equation}
In general, for two vectors ${\bf a}$ and ${\bf b}$ the geometric product ${{\bf N}({\bf a}_r,\,{\bf a}_d,\,0,\,\lambda){\bf N}({\bf b}_r,\,{\bf b}_d,\,0,\,\lambda)}$ is highly non-trivial, as we saw in (\ref{neoprod}):
\begin{equation}
{\bf N}({\bf a}_r,\,{\bf a}_d,\,0,\,\lambda)\,{\bf N}({\bf b}_r,\,{\bf b}_d,\,0,\,\lambda)\,=\,-\,{\bf a}_r\cdot{\bf b}_r\,-\,{\bf a}_d\cdot{\bf b}_d\,-\,{\bf N}({\bf a}_r\times{\bf b}_r + {\bf a}_d\times{\bf b}_d,\,{\bf a}_r\times{\bf b}_d + {\bf a}_d\times{\bf b}_r,\,{\bf a}_r\cdot{\bf b}_d + {\bf a}_d\cdot{\bf b}_r,\,\lambda)\,. \label{bi-expan}
\end{equation}
Unlike the general case, however, since we wish to identify the external vectors ${{\bf a}\leftrightarrow{\bf a}_r+{\bf a}_d\,\varepsilon_+\,}$ and ${{\bf b}\leftrightarrow{\bf b}_r+{\bf b}_d\,\varepsilon_+\,}$ with the measurement directions within ${{\mathbb E}^3}$, the following constraints induced by their scalar product naturally hold: 
\begin{equation}
{\bf a}\cdot{\bf b}\,:=\,\frac{1}{2}\{{\bf a}{\bf b}\,+\,{\bf b}{\bf a}\}\,=\,({\bf a}_r\cdot{\bf b}_r\,+\,{\bf a}_d\cdot{\bf b}_d)\,+\,({\bf a}_r\cdot{\bf b}_d\,+\,{\bf a}_d\cdot{\bf b}_r)\,\varepsilon_+\;\Longrightarrow\;
\begin{cases}
{\bf a}_r\cdot{\bf b}_d\,=\,{\bf a}_d\cdot{\bf b}_r\,=\,0 \\
\;\;\;\;\;\;\;\;\;\;\,\text{and} \\
{\bf a}_r\cdot{\bf b}_r\,=\,{\bf a}_d\cdot{\bf b}_d\,=\,\frac{1}{2}\cos\theta_{{\bf a}{\bf b}}\,, \\
\end{cases} \label{new-dotp} 
\end{equation}
which are consistent with ${{\bf a}\cdot{\bf b}=1}$ for the special case ${{\bf a}={\bf b}}$ and the normalization conditions for ${{\bf a}_r}$ and ${{\bf b}_d}$, giving    
\begin{equation}
{\bf N}({\bf a}_r,\,{\bf a}_d,\,0,\,\lambda)\,{\bf N}({\bf b}_r,\,{\bf b}_d,\,0,\,\lambda)\,=\,-\,{\bf a}_r\cdot{\bf b}_r\,-\,{\bf a}_d\cdot{\bf b}_d \,-\,{\bf N}\left({\bf a}_r\times{\bf b}_r + {\bf a}_d\times{\bf b}_d,\,{\bf a}_r\times{\bf b}_d + {\bf a}_d\times{\bf b}_r,\,0,\,\lambda\right). \label{bi-0-expan}
\end{equation}
Labeling the experimental trials with index ${k}$, we can now define the measurement functions (\ref{theoequa}) as maps of the form
\begin{equation}
S^7\ni\pm\,1\,=\,{\mathscr N}({\bf n},\,\lambda^k): {\rm I\!R}^3\!\times\left\{\,\lambda^k\right\}\longrightarrow \, S^7 \hookrightarrow{\rm I\!R}^8. \label{mesfun}
\end{equation}
These maps can be realized for the freely chosen measurement directions, specified by the vectors such as ${\bf a}$ and ${\bf b}$, as
\begin{align}
S^7\ni{\mathscr A}({\bf a}\,,\,\lambda^k):=&\lim_{\,\substack{{\bf s}_{r1}\,\rightarrow\;{\bf a}_r \\ \,{\bf s}_{d1}\,\rightarrow\;{\bf a}_d}}\left\{\,\pm\,{\bf D}({\bf a}_r,\,{\bf a}_d,\,0)\,{\bf N}({\bf s}_{r1},\,{\bf s}_{d1},\,0,\,\lambda^k)\,\right\}=\,
\begin{cases}
\,\mp\,1\;\;\;\;\;{\rm if} &\lambda^k\,=\,+\,1 \\
\,\pm\,1\;\;\;\;\;{\rm if} &\lambda^k\,=\,-\,1
\end{cases} \Bigg\} \nonumber \\
&\;\,\text{together with}\,\;\Bigl\langle\, {\mathscr A}({\bf a}\,,\,\lambda^k) \,\Bigr\rangle\,=\,0  \label{sevendone}
\end{align}
and
\begin{align}
S^7\ni{\mathscr B}({\bf b}\,,\,\lambda^k):=&\lim_{\,\substack{{\bf s}_{r2}\,\rightarrow\;{\bf b}_r \\ \,{\bf s}_{d2}\,\rightarrow\;{\bf b}_d}}\left\{\,\mp\,{\bf D}({\bf b}_r,\,{\bf b}_d,\,0)\,{\bf N}({\bf s}_{r2},\,{\bf s}_{d2},\,0,\,\lambda^k)\,\right\}=\,
\begin{cases}
\,\pm\,1\;\;\;\;\;{\rm if} &\lambda^k\,=\,+\,1 \\
\,\mp\,1\;\;\;\;\;{\rm if} &\lambda^k\,=\,-\,1
\end{cases} \Bigg\} \nonumber \\
&\;\,\text{together with}\,\;\Bigl\langle\, {\mathscr B}({\bf b}\,,\,\lambda^k) \,\Bigr\rangle\,=\,0\,. \label{sevenundone}
\end{align}
Here we have assumed that orientation ${\lambda=\pm1}$ of ${S^7}$ is a fair coin. Evidently, the functions ${{\mathscr A}({\bf a}\,,\,\lambda^k)}$ and ${{\mathscr B}({\bf b}\,,\,\lambda^k)}$ define local, realistic, and deterministically determined measurement outcomes \cite{IJTP,local}. Apart from the common cause ${\lambda^k}$ originating in the overlap of the backward lightcones of ${{\mathscr A}({\bf a}\,,\,\lambda^k)}$ and ${{\mathscr B}({\bf b}\,,\,\lambda^k)}$, the event ${{\mathscr A}=\pm1}$ depends only on a freely chosen measurement direction ${\bf a}$ \cite{Bell-1964}. And likewise, apart from the common cause ${\lambda^k}$, the event ${{\mathscr B}=\pm1}$ depends only on a freely chosen measurement direction ${\bf b}$. In particular, the function ${{\mathscr A}({\bf a}\,,\,\lambda^k)}$ does not depend on either ${\bf b}$ or ${\mathscr B}$, and the function ${{\mathscr B}({\bf b}\,,\,\lambda^k)}$ does not depend on either ${\bf a}$ or ${\mathscr A}$. This leads us to the following remarkable theorem.

\subsection{Quantum Correlations from the Algebra of Euclidean Primitives} \label{theo}

\begin{theorem}\label{T31} {\sl Every quantum mechanical correlation can be understood as a classical, local, realistic, and deterministic correlation among a set of points of ${S^7}$ constructed above, represented by maps of the form defined in (\ref{sevendone}) and (\ref{sevenundone})}.
\end{theorem}
{\parindent 0pt
{\bf Proof}:} Recall that -- as von Neumann recognized in his classic analysis \cite{von} -- regardless of the model of physics one is concerned with -- whether it is the quantum mechanical model or a hidden variable model -- it is sufficient to consider expectation values of the observables measured in possible states of the physical systems, since probabilities are but expectation values of the indicator random variables. Thus, probability ${P(E)}$ of event ${E}$ is expectation value ${{\cal E}(\mathbb{1}_{\!E})}$,
\begin{equation}
P(E)=\,{\cal E}(\mathbb{1}_{\!E})\,,
\end{equation}
of the indicator random variable ${\mathbb{1}_{\!E}}$ defined as 
\begin{equation}
\mathbb{1}_{\!E}:=
\begin{cases}
\,1 & \text{if ${E}$ occurs} \\
\,0 & \;\text{otherwise}\,.
\end{cases}
\end{equation}
Conversely, the expectation value of ${\mathbb{1}_{\!E}}$ is
\begin{equation}
{\cal E}(\mathbb{1}_{\!E})\,=\,\frac{1\times P(E)\,+\,0\times\{1-P(E)\}}{P(E)\,+\,\{1-P(E)\}}\,=\,P(E)\,.
\end{equation}
Thus every statement involving probabilities can be translated into a statement involving expectation values, and vice versa. In what follows we shall therefore work exclusively with expectation values, because our primary goal here is to trace the origins of the quantum correlations to the algebraic and geometrical properties of the Euclidean primitives.

To that end, consider an arbitrary quantum state ${|\Psi\rangle\in{\cal H}}$ of a system, where ${\cal H}$ is a Hilbert space of arbitrary dimensions -- not necessarily finite. Apart from their usual quantum mechanical meanings, we impose no restrictions on either ${|\Psi\rangle}$ or ${\cal H}$. In particular, the state ${|\Psi\rangle}$ can be as entangled as one may wish \cite{disproof}. Next, consider a self-adjoint operator ${{\cal\widehat O}({\bf n}^1,\,{\bf n}^2,\,{\bf n}^3,\,{\bf n}^4,\,{\bf n}^5,\,\dots\,)}$ on this Hilbert space, parameterized by arbitrary number of local contexts ${{\bf n}^1,\,}$ ${{\bf n}^2,\,{\bf n}^3,\,{\bf n}^4,\,{\bf n}^5,\,}$ {\it etc.} The quantum mechanically expected value of this observable in the state ${|\Psi\rangle}$ is then defined by:
\begin{equation}
{\cal E}_{{\!}_{Q.M.}}({\bf n}^1,\,{\bf n}^2,\,{\bf n}^3,\,{\bf n}^4,\,{\bf n}^5,\,\dots\,)\,
=\,\langle\Psi|\;{\cal\widehat O}({\bf n}^1,\,{\bf n}^2,\,{\bf n}^3,\,{\bf n}^4,\,{\bf n}^5,\,\dots\,)\,|\Psi\rangle\,. \label{gen-q-state-a}
\end{equation}
More generally, if the system is in a mixed state, then its quantum mechanically expected value can be expressed as
\begin{equation}
{\cal E}_{{\!}_{Q.M.}}({\bf n}^1,\,{\bf n}^2,\,{\bf n}^3,\,{\bf n}^4,\,{\bf n}^5,\,\dots\,)\,
=\,\text{Tr}\left\{{\cal\widehat W}\;{\cal\widehat O}({\bf n}^1,\,{\bf n}^2,\,{\bf n}^3,\,{\bf n}^4,\,{\bf n}^5,\,\dots\,)\right\},\label{gen-q-state-b}
\end{equation}
where ${\cal\widehat W}$ is a statistical operator of unit trace representing the state of the system. Setting ${{\bf n}^1={\bf a}\longleftrightarrow{\bf a}_r+{\bf a}_d\,\varepsilon_+\,}$, ${{\bf n}^2={\bf b}\longleftrightarrow{\bf b}_r+{\bf b}_d\,\varepsilon_+\,}$, {\it etc.}, the corresponding local-realistic expectation value for the same system can be written${\;}$as    
\begin{equation}
{\cal E}_{{\!}_{L.R.}}({\bf a},\,{\bf b},\,{\bf c},\,{\bf d},\,\dots\,)\,=\int_{\Lambda}{\mathscr A}({\bf a}\,,\,\lambda)\;{\mathscr B}({\bf b}\,,\,\lambda)\;{\mathscr C}({\bf c}\,,\,\lambda)\;{\mathscr D}({\bf d}\,,\,\lambda)\,\dots\;\rho(\lambda)\,d\lambda\,,\label{locnest}
\end{equation}
where the binary measurement functions ${{\mathscr N}\!({\bf n}\,,\,\lambda^k)}$ are defined in Eq.${\,}$(\ref{mesfun}) and the overall probability${\;}$distribution${\;\rho(\lambda)}$,
\begin{equation}
\text{with}\;\;\int_{\Lambda}\rho(\lambda)\;d\lambda\,=\,1\,\;\;\;\text{for all}\;\;\;\lambda\in\Lambda\,,
\end{equation}
is in general a continuous function of ${\lambda}$. Since in our framework ${\lambda=\pm1}$ is a fair coin, the above integral simplifies${\;}$to
\begin{equation}
{\cal E}_{{\!}_{L.R.}}({\bf a},\,{\bf b},\,{\bf c},\,{\bf d},\,\dots\,)\,=\!\!\lim_{\,m\,\rightarrow\,\infty}\!\left[\frac{1}{m}\sum_{k\,=\,1}^{m}\,{\mathscr A}({\bf a}\,,\,\lambda^k)\;{\mathscr B}({\bf b}\,,\,\lambda^k)\;{\mathscr C}({\bf c}\,,\,\lambda^k)\;{\mathscr D}({\bf d}\,,\,\lambda^k)\,\dots\,\right]. \label{b9a}
\end{equation}
We shall soon prove, however, that --- thanks to the definitions like (\ref{sevendone}) --- this average is geometrically equivalent to
\begin{equation}
{\cal E}_{{\!}_{L.R.}}({\bf a},\,{\bf b},\,{\bf c},\,{\bf d},\,\dots\,)\,=\!\!\lim_{\,m\,\rightarrow\,\infty}\!\left[\frac{1}{m}\sum_{k\,=\,1}^{m}{\bf N}({\bf a}_r,\,{\bf a}_d,\,0,\,\lambda^k)\,{\bf N}({\bf b}_r,\,{\bf b}_d,\,0,\,\lambda^k)\,{\bf N}({\bf c}_r,\,{\bf c}_d,\,0,\,\lambda^k)\,{\bf N}({\bf d}_r,\,{\bf d}_d,\,0,\,\lambda^k)\,\dots\,\right]. \label{a9b}
\end{equation}
Moreover, since as we saw in subsection \ref{1D} the representation space ${{\cal K}^{\lambda}}$ defined in (\ref{spring}), with or without the constraints in (\ref{new-dotp}), remains closed under multiplication, the product appearing in the expectation (\ref{a9b}) is equivalent to the product  
\begin{equation}
{\bf N}({\bf x}_r,\,{\bf x}_d,\,0,\,\lambda)\,{\bf N}({\bf y}_r,\,{\bf y}_d,\,0,\,\lambda)\,=\,-\,{\bf x}_r\cdot{\bf y}_r\,-\,{\bf x}_d\cdot{\bf y}_d\,-\,{\bf N}\left({\bf x}_r\times{\bf y}_r + {\bf x}_d\times{\bf y}_d,\,{\bf x}_r\times{\bf y}_d + {\bf x}_d\times{\bf y}_r,\,0,\,\lambda\right), \label{gen-bi-expan}
\end{equation}
for some vectors ${\bf x}$ and ${\bf y}$, depending in general on the measurement directions ${\bf a}$, ${\bf b}$, ${\bf c}$, ${\bf d}$, {\it etc}. Consequently we have
\begin{align}
{\cal E}_{{\!}_{L.R.}}({\bf a},\,{\bf b},\,{\bf c},\,{\bf d},\,\dots\,)&=\!\!\lim_{\,m\,\rightarrow\,\infty}\!\left[\frac{1}{m}\sum_{k\,=\,1}^{m}\,{\mathscr A}({\bf a}\,,\,\lambda^k)\;{\mathscr B}({\bf b}\,,\,\lambda^k)\;{\mathscr C}({\bf c}\,,\,\lambda^k)\;{\mathscr D}({\bf d}\,,\,\lambda^k)\;\dots\,\right] \label{g109} \\
&=\!\!\lim_{\,m\,\rightarrow\,\infty}\!\left[\frac{1}{m}\sum_{k\,=\,1}^{m}\!{\bf N}({\bf a}_r,\,{\bf a}_d,\,0,\,\lambda^k)\,{\bf N}({\bf b}_r,\,{\bf b}_d,\,0,\,\lambda^k)\,{\bf N}({\bf c}_r,\,{\bf c}_d,\,0,\,\lambda^k)\,{\bf N}({\bf d}_r,\,{\bf d}_d,\,0,\,\lambda^k)\dots\right] \label{g110}\\
&=\!\!\lim_{\,m\,\rightarrow\,\infty}\!\left[\frac{1}{m}\sum_{k\,=\,1}^{m}{\bf N}({\bf x}_r,\,{\bf x}_d,\,0,\,\lambda^k)\,{\bf N}({\bf y}_r,\,{\bf y}_d,\,0,\,\lambda^k)\,\right] \label{g111}\\
&=-\,{\bf x}_r\cdot{\bf y}_r-{\bf x}_d\cdot{\bf y}_d-\!\!\lim_{\,m\,\rightarrow\,\infty}\left[\frac{1}{m}\sum_{k\,=\,1}^{m}\!{\bf N}\left({\bf x}_r\times{\bf y}_r + {\bf x}_d\times{\bf y}_d,\,{\bf x}_r\times{\bf y}_d + {\bf x}_d\times{\bf y}_r,\,0,\,\lambda^k\right)\right] \label{g112} \\
&=-\,{\bf x}_r\cdot{\bf y}_r-{\bf x}_d\cdot{\bf y}_d-\!\!\lim_{\,m\,\rightarrow\,\infty}\left[\frac{1}{m}\sum_{k\,=\,1}^{m}\lambda^k\right]{\bf D}\left({\bf x}_r\times{\bf y}_r + {\bf x}_d\times{\bf y}_d,\,{\bf x}_r\times{\bf y}_d + {\bf x}_d\times{\bf y}_r,\,0\right) \label{g113} \\
&=\,-\,\cos\theta_{{\bf x}{\bf y}}({\bf a},\,{\bf b},\,{\bf c},\,{\bf d},\,\dots\,)\,-\,0\,, \label{g114}
\end{align}
because ${\lambda^k}$, as in (\ref{ori-rela}), is a fair coin. Here Eq.~(\ref{g113}) follows from Eq.~(\ref{g112}) by using Eq.~(\ref{ori-rela}), which now takes the form
\begin{equation}
{\bf N}\big({\bf x}_r\times{\bf y}_r + {\bf x}_d\times{\bf y}_d,\,{\bf x}_r\times{\bf y}_d + {\bf x}_d\times{\bf y}_r,\,0,\,\lambda^k\big)\,=\,\lambda^k\,
{\bf D}\left({\bf x}_r\times{\bf y}_r + {\bf x}_d\times{\bf y}_d,\,{\bf x}_r\times{\bf y}_d + {\bf x}_d\times{\bf y}_r,\,0\right),\tag{\ref{ori-rela}}
\end{equation}
together with ${\lambda^k=\pm1}$. We can now identify the above local-realistic expectation with its quantum mechanical counterpart:
\begin{equation}
\langle\Psi|\;{\cal\widehat O}({\bf a},\,{\bf b},\,{\bf c},\,{\bf d},\,\dots\,)\,|\Psi\rangle\,=\,{\cal E}_{{\!}_{L.R.}}({\bf a},\,{\bf b},\,{\bf c},\,{\bf d},\,\dots\,)\,=\,-\,\cos\theta_{{\bf x}{\bf y}}({\bf a},\,{\bf b},\,{\bf c},\,{\bf d},\,\dots\,)\,.
\end{equation}
This identification proves our main theorem: Every quantum mechanical correlation can be understood as a classical, local, deterministic and realistic correlation among a set of points of the representation space ${S^7\subset{\cal K}^{\lambda}}$ described above (cf. Theorem \ref{T31}). 

It is instructive to evaluate the sum in Eq.~(\ref{g111}) somewhat differently to bring out the fundamental role played by the orientation ${\lambda^k}$ in the derivation of the strong correlations (\ref{g114}). Instead of assuming ${\lambda^k=\pm1}$ to be an orientation of ${S^7}$ as our starting point, we may view it as specifying the ordering relation between ${{\bf N}({\bf x}_r,\,{\bf x}_d,\,0,\,{\lambda}^k=\pm1)}$ and ${{\bf N}({\bf y}_r,\,{\bf y}_d,\;0,\,{\lambda}^k=\pm1)}$ and the corresponding detectors ${{\bf D}({\bf x}_r,\,{\bf x}_d,\,0)}$ and ${{\bf D}({\bf y}_r,\,{\bf y}_d,\,0)}$ with 50/50 chance of occurring, and only subsequently identify it with the orientation of ${S^7}$. Then, using the relations (\ref{ori-rela}) and (\ref{gen-bi-expan}), the sum in Eq.${\,}$(\ref{g111}) can be evaluated directly by recognizing that in the right and left oriented ${S^7}$ the following geometrical relations hold:
\begin{align}
{\bf N}({\bf x}_r,\,{\bf x}_d,\,0,\,{\lambda}^k=+1)\;{\bf N}(&{\bf y}_r,\,{\bf y}_d,\;0,\,{\lambda}^k=+1) \notag \\
\,&=\,-\,{\bf x}_r\cdot{\bf y}_r\,-\,{\bf x}_d\cdot{\bf y}_d\,-\,{\bf N}\left({\bf x}_r\times{\bf y}_r + {\bf x}_d\times{\bf y}_d,\,{\bf x}_r\times{\bf y}_d + {\bf x}_d\times{\bf y}_r,\,0,\,\lambda^k=+1\right) \nonumber \\
\,&=\,-\,{\bf x}_r\cdot{\bf y}_r\,-\,{\bf x}_d\cdot{\bf y}_d\,-\,{\bf D}\left({\bf x}_r\times{\bf y}_r + {\bf x}_d\times{\bf y}_d,\,{\bf x}_r\times{\bf y}_d + {\bf x}_d\times{\bf y}_r,\,0\right) \nonumber \\
\,&=\,{\bf D}({\bf x}_r,\,{\bf x}_d,\,0)\;{\bf D}({\bf y}_r,\,{\bf y}_d,\,0) \label{grhspen}
\end{align}
and
\begin{align}
{\bf N}({\bf x}_r,\,{\bf x}_d,\,0,\,{\lambda}^k=-1)\;{\bf N}(&{\bf y}_r,\,{\bf y}_d,\;0,\,{\lambda}^k=-1) \notag \\
\,&=\,-\,{\bf x}_r\cdot{\bf y}_r\,-\,{\bf x}_d\cdot{\bf y}_d\,-\,{\bf N}\left({\bf x}_r\times{\bf y}_r + {\bf x}_d\times{\bf y}_d,\,{\bf x}_r\times{\bf y}_d + {\bf x}_d\times{\bf y}_r,\,0,\,\lambda^k=-1\right) \nonumber \\
\,&=\,-\,{\bf x}_r\cdot{\bf y}_r\,-\,{\bf x}_d\cdot{\bf y}_d\,+\,{\bf D}\left({\bf x}_r\times{\bf y}_r + {\bf x}_d\times{\bf y}_d,\,{\bf x}_r\times{\bf y}_d + {\bf x}_d\times{\bf y}_r,\,0\right) \nonumber \\
\,&=\,-\,{\bf y}_r\cdot{\bf x}_r\,-\,{\bf y}_d\cdot{\bf x}_d\,-\,{\bf D}\left({\bf y}_r\times{\bf x}_r + {\bf y}_d\times{\bf x}_d,\,{\bf y}_r\times{\bf x}_d + {\bf y}_d\times{\bf x}_r,\,0\right) \nonumber \\
\,&=\,{\bf D}({\bf y}_r,\,{\bf y}_d,\,0)\;{\bf D}({\bf x}_r,\,{\bf x}_d,\,0). \label{grhsden}
\end{align}
Changes in the orientation ${\lambda^k}$ thus alternates the algebraic order of ${{\bf N}({\bf x}_r,\,{\bf x}_d,\,0,\,{\lambda}^k=\pm1)}$ and ${{\bf N}({\bf y}_r,\,{\bf y}_d,\;0,\,{\lambda}^k=\pm1)}$ {\it relative} to the algebraic order of the detectors ${{\bf D}({\bf x}_r,\,{\bf x}_d,\,0)}$ and ${{\bf D}({\bf y}_r,\,{\bf y}_d,\,0)}$. Consequently, the sum (\ref{g111}) reduces to
\begin{align}
{\cal E}_{{\!}_{L.R.}}({\bf a},&\;{\bf b},\,{\bf c},\,{\bf d},\,\dots\,)\,=\!\lim_{\,m\,\rightarrow\,\infty}\!\left[\frac{1}{m}\sum_{k\,=\,1}^{m}{\bf N}({\bf x}_r,\,{\bf x}_d,\,0,\,\lambda^k)\,{\bf N}({\bf y}_r,\,{\bf y}_d,\,0,\,\lambda^k)\,\right] \nonumber \\
&=\,\frac{1}{2}\{\,{\bf N}({\bf x}_r,\,{\bf x}_d,\,0,\,\lambda^k=+1)\;{\bf N}({\bf y}_r,\,{\bf y}_d,\,0,\,\lambda^k=+1)\}\,+\,\frac{1}{2}\{\,{\bf N}({\bf x}_r,\,{\bf x}_d,\,0,\,\lambda^k=-1)\;{\bf N}({\bf y}_r,\,{\bf y}_d,\,0,\,\lambda^k=-1)\} \nonumber \\
&=\,\frac{1}{2}\{\,{\bf D}({\bf x}_r,\,{\bf x}_d,\,0)\;{\bf D}({\bf y}_r,\,{\bf y}_d,\,0)\}\,+\,\frac{1}{2}\{\,{\bf D}({\bf y}_r,\,{\bf y}_d,\,0)\;{\bf D}({\bf x}_r,\,{\bf x}_d,\,0)\} \nonumber \\
&=\,-\,\frac{1}{2}\{{\bf x}_r{\bf y}_r\,+\,{\bf y}_r{\bf x}_r\}\,-\,\frac{1}{2}\{{\bf x}_d{\bf y}_d\,+\,{\bf y}_d{\bf x}_d\}\,=\,-\,{\bf x}_r\cdot{\bf y}_r\,-\,{\bf x}_d\cdot{\bf y}_d\,=\,-\,{\bf x}\cdot{\bf y} \nonumber \\
&=\,-\,\cos\theta_{{\bf x}{\bf y}}({\bf a},\,{\bf b},\,{\bf c},\,{\bf d},\,\dots\,)\,, \label{calcoll}
\end{align}
because the orientation ${\lambda^k}$ of ${S^7}$ is a fair coin. Here ${{\bf x}\cdot{\bf y}=\frac{1}{2}\{{\bf x}{\bf y}+{\bf y}{\bf x}\}}$ is the standard definition of the inner product.

Evidently the above method of calculating suggests that a given initial state ${\lambda}$ of the physical system can indeed be viewed as specifying an ordering relation between ${{\bf N}({\bf n}_r,\,{\bf n}_d,\,0,\,\lambda)}$ and the detectors ${{\bf D}({\bf n}_r,\,{\bf n}_d,\,0)}$ that measure${\;}$it:
\begin{equation}
{\bf N}({\bf x}_r,\,{\bf x}_d,\,0,\,\lambda^k=+1)\;{\bf N}({\bf y}_r,\,{\bf y}_d,\,0,\,\lambda^k=+1)\,=\,{\bf D}({\bf x}_r,\,{\bf x}_d,\,0)\;{\bf D}({\bf y}_r,\,{\bf y}_d,\,0) \label{pair1}
\end{equation}
or
\begin{equation}
{\bf N}({\bf x}_r,\,{\bf x}_d,\,0,\,\lambda^k=-1)\;{\bf N}({\bf y}_r,\,{\bf y}_d,\,0,\,\lambda^k=-1)\,=\,{\bf D}({\bf y}_r,\,{\bf y}_d,\,0)\;{\bf D}({\bf x}_r,\,{\bf x}_d,\,0). \label{pair2}
\end{equation}
Then, using the right-hand sides of the Eqs.${\,}$(\ref{grhspen}) and (\ref{grhsden}), the above pair can be reduced to the combined relation 
\begin{equation}
{\bf N}\big({\bf x}_r\times{\bf y}_r + {\bf x}_d\times{\bf y}_d,\,{\bf x}_r\times{\bf y}_d \,+\,{\bf x}_d\times{\bf y}_r,\,0,\,\lambda^k\big) \,=\,\lambda^k\,{\bf D}\left({\bf x}_r\times{\bf y}_r + {\bf x}_d\times{\bf y}_d,\,{\bf x}_r\times{\bf y}_d + {\bf x}_d\times{\bf y}_r,\,0\right),
\end{equation}
which is identical to the relation (\ref{ori-rela}) for normalized vectors. We have thus proved that the ordering relations (\ref{pair1}) and (\ref{pair2}) between ${\,{\bf N}({\bf n}_r,\,{\bf n}_d,\,0,\,\lambda)\,}$ and ${\,{\bf D}({\bf n}_r,\,{\bf n}_d,\,0)\,}$ are equivalent to the alternatively possible orientations of ${S^7}$.

\subsubsection{Special Case of a Two-level System Entangled in the Singlet State}\label{twosect}

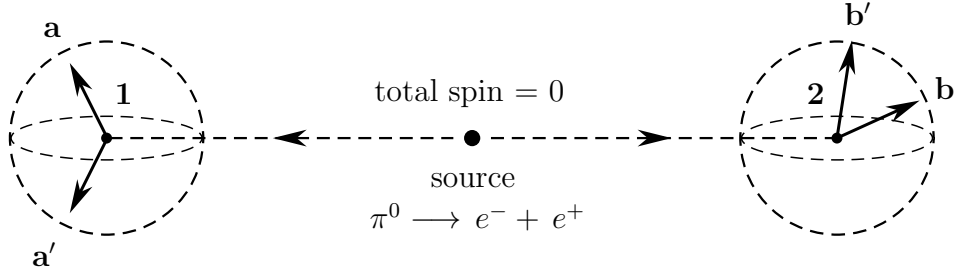
\begin{figure}
\hrule
\scalebox{1}{
\begin{pspicture}(1.2,-2.5)(4.2,2.5)

\psline[linewidth=0.1mm,dotsize=3pt 4]{*-}(-2.51,0)(-2.5,0)

\psline[linewidth=0.1mm,dotsize=3pt 4]{*-}(7.2,0)(7.15,0)

\psline[linewidth=0.4mm,arrowinset=0.3,arrowsize=3pt 3,arrowlength=2]{->}(-2.5,0)(-3,1)

\psline[linewidth=0.4mm,arrowinset=0.3,arrowsize=3pt 3,arrowlength=2]{->}(-2.5,0)(-3,-1)

\psline[linewidth=0.4mm,arrowinset=0.3,arrowsize=3pt 3,arrowlength=2]{->}(7.2,0)(8.3,0.5)

\psline[linewidth=0.4mm,arrowinset=0.3,arrowsize=3pt 3,arrowlength=2]{->}(7.2,0)(7.4,1.3)

\put(-2.4,+0.45){{\large ${\bf 1}$}}

\put(6.8,+0.45){{\large ${\bf 2}$}}

\put(-3.35,1.35){{\large ${\bf a}$}}

\put(-3.5,-1.7){{\large ${\bf a'}$}}

\put(8.5,0.52){{\large ${\bf b}$}}

\put(7.3,1.5){{\large ${\bf b'}$}}

\put(1.8,-0.65){\large source}

\put(0.99,-1.2){\large ${\pi^0\longrightarrow\,e^{-}+\,e^{+}\,}$}

\put(1.05,0.5){\large total spin = 0}

\psline[linewidth=0.3mm,linestyle=dashed](-2.47,0)(2.1,0)

\psline[linewidth=0.4mm,arrowinset=0.3,arrowsize=3pt 3,arrowlength=2]{->}(-0.2,0)(-0.3,0)

\psline[linewidth=0.3mm,linestyle=dashed](2.6,0)(7.2,0)

\psline[linewidth=0.4mm,arrowinset=0.3,arrowsize=3pt 3,arrowlength=2]{->}(4.9,0)(5.0,0)

\psline[linewidth=0.1mm,dotsize=5pt 4]{*-}(2.35,0)(2.4,0)

\pscircle[linewidth=0.3mm,linestyle=dashed](7.2,0){1.3}

\psellipse[linewidth=0.2mm,linestyle=dashed](7.2,0)(1.28,0.3)

\pscircle[linewidth=0.3mm,linestyle=dashed](-2.51,0){1.3}

\psellipse[linewidth=0.2mm,linestyle=dashed](-2.51,0)(1.28,0.3)

\end{pspicture}}
\hrule
\caption{A spin-less neutral pion decays into an electron-positron pair (such a photon-less decay is quite rare but not impossible, and will suffice for our theoretical purposes here). Measurements of spin components on each separated fermion are performed at remote stations ${\bf 1}$ and ${\bf 2}$, providing binary outcomes along arbitrary directions such as ${\bf a}$ and ${\bf b}$.}
\vspace{5pt}
\hrule
\label{fig6}
\end{figure}

Now, to complete the above proof of the Theorem \ref{T31} we must prove the step from Eq.~(\ref{b9a}) to Eq.~(\ref{a9b}). To that end, let us first consider observations of the spins of only two spin-${\frac{1}{2}}$ particles produced in a decay of a single spinless particle as shown in Fig.${\,}$\ref{fig6}. After the decay the two emerging spin-${\frac{1}{2}}$ particles move freely in opposite directions, subject to spin measurements along freely chosen unit directions ${\bf a}$ and ${\bf b}$, which may be located at a spacelike distance from one another \cite{local}. Since initially the emerging pair has zero net spin, its quantum mechanical state is described by the entangled singlet state 
\begin{equation}
|\Psi_{\bf z}\rangle=\frac{1}{\sqrt{2}}\Bigl\{|{\bf z},\,+\rangle_1\otimes
|{\bf z},\,-\rangle_2\,-\,|{\bf z},\,-\rangle_1\otimes|{\bf z},\,+\rangle_2\Bigr\}\,,
\label{single}
\end{equation}
with ${{\boldsymbol\sigma}\cdot{\bf z}\,|{\bf z},\,\pm\rangle\,=\,\pm\,|{\bf z},\,\pm\rangle}$ describing the eigenstates of the Pauli spin ``vector" ${\boldsymbol\sigma}$ in which the particles have spin ``up" or ``down" along ${\bf z}$-axis, in the units of ${\hbar=2}$. Our interest lies in comparing the quantum mechanical predictions,
\begin{equation}
{\cal E}^{\Psi_{\bf z}}_{{\!}_{Q.M.}}({\bf a},\,{\bf b})\,=\,
\langle\Psi_{\bf z}|\,{\boldsymbol\sigma}_1\cdot{\bf a}\,\otimes\,
{\boldsymbol\sigma}_2\cdot{\bf b}\,|\Psi_{\bf z}\rangle\,=\,-\,\cos\theta_{{\bf a}{\bf b}}\,, \label{twoobserve}
\end{equation}
together with
\begin{equation}
{\cal E}^{\Psi_{\bf z}}_{{\!}_{Q.M.}}({\bf a})\,=\,\langle\Psi_{\bf z}|\,{\boldsymbol\sigma}_1\cdot{\bf a}\otimes\dbl\,|\Psi_{\bf z}\rangle\,=\,0\;\;\;{\rm and}\;\;\;{\cal E}^{\Psi_{\bf z}}_{{\!}_{Q.M.}}({\bf b})\,=\,\langle\Psi_{\bf z}|\,\dbl\otimes{\boldsymbol\sigma}_2
\cdot{\bf b}\,|\Psi_{\bf z}\rangle\,=\,0\,, \label{singof}
\end{equation}
of spin correlations between the two subsystems, with those derived within our locally causal framework, regardless of the relative distance between the two remote locations represented by the unit detection vectors ${\bf a}$ and ${\bf b}$. Here ${\dbl}$ is the identity matrix. The corresponding locally causal description of this emblematic system within our framework thus involves only two contexts, ${{\bf n}^1={\bf a}\longleftrightarrow{\bf a}_r+{\bf a}_d\,\varepsilon_+\,}$ and ${{\bf n}^2={\bf b}\longleftrightarrow{\bf b}_r+{\bf b}_d\,\varepsilon_+\,}$, with measurement results defined by the functions 
\begin{align}
S^7\ni{\mathscr A}({\bf a}\,,\,\lambda^k):=&\lim_{\,\substack{{\bf s}_{r1}\,\rightarrow\;{\bf a}_r \\ {\bf s}_{d1}\,\rightarrow\;{\bf a}_d}}\!\left\{\,-\,{\bf D}({\bf a}_r,\,{\bf a}_d,\,0)\,{\bf N}({\bf s}_{r1},\,{\bf s}_{d1},\,0,\,\lambda^k)\,\right\}=\,                  
\begin{cases}
\,+\,1\;\;\;\;\;{\rm if} &\lambda^k\,=\,+\,1 \\
\,-\,1\;\;\;\;\;{\rm if} &\lambda^k\,=\,-\,1
\end{cases} \Bigg\} \nonumber \\
&\;\,\text{together with}\,\;\Bigl\langle\, {\mathscr A}({\bf a}\,,\,\lambda^k) \,\Bigr\rangle\,=\,0\, \label{onedone}
\end{align}
and
\begin{align}
S^7\ni{\mathscr B}({\bf b}\,,\,\lambda^k):=&\lim_{\,\substack{{\bf s}_{r2}\,\rightarrow\;{\bf b}_r \\ {\bf s}_{d2}\,\rightarrow\;{\bf b}_d}}\!\left\{\,+\,{\bf N}({\bf s}_{r2},\,{\bf s}_{d2},\,0,\,\lambda^k)\,{\bf D}({\bf b}_r,\,{\bf b}_d,\,0)\,\right\}=\,                  
\begin{cases}
\,-\,1\;\;\;\;\;{\rm if} &\lambda^k\,=\,+\,1 \\
\,+\,1\;\;\;\;\;{\rm if} &\lambda^k\,=\,-\,1
\end{cases} \Bigg\} \nonumber \\
&\;\,\text{together with}\,\;\Bigl\langle\, {\mathscr B}({\bf b}\,,\,\lambda^k) \,\Bigr\rangle\,=\,0\,, \label{twodone}
\end{align}
where ${{\bf s}_1\longleftrightarrow{\bf s}_{r1}+{\bf s}_{d1}\,\varepsilon_+\,}$ and ${{\bf s}_2\longleftrightarrow{\bf s}_{r2}+{\bf s}_{d2}\,\varepsilon_+\,}$ represent the directions of the two spins emerging from the source.

Next, recalling that physically all bivectors ${{\boldsymbol \xi}({\bf n})\in S^2 \subset S^3}$ represent spins \cite{disproof,IJTP}, we require that the total spin-zero angular momentum for the initial or ``complete" state associated with the above measurement functions is conserved,
\begin{equation}
\text{total real spin}\,:=\,-\,\lambda\,{\boldsymbol \xi}({\bf s}_{r1})\,+\,\lambda\,{\boldsymbol \xi}({\bf s}_{r2})\,=\,0\;\;\Longleftrightarrow\;\; {\bf s}_{r1}=\,{\bf s}_{r2}\,\equiv\,{\bf s}_r \label{656a}
\end{equation}
and
\begin{equation}
\text{total dual spin}\,:=\,-\,\lambda\,{\boldsymbol \xi}({\bf s}_{d1})\,+\,\lambda\,{\boldsymbol \xi}({\bf s}_{d2})\,=\,0\;\;\Longleftrightarrow\;\; {\bf s}_{d1}=\,{\bf s}_{d2}\,\equiv\,{\bf s}_d\,, \label{656}
\end{equation}
just as it is in the EPR-Bohm type experiment depicted in Fig.${\,}$\ref{fig6}. For ${{\bf N}({\bf s}_{r},\,{\bf s}_{d},\,0,\,\lambda^k)}$ this is equivalent to the condition
\begin{equation}
-\,{\bf N}({\bf s}_{r1},\,{\bf s}_{d1},\,0,\,\lambda^k)\,+\,{\bf N}({\bf s}_{r2},\,{\bf s}_{d2},\,0,\,\lambda^k)\,=\,0\;\;\Longleftrightarrow\;\;{\bf N}({\bf s}_{r1},\,{\bf s}_{d1},\,0,\,\lambda^k)\,=\,{\bf N}({\bf s}_{r2},\,{\bf s}_{d2},\,0,\,\lambda^k)\,. \label{756}
\end{equation}
In the light of the product rule (\ref{bi-0-expan}) for anti-symmetric elements, the above condition is also equivalent to the condition
\begin{equation}
{\bf N}({\bf s}_{r1},\,{\bf s}_{d1},\,0,\,\lambda^k)\,{\bf N}({\bf s}_{r2},\,{\bf s}_{d2},\,0,\,\lambda^k)\,=\,\left\{{\bf N}({\bf s}_{r},\,{\bf s}_{d},\,0,\,\lambda^k)\right\}^2\,=\,{\bf N}^2({\bf s}_{r},\,{\bf s}_{d},\,0,\,\lambda^k)\,=\,-1\,. \label{75699}
\end{equation}
In the next subsection we will derive this condition geometrically as a natural consequence of the twist in the Hopf bundle of ${S^3}$. Here it leads to the following statistical equivalence, which can be viewed also as a geometrical identity:
\begin{equation}
\lim_{\,m\,\rightarrow\,\infty}\!\left[\frac{1}{m}\sum_{k\,=\,1}^{m}\,{\mathscr A}({\bf a}\,,\,\lambda^k)\;{\mathscr B}({\bf b}\,,\,\lambda^k)\,\right]\,\equiv\!\lim_{\,m\,\rightarrow\,\infty}\!\left[\frac{1}{m}\sum_{k\,=\,1}^{m}\,{\bf N}({\bf a}_r,\,{\bf a}_d,\,0,\,\lambda^k)\,{\bf N}({\bf b}_r,\,{\bf b}_d,\,0,\,\lambda^k)\,\right].
\end{equation}
Given the definitions (\ref{onedone}) and (\ref{twodone}), there are more than one ways to prove this identity. In the following we will use one such way. But it can also be proved by simply taking the limits in (\ref{onedone}) and (\ref{twodone}) while maintaining (\ref{756}), and then using Eq.~(\ref{ori-rela}). Then the computation of correlations between ${{\mathscr A}({\bf a},\,\lambda^k)=\pm1\,}$ and ${{\mathscr B}({\bf b},\,\lambda^k)=\pm1\,}$ works out as
\begin{align}
&{\cal E}^{\rm EPR}_{{\!}_{L.R.}}({\bf a},\,{\bf b})=\!\!\lim_{\,m\,\rightarrow\,\infty}\!\left[\frac{1}{m}\sum_{k\,=\,1}^{m}\,{\mathscr A}({\bf a}\,,\,\lambda^k)\;{\mathscr B}({\bf b}\,,\,\lambda^k)\right] \label{357} \\
&\;\;\;\;=\!\!\lim_{\,m\,\rightarrow\,\infty}\!\Bigg[\frac{1}{m}\sum_{k\,=\,1}^{m}\!\Bigg\{\!\lim_{\substack{{\bf s}_{r1}\,\rightarrow\,{\bf a}_r \\ {\bf s}_{d1}\,\rightarrow\,{\bf a}_d}}\!\Big[-{\bf D}({\bf a}_r,\,{\bf a}_d,\,0)\,{\bf N}({\bf s}_{r1},\,{\bf s}_{d1},\,0,\,\lambda^k)\,\Big]\!\Bigg\} \nonumber \\
&\;\;\;\;\;\;\;\;\;\;\;\;\;\;\;\;\;\;\;\;\;\;\;\;\;\;\;\;\;\;\;\;\;\;\;\;\;\;\;\;\;\;\;\;\;\;\;\;\;\;\;\;\;\;\;\;\;\;\;\;\;\;\;\;\;\;\;\;\;\;\;\;\;\;\;\;\;\;\;\;\;\;\;\;\;\;\;\;\;\;\;\;\;\Bigg\{\!\lim_{\substack{{\bf s}_{r2}\,\rightarrow\,{\bf b}_r \\ {\bf s}_{d2}\,\rightarrow\,{\bf b}_d}}\!\Big[\,{\bf N}({\bf s}_{r2},\,{\bf s}_{d2},\,0,\,\lambda^k)\,{\bf D}({\bf b}_r,\,{\bf b}_d,\,0)\,\Big]\!\Bigg\}\!\Bigg]\label{358} \\
&\;\;\;\;=\!\!\lim_{\,m\,\rightarrow\,\infty}\!\Bigg[\frac{1}{m}\sum_{k\,=\,1}^{m}\!\Bigg\{\!\lim_{\substack{{\bf s}_{r1}\,\rightarrow\,{\bf a}_r \\ {\bf s}_{d1}\,\rightarrow\,{\bf a}_d}}\;\lim_{\substack{{\bf s}_{r2}\,\rightarrow\,{\bf b}_r \\ {\bf s}_{d2}\,\rightarrow\,{\bf b}_d}}\!\Big[-{\bf D}({\bf a}_r,\,{\bf a}_d,\,0) \nonumber \\
&\;\;\;\;\;\;\;\;\;\;\;\;\;\;\;\;\;\;\;\;\;\;\;\;\;\;\;\;\;\;\;\;\;\;\;\;\;\;\;\;\;\;\;\;\;\;\;\;\;\;\;\;\;\;\;\;\;\;\;\;\;\;\;\;\;\;\;\;\;\big\{\,{\bf N}({\bf s}_{r1},\,{\bf s}_{d1},\,0,\,\lambda^k)\,{\bf N}({\bf s}_{r2},\,{\bf s}_{d2},\,0,\,\lambda^k)\,\big\}\,{\bf D}({\bf b}_r,\,{\bf b}_d,\,0)\,\Big]\!\Bigg\}\!\Bigg]\label{35991} \\
&\;\;\;\;=\!\!\lim_{\,m\,\rightarrow\,\infty}\!\left[\frac{1}{m}\sum_{k\,=\,1}^{m}\!\Bigg\{\!\lim_{\substack{{\bf s}_{r1}\,\rightarrow\,{\bf a}_r \\ {\bf s}_{d1}\,\rightarrow\,{\bf a}_d}}\;\lim_{\substack{{\bf s}_{r2}\,\rightarrow\,{\bf b}_r \\ {\bf s}_{d2}\,\rightarrow\,{\bf b}_d}}\Big[-\lambda^k\,{\bf N}({\bf a}_r,\,{\bf a}_d,\,0,\,\lambda^k)\,\left\{\,-1\,\right\}\,\lambda^k\,{\bf N}({\bf b}_r,\,{\bf b}_d,\,0,\,\lambda^k)\,\Big]\!\Bigg\}\!\right]\label{35992} \\
&\;\;\;\;=\!\!\lim_{\,m\,\rightarrow\,\infty}\!\left[\frac{1}{m}\sum_{k\,=\,1}^{m}\!\Bigg\{\!\lim_{\substack{{\bf s}_{r1}\,\rightarrow\,{\bf a}_r \\ {\bf s}_{d1}\,\rightarrow\,{\bf a}_d}}\;\lim_{\substack{{\bf s}_{r2}\,\rightarrow\,{\bf b}_r \\ {\bf s}_{d2}\,\rightarrow\,{\bf b}_d}}\Big[+\left(\lambda^k\right)^2\,{\bf N}({\bf a}_r,\,{\bf a}_d,\,0,\,\lambda^k)\,{\bf N}({\bf b}_r,\,{\bf b}_d,\,0,\,\lambda^k)\,\Big]\!\Bigg\}\!\right]\label{35993} \\
&\;\;\;\;=\!\!\lim_{\,m\,\rightarrow\,\infty}\!\left[\frac{1}{m}\sum_{k\,=\,1}^{m}{\bf N}({\bf a}_r,\,{\bf a}_d,\,0,\,\lambda^k)\,{\bf N}({\bf b}_r,\,{\bf b}_d,\,0,\,\lambda^k)\,\right] \label{362}\\
&\;\;\;\;=\,-\,{\bf a}_r\cdot{\bf b}_r -\,{\bf a}_d\cdot{\bf b}_d\,-\!\!\lim_{\,m\,\rightarrow\,\infty}\left[\frac{1}{m}\sum_{k\,=\,1}^{m}{\bf N}\left({\bf a}_r\times{\bf b}_r + {\bf a}_d\times{\bf b}_d,\,{\bf a}_r\times{\bf b}_d + {\bf a}_d\times{\bf b}_r,\,0,\,\lambda^k\right)\right] \label{363} \\
&\;\;\;\;=\,-\,{\bf a}_r\cdot{\bf b}_r -\,{\bf a}_d\cdot{\bf b}_d\,-\!\!\lim_{\,m\,\rightarrow\,\infty}\left[\frac{1}{m}\sum_{k\,=\,1}^{m}\,\lambda^k\right]{\bf D}\left({\bf a}_r\times{\bf b}_r + {\bf a}_d\times{\bf b}_d,\,{\bf a}_r\times{\bf b}_d + {\bf a}_d\times{\bf b}_r,\,0\right) \label{364} \\
&\;\;\;\;=\,-\,\cos\theta_{{\bf a}{\bf b}}\,-\,0\,. \label{365}
\end{align}
Here Eq.${\,}$(\ref{358}) follows from Eq.${\,}$(\ref{357}) by substituting the functions ${{\mathscr A}({\bf a},\,\lambda^k)}$ and ${{\mathscr B}({\bf b},\,\lambda^k)}$ from their definitions (\ref{onedone}) and (\ref{twodone}); Eq.${\,}$(\ref{35991}) follows from Eq.${\,}$(\ref{358}) by using the ``product of limits equal to limits of product" rule [which can be verified by noting that the same multivector results from the limits in Eqs.${\,}$(\ref{358}) and (\ref{35991})]; Eq.${\,}$(\ref{35992}) follows from Eq.${\,}$(\ref{35991}) by using (i) the relations (\ref{ori-rela}) [thus rewriting all anti-symmetric elements in the same bases], (ii) the associativity of the geometric product, and (iii) the consequence (\ref{75699}) of the conservation of the spin angular momenta in ${S^3}$; Eq.${\,}$(\ref{35993}) follows from Eq.${\,}$(\ref{35992}) by recalling that scalars ${\lambda^k}$ commute with the elements of all grades; Eq.${\,}$(\ref{362}) follows from Eq.${\,}$(\ref{35993}) because ${\lambda^2=+1}$, and by removing the superfluous limit operations; Eq.${\,}$(\ref{363}) follows from Eq.${\,}$(\ref{362}) by using the geometric product (\ref{bi-0-expan}); Eq.${\,}$(\ref{364}) follows from Eq.${\,}$(\ref{363}) by using the relations (\ref{ori-rela}); and finally Eq.${\,}$(\ref{365}) follows from Eq.${\,}$(\ref{364}) by using Eq.${\,}$(\ref{new-dotp}) and because the scalar coefficient of ${\bf D}$ vanishes in the ${m\rightarrow\infty}$ limit since ${\lambda^k}$ is a fair coin.${\;}$This proves that singlet correlations\textsuperscript{\ref{Hardy}} are correlations among the scalar points of a quaternionic${\;S^3}$.

As we did above for the general case, let us again evaluate the sum in Eq.${\,}$(\ref{362}) somewhat differently to bring out the crucial role played by ${\lambda^k}$ in the derivation of the correlations (\ref{365}). Using the relations (\ref{ori-rela}) and (\ref{bi-0-expan}), the sum (\ref{362}) can be evaluated directly by recognizing that in the right and left oriented ${S^7}$ the following geometrical relations hold:
\begin{align}
{\bf N}({\bf a}_r,\,{\bf a}_d,\,0,\,{\lambda}^k=+1)\;{\bf N}(&{\bf b}_r,\,{\bf b}_d,\;0,\,{\lambda}^k=+1)\, \nonumber \\
\,&=\,-\,{\bf a}_r\cdot{\bf b}_r\,-\,{\bf a}_d\cdot{\bf b}_d\,-\,{\bf N}\left({\bf a}_r\times{\bf b}_r + {\bf a}_d\times{\bf b}_d,\,{\bf a}_r\times{\bf b}_d + {\bf a}_d\times{\bf b}_r,\,0,\,\lambda^k=+1\right) \nonumber \\
\,&=\,-\,{\bf a}_r\cdot{\bf b}_r\,-\,{\bf a}_d\cdot{\bf b}_d\,-\,{\bf D}\left({\bf a}_r\times{\bf b}_r + {\bf a}_d\times{\bf b}_d,\,{\bf a}_r\times{\bf b}_d + {\bf a}_d\times{\bf b}_r,\,0\right) \nonumber \\
\,&=\,{\bf D}({\bf a}_r,\,{\bf a}_d,\,0)\;{\bf D}({\bf b}_r,\,{\bf b}_d,\,0) \label{rhspen}
\end{align}
and
\begin{align}
{\bf N}({\bf a}_r,\,{\bf a}_d,\,0,\,{\lambda}^k=-1)\;{\bf N}(&{\bf b}_r,\,{\bf b}_d,\;0,\,{\lambda}^k=-1)\, \nonumber \\ 
\,&=\,-\,{\bf a}_r\cdot{\bf b}_r\,-\,{\bf a}_d\cdot{\bf b}_d\,-\,{\bf N}\left({\bf a}_r\times{\bf b}_r + {\bf a}_d\times{\bf b}_d,\,{\bf a}_r\times{\bf b}_d + {\bf a}_d\times{\bf b}_r,\,0,\,\lambda^k=-1\right) \nonumber \\
\,&=\,-\,{\bf a}_r\cdot{\bf b}_r\,-\,{\bf a}_d\cdot{\bf b}_d\,+\,{\bf D}\left({\bf a}_r\times{\bf b}_r + {\bf a}_d\times{\bf b}_d,\,{\bf a}_r\times{\bf b}_d + {\bf a}_d\times{\bf b}_r,\,0\right) \nonumber \\
\,&=\,-\,{\bf b}_r\cdot{\bf a}_r\,-\,{\bf b}_d\cdot{\bf a}_d\,-\,{\bf D}\left({\bf b}_r\times{\bf a}_r + {\bf b}_d\times{\bf a}_d,\,{\bf b}_r\times{\bf a}_d + {\bf b}_d\times{\bf a}_r,\,0\right) \nonumber \\
\,&=\,{\bf D}({\bf b}_r,\,{\bf b}_d,\,0)\;{\bf D}({\bf a}_r,\,{\bf a}_d,\,0). \label{rhsden}
\end{align}
Changes in ${\lambda^k}$ thus alternates the {\it relative} order of ${{\bf D}({\bf a}_r,\,{\bf a}_d,\,0)\;{\bf D}({\bf b}_r,\,{\bf b}_d,\,0)}$. As a result, the sum (\ref{362}) reduces to
\begin{align}
{\cal E}^{\rm EPR}_{{\!}_{L.R.}}({\bf a},\,{\bf b})\,&=\!\lim_{\,m\,\rightarrow\,\infty}\!\left[\frac{1}{m}\sum_{k\,=\,1}^{m}{\bf N}({\bf a}_r,\,{\bf a}_d,\,0,\,\lambda^k)\,{\bf N}({\bf b}_r,\,{\bf b}_d,\,0,\,\lambda^k)\,\right] \nonumber \\
&=\frac{1}{2}\{{\bf N}({\bf a}_r,\,{\bf a}_d,\,0,\,\lambda^k=+1)\,{\bf N}({\bf b}_r,\,{\bf b}_d,\,0,\,\lambda^k=+1)\}+\frac{1}{2}\{{\bf N}({\bf a}_r,\,{\bf a}_d,\,0,\,\lambda^k=-1)\,{\bf N}({\bf b}_r,\,{\bf b}_d,\,0,\,\lambda^k=-1)\} \nonumber \\
&=\,\frac{1}{2}\{\,{\bf D}({\bf a}_r,\,{\bf a}_d,\,0)\;{\bf D}({\bf b}_r,\,{\bf b}_d,\,0)\}\,+\,\frac{1}{2}\{\,{\bf D}({\bf b}_r,\,{\bf b}_d,\,0)\;{\bf D}({\bf a}_r,\,{\bf a}_d,\,0)\} \nonumber \\
&=\,-\,\frac{1}{2}\{{\bf a}_r{\bf b}_r\,+\,{\bf b}_r{\bf a}_r\}\,-\,\frac{1}{2}\{{\bf a}_d{\bf b}_d\,+\,{\bf b}_d{\bf a}_d\} \nonumber \\
&=\,-\,{\bf a}_r\cdot{\bf b}_r\,-\,{\bf a}_d\cdot{\bf b}_d\,=\,-\,{\bf a}\cdot{\bf b}\,=\,-\,\cos\theta_{{\bf a}{\bf b}}\,, \label{365b}
\end{align}
because the orientation ${\lambda^k}$ of ${S^7}$ is a fair coin. Here ${{\bf a}\cdot{\bf b}=\frac{1}{2}\{{\bf a}{\bf b}+{\bf b}{\bf a}\}}$ is the standard definition of the inner product.

The above method of calculating the correlations suggests that a given initial state ${\lambda}$ of the physical system can be viewed also as specifying an ordering relation between ${{\bf N}({\bf n}_r,\,{\bf n}_d,\,0,\,\lambda)}$ and the detectors ${{\bf D}({\bf n}_r,\,{\bf n}_d,\,0)}$ that measure${\;}$it:
\begin{equation}
{\bf N}({\bf a}_r,\,{\bf a}_d,\,0,\,\lambda^k=+1)\;{\bf N}({\bf b}_r,\,{\bf b}_d,\,0,\,\lambda^k=+1)\,=\,{\bf D}({\bf a}_r,\,{\bf a}_d,\,0)\;{\bf D}({\bf b}_r,\,{\bf b}_d,\,0) \label{pair11}
\end{equation}
or
\begin{equation}
\text{\;}{\bf N}({\bf a}_r,\,{\bf a}_d,\,0,\,\lambda^k=-1)\;{\bf N}({\bf b}_r,\,{\bf b}_d,\,0,\,\lambda^k=-1)\,=\,{\bf D}({\bf b}_r,\,{\bf b}_d,\,0)\;{\bf D}({\bf a}_r,\,{\bf a}_d,\,0). \label{pair22}
\end{equation}
Then, using the right-hand sides of the Eqs.${\,}$(\ref{rhspen}) and (\ref{rhsden}), the above pair can be reduced to the combined relation 
\begin{equation}
{\bf N}\big({\bf a}_r\times{\bf b}_r + {\bf a}_d\times{\bf b}_d,\,{\bf a}_r\times{\bf b}_d + {\bf a}_d\times{\bf b}_r,\,0,\,\lambda^k\big)\,=\,\lambda^k\,{\bf D}\left({\bf a}_r\times{\bf b}_r + {\bf a}_d\times{\bf b}_d,\,{\bf a}_r\times{\bf b}_d + {\bf a}_d\times{\bf b}_r,\,0\right),
\end{equation}
which is equivalent to the relation (\ref{ori-rela}) for normalized vectors. We have thus proved that the ordering relations (\ref{pair11}) and (\ref{pair22}) between ${\,{\bf N}({\bf n}_r,\,{\bf n}_d,\,0,\,\lambda)\,}$ and ${\,{\bf D}({\bf n}_r,\,{\bf n}_d,\,0)\,}$ are equivalent to the alternatively possible orientations of ${S^7}$.

\subsubsection{Conservation of the Initial Spin-${0}$ from the Twist in the Hopf Bundle of ${S^3}$}

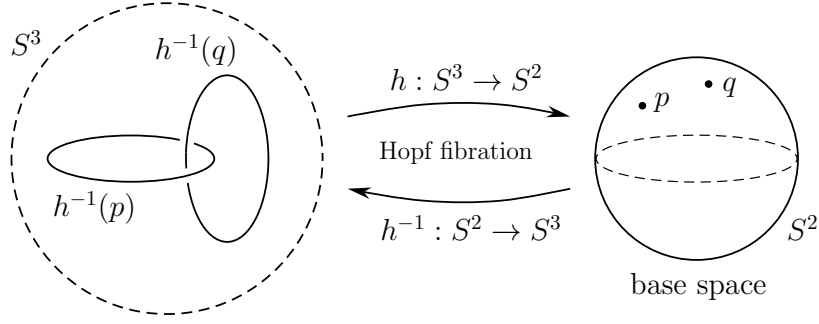
\begin{figure}
\hrule
\scalebox{0.8}{
\begin{pspicture}(0.3,-3.9)(4.5,3.1)

\pscircle[linewidth=0.3mm,linestyle=dashed](-1.8,-0.45){2.6}

\psellipse[linewidth=0.3mm](-0.8,-0.45)(0.7,1.4)

\psellipse[linewidth=0.3mm,border=3pt](-2.4,-0.45)(1.4,0.4)

\pscurve[linewidth=0.3mm,border=3pt](-1.485,-0.35)(-1.48,-0.25)(-1.45,0.0)

\pscircle[linewidth=0.3mm](7.0,-0.45){1.7}

\psellipse[linewidth=0.2mm,linestyle=dashed](7.0,-0.45)(1.68,0.4)

\put(-4.4,1.27){{\Large ${S^3}$}}

\put(-2.0,1.2){{\Large ${h^{-1}(q)}$}}

\put(-3.7,-1.4){{\Large ${h^{-1}(p)}$}}

\put(7.43,0.67){{\Large ${q}$}}

\psdot*(7.2,0.79)

\put(6.3,0.43){{\Large ${p}$}}

\psdot*(6.1,0.43)

\put(8.5,-1.8){{\Large ${S^2}$}}

\put(5.9,-2.7){\Large base space}

\put(1.9,0.7){\Large ${h:S^3\rightarrow S^2}$}

\pscurve[linewidth=0.3mm,arrowinset=0.3,arrowsize=3pt 3,arrowlength=2]{->}(1.2,0.25)(2.47,0.45)(3.74,0.45)(4.9,0.25)

\put(1.73,-0.45){\large Hopf fibration}

\pscurve[linewidth=0.3mm,arrowinset=0.3,arrowsize=3pt 3,arrowlength=2]{->}(4.9,-0.95)(3.74,-1.15)(2.47,-1.15)(1.2,-0.95)

\put(1.75,-1.8){\Large ${h^{-1}:S^2\rightarrow S^3}$}

\end{pspicture}}
\hrule
\caption{The tangled web of linked Hopf circles depicting the geometrical and topological non-trivialities of the 3-sphere.}
\vspace{5pt}
\label{fig-8}
\hrule
\end{figure}

Note that, apart from the initial state ${\lambda^k}$, the only other assumption used in the above derivation is that of the conservation of spin angular momentum (\ref{75699}). These two assumptions are necessary and sufficient to dictate the singlet correlations:
\begin{equation}
{\cal E}^{\rm EPR}_{{\!}_{L.R.}}({\bf a},\,{\bf b})\,=\!\lim_{\,m\,\rightarrow\,\infty}\left[\frac{1}{m}\sum_{k\,=\,1}^{m}\,{\mathscr A}({\bf a}\,,\,\lambda^k)\;{\mathscr B}({\bf b}\,,\,\lambda^k)\right] =\,-\,\cos\theta_{{\bf a}{\bf b}}\,.
\end{equation}
The conservation of spin, however, can be understood in terms of the twist in the Hopf bundle of ${S^3 \cong {\rm SU(2)}}$. Recall that locally (in the topological sense) ${S^3}$ can be written as a product ${S^2\times S^1}$, but globally it has no cross-section \cite{Ryder}. It can be viewed also as a principal U(1) bundle over ${S^2}$, with the points of its base space ${S^2}$ being the elements of the Lie algebra su(2), which are pure quaternions, or bivectors \cite{disproof,local,Eguchi}. The product of two such bivectors are in general non-pure quaternions of the form (\ref{new3}), and are elements of the group SU(2) itself. That is to say, they are points of the bundle space ${S^3}$, whose elements are the preimages of the points of the base space ${S^2}$ \cite{Ryder}. These preimages are 1-spheres, ${S^1}$, called Hopf circles, or Clifford parallels \cite{Penrose}. Since these 1-spheres are the fibers of the bundle, they do not share a single point in common. Each circle threads through every other circle in the bundle as shown in Fig.~\ref{fig-8}, making them linked together in a highly non-trivial configuration. This configuration can be quantified by the following relation among the fibers \cite{Eguchi}:
\begin{equation}
e^{i\psi_-}\,=\,e^{i\phi}\,e^{i\psi_+}\,, \label{72-33}
\end{equation}
where ${e^{i\psi_-}}$ and ${e^{i\psi_+}}$, respectively, are the U(1) fiber coordinates above the two hemispheres ${H_-}$ and ${H_+}$ of the base space ${S^2}$, with spherical coordinates ${(0\leqslant\theta < \pi,\;0\leqslant\phi < 2\pi)}$; ${\phi}$ is the angle parameterizing a thin strip ${H_-\cap H_+}$ around the equator of ${S^2}$ [${\theta\sim\frac{\pi}{2}}$]; and ${e^{i\phi}}$ is the transition function that glues the two sections ${H_-}$ and ${H_+}$ together, thus constituting the 3-sphere. It is evident from Eq.${\,}$(\ref{72-33}) that the fibers match perfectly at the angle ${\phi=0}$ (modulo ${2\pi}$), but differ from each other at all intermediate angles ${\phi}$. For example, ${e^{i\psi_-}}$ and ${e^{i\psi_+}}$ differ by a minus sign at the angle ${\phi=\pi}$. Now to derive the conservation of spin (\ref{75699}), we rewrite the exponential relation (\ref{72-33}) in our notation as
\begin{equation}
\left\{-\,{\boldsymbol \xi}({\bf a}_{r})\,{\boldsymbol \xi}({\bf s}_{r1})\right\}\,=\,\left\{{\boldsymbol \xi}({\bf a}_{r})\,{\boldsymbol \xi}({\bf b}_{r})\right\}\, \left\{{\boldsymbol \xi}({\bf s}_{r2})\,{\boldsymbol \xi}({\bf b}_{r})\right\}\label{10900}
\end{equation}
by identifying the angles ${\eta_{{\bf a}_r{\bf s}_{r1}}}$ and ${\eta_{{\bf s}_{r2}{\bf b}_r}}$ between ${{\bf a}_r}$ and ${{\bf s}_{r1}}$ and ${{\bf s}_{r2}}$ and ${{\bf b}_r}$ with the fibers ${\psi_-}$ and ${\psi_+\,}$, and the angle ${\eta_{{\bf a}_r{\bf b}_r}}$ between ${{\bf a}_r}$ and ${{{\bf b}_r}}$ with the generator of the transition function ${e^{i\phi}}$ on the equator of ${S^2}$. Here we have used the sign conventions to match the sign conventions in our definitions (\ref{onedone}) and (\ref{twodone}) and the correlations (\ref{365}). The above representation of Eq.(\ref{72-33}) is not as unusual as it may appear at first sight once we recall that geometric products of the bivectors appearing in it are all non-pure quaternions, which can be parameterized to take the exponential form
\begin{equation}
-\,{\boldsymbol\xi}({\bf u})\,{\boldsymbol\xi}({\bf v})\,=\,-\,(\lambda\,I\cdot{\bf u})\,(\lambda\,I\cdot{\bf v}) \,=\, \cos(\,\eta_{{\bf u}{\bf v}})\,+\,\frac{{\bf u}\wedge{\bf v}}{||{\bf u}\wedge{\bf v}||}\,\sin(\,\eta_{{\bf u}{\bf v}})\,=\,\exp{\left\{\frac{{\bf u}\wedge{\bf v}}{||{\bf u}\wedge{\bf v}||}\;\eta_{{\bf u}{\bf v}}\right\}}\,.
\end{equation}
Multiplying both sides of Eq.${\,}$(\ref{10900}) from the left with ${{\boldsymbol \xi}({\bf a}_{r})}$ and noting that all unit bivectors square to ${-1}$, we obtain 
\begin{equation}
{\boldsymbol \xi}({\bf s}_{r1})\,=\,-\,{\boldsymbol \xi}({\bf b}_{r})\,{\boldsymbol \xi}({\bf s}_{r2})\,{\boldsymbol \xi}({\bf b}_{r})\,.
\end{equation}
Multiplying the numerator and denominator on the RHS of this similarity relation with ${-\,{\boldsymbol \xi}({\bf b}_{r})}$ from the right and ${{\boldsymbol \xi}({\bf b}_{r})}$ from the left then leads to the conservation of the spin angular momentum, just as we have specified in Eq.${\,}$(\ref{656a}):
\begin{equation}
\lambda\,{\boldsymbol \xi}({\bf s}_{r1})\,=\,\lambda\,{\boldsymbol \xi}({\bf s}_{r2})\;\Longleftrightarrow\;{\bf s}_{r1}\,=\,{\bf s}_{r2}\,. \label{conspin1}
\end{equation}
Similarly, we can derive analogous conservation law for the zero spin within the dual 3-sphere, as specified in Eq.${\,}$(\ref{656}):
\begin{equation}
\lambda\,{\boldsymbol \xi}({\bf s}_{d1})\,=\,\lambda\,{\boldsymbol \xi}({\bf s}_{d2})\;\Longleftrightarrow\;{\bf s}_{d1}\,=\,{\bf s}_{d2}\,. \label{conspin2}
\end{equation}
Given the conservation laws derived in Eqs.${\,}$(\ref{conspin1}) and (\ref{conspin2}), we can combine them to arrive at the net condition (\ref{75699}):
\begin{equation}
{\bf N}({\bf s}_{r1},\,{\bf s}_{d1},\,0,\,\lambda^k)\,{\bf N}({\bf s}_{r2},\,{\bf s}_{d2},\,0,\,\lambda^k)\,=\,-1\,, \label{hopfs3}
\end{equation}
which was used in Eq.${\,}$(\ref{35992}) to derive the strong correlations (\ref{365}). We have thus shown that the conservation of spin angular momentum is not an additional assumption, but follows from the very geometry and topology of the 3-sphere.

In fact it is not difficult to see from the twist in the Hopf bundle of ${S^3}$, captured in Eq.${\,}$(\ref{10900}), that if we set ${{\bf a}_r={\bf b}_r}$ (or equivalently ${\eta_{{\bf a}_r{\bf b}_r}=0}$) for all fibers, then ${S^3}$ reduces to the trivial bundle ${S^2\times S^1}$, since then the fiber coordinates ${\eta_{{\bf a}_r{\bf s}_{r1}}}$ and ${\eta_{{\bf s}_{r2}{\bf b}_r}}$ would match up exactly on the equator of ${S^2}$ [${\theta\sim\frac{\pi}{2}}$]. In general, however, for ${{\bf a}_r\not={\bf b}_r}$, ${S^3\not=S^2\times S^1}$. For example, when ${{\bf a}_r=-\,{\bf b}_r}$ (or equivalently when ${\eta_{{\bf a}_r{\bf b}_r}=\pi}$) there will be a sign difference between the fibers at that point of the equator \cite{Ryder,Eguchi}. That in turn  would produce a twist in the bundle analogous to the twist in a M\"obius strip. It is this non-trivial twist in the ${S^3}$ bundle that is responsible for the observed sign flips in the product ${\mathscr{A}\mathscr{B}}$ of measurement results, from ${\mathscr{A}\mathscr{B}=-1}$ for ${{\bf a}_r={\bf b}_r}$ to ${\mathscr{A}\mathscr{B}=+1}$ for ${{\bf a}_r=-\,{\bf b}_r}$, as evident from the correlations (\ref{365}). In the appendix of the first chapter of Ref.~\cite{disproof} this is illustrated in a toy model of Alice and Bob in a M\"obius world. But while the twist in a M\"obius strip is in the ${S^1}$ worth of parallel lines that make up the untwisted cylinder, the twist in ${S^3}$ is in the arrangement of the ${S^2}$ worth of circles that make up that 3-sphere (cf. Fig.~\ref{fig-8}) \cite{Eguchi}. 

\subsubsection{The General Case of Arbitrarily Entangled Quantum State}

We now proceed to generalize the above two-particle case\footnote{\label{Hardy}It is worth noting here that correlations predicted by the two-level systems can be reproduced also within a quaternionic 3-sphere model without any reference to the general 7-sphere framework presented here, as we have shown elsewhere \cite{local}. In fact, even highly non-trivial Hardy-type correlations can also be reproduced within the quaternionic 3-sphere model, as we have demonstrated in Chapter 6 of Ref.~\cite{disproof}. The 3-sphere framework, however, is rather restrictive. It can accommodate the singlet correlations and Hardy-type correlations, but cannot reproduce more intricate correlations, such as, for example, those predicted by the rotationally non-invariant GHZ states [cf. subsection \ref{SecGHZ}]. On the other hand, the 7-sphere framework is more general and comprehensive. This is because ${S^7}$ is made of ${S^4}$ worth of 3-spheres, with a highly non-trivial twist in the bundle. In other words, in the language of Hopf fibration, ${S^7}$ is fibrated by ${S^3}$ over the base manifold ${S^4}$. Thus each of the many fibers of ${S^7}$ that make it up is itself an ${S^3}$. It is therefore not surprising that ${S^7}$ framework presented in this paper is more complete and is able to reproduce quantum correlations more comprehensively (cf. appendix \ref{ApA}). Moreover, as we saw in section \ref{222}, the algebraic and geometrical properties of the physical space are captured more completely by the octonion-like representation space ${S^7}$ constructed in Eq.~(\ref{spring}), and not by the 3-dimensional conformal physical space ${S^3}$ of Eq.~(\ref{3-sphere}) itself.} to the most general case of arbitrarily entangled quantum state considered in (\ref{gen-q-state-a}). To this end, let us consider any arbitrary number of measurement results corresponding to those in (\ref{gen-q-state-a}) and (\ref{gen-q-state-b}):
\begin{equation}
{\mathscr A}({\bf a}\,,\,\lambda^k)\;{\mathscr B}({\bf b}\,,\,\lambda^k)\;{\mathscr C}({\bf c}\,,\,\lambda^k)\;{\mathscr D}({\bf d}\,,\,\lambda^k)\;{\mathscr E}({\bf e}\,,\,\lambda^k)\;{\mathscr F}({\bf f}\,,\,\lambda^k)\;{\mathscr G}({\bf g}\,,\,\lambda^k)\,\dots\,, \label{meform}
\end{equation}
with each pair such as ${({\mathscr C},\,{\mathscr D})}$ defined for the contexts such as ${{\bf n}^3={\bf c}\longleftrightarrow{\bf c}_r+{\bf c}_d\,\varepsilon_+\,}$ and ${{\bf n}^4={\bf d}\longleftrightarrow{\bf d}_r+{\bf d}_d\,\varepsilon_+\,}$:
\begin{align}
S^7\ni{\mathscr C}({\bf c}\,,\,\lambda^k):=\!\!&\lim_{\,\substack{{\bf t}_{r1}\,\rightarrow\;{\bf c}_r \\ {\bf t}_{d1}\,\rightarrow\;{\bf c}_d}}\!\left\{\,-\,{\bf D}({\bf c}_r,\,{\bf c}_d,\,0)\,{\bf N}({\bf t}_{r1},\,{\bf t}_{d1},\,0,\,\lambda^k)\,\right\}=\,
\begin{cases}
\,+\,1\;\;\;\;\;{\rm if} &\lambda^k\,=\,+\,1 \\
\,-\,1\;\;\;\;\;{\rm if} &\lambda^k\,=\,-\,1
\end{cases} \Bigg\} \nonumber \\
&\,\;\text{together with}\,\;\langle\, {\mathscr C}({\bf c}\,,\,\lambda^k) \,\rangle\,=\,0
\end{align}
and
\begin{align}
S^7\ni{\mathscr D}({\bf d}\,,\,\lambda^k):=\!\!&\lim_{\,\substack{{\bf t}_{r2}\,\rightarrow\;{\bf d}_r \\ {\bf t}_{d2}\,\rightarrow\;{\bf d}_d}}\!\left\{\,+\,{\bf N}({\bf t}_{r2},\,{\bf t}_{d2},\,0,\,\lambda^k)\,{\bf D}({\bf d}_r,\,{\bf d}_d,\,0)\,\right\}=\,
\begin{cases}
\,-\,1\;\;\;\;\;{\rm if} &\lambda^k\,=\,+\,1 \\
\,+\,1\;\;\;\;\;{\rm if} &\lambda^k\,=\,-\,1
\end{cases} \Bigg\} \nonumber \\
&\,\;\text{together with}\,\;\langle\, {\mathscr D}({\bf d}\,,\,\lambda^k) \,\rangle\,=\,0\,. \label{tttdone}
\end{align}
If the number of measurement results happens to be odd instead of even, then the product of an even number of results can be first evaluated, and then that factor can be paired with the remaining result, as done in Eq.${\,}$(\ref{oddeven}) below.

It is important to recall here the elementary fact that any experiment of any kind in physics can always be reduced to a series of questions with ``yes"/~``no" answers, represented by binary measurement outcomes of the form (\ref{meform}) to (\ref{tttdone}). Therefore the measurement framework we have developed here is completely general and applicable to any physical experiment. 

Now, as in the EPR-Bohm type experiment with a singlet state discussed above [cf. Fig.${\,}$\ref{fig6} and Eqs.${\,}$(\ref{756}), (\ref{75699}) and (\ref{hopfs3})], for each pair of measurement outcomes such as (\ref{tttdone}) the twist in the Hopf bundle of ${S^3}$ dictates the condition
\begin{equation}
-\,{\bf N}({\bf t}_{r1},\,{\bf t}_{d1},\,0,\,\lambda^k)\,+\,{\bf N}({\bf t}_{r2},\,{\bf t}_{d2},\,0,\,\lambda^k)\,=\,0\;\;\Longleftrightarrow\;\;{\bf N}({\bf t}_{r1},\,{\bf t}_{d1},\,0,\,\lambda^k)\,=\,{\bf N}({\bf t}_{r2},\,{\bf t}_{d2},\,0,\,\lambda^k)\,, \label{7756}
\end{equation}
or equivalently the condition
\begin{equation}
{\bf N}({\bf t}_{r1},\,{\bf t}_{d1},\,0,\,\lambda^k)\,{\bf N}({\bf t}_{r2},\,{\bf t}_{d2},\,0,\,\lambda^k)\,=\,\left\{{\bf N}({\bf t}_{r},\,{\bf t}_{d},\,0,\,\lambda^k)\right\}^2\,=\,{\bf N}^2({\bf t}_{r},\,{\bf t}_{d},\,0,\,\lambda^k)\,=\,-1\,. \label{775699}
\end{equation}
Consequently, by following the steps analogous to those in Eqs.${\,}$(\ref{onedone}) to (\ref{362}), we arrive at the geometrical equivalence
\begin{equation}
\lim_{\,m\,\rightarrow\,\infty}\!\left[\frac{1}{m}\sum_{k\,=\,1}^{m}\,{\mathscr C}({\bf c}\,,\,\lambda^k)\;{\mathscr D}({\bf d}\,,\,\lambda^k)\,\right]\,\equiv\!\lim_{\,m\,\rightarrow\,\infty}\!\left[\frac{1}{m}\sum_{k\,=\,1}^{m}\,{\bf N}({\bf c}_r,\,{\bf c}_d,\,0,\,\lambda^k)\,{\bf N}({\bf d}_r,\,{\bf d}_d,\,0,\,\lambda^k)\,\right]
\end{equation}
for each pair ${({\mathscr C},\,{\mathscr D})}$ of measurement outcomes. As a result, the correlations among the outcomes (\ref{meform}) take the form
\begin{align}
&{\cal E}_{{\!}_{L.R.}}({\bf a},\,{\bf b},\,{\bf c},\,{\bf d},\,\dots\,)=\!\!\lim_{\,m\,\rightarrow\,\infty}\!\left[\frac{1}{m}\sum_{k\,=\,1}^{m}\,{\mathscr A}({\bf a}\,,\,\lambda^k)\;{\mathscr B}({\bf b}\,,\,\lambda^k)\;{\mathscr C}({\bf c}\,,\,\lambda^k)\;{\mathscr D}({\bf d}\,,\,\lambda^k)\;\dots\,\right] \label{gg109} \\
&\;\;\;\;=\!\!\lim_{\,m\,\rightarrow\,\infty}\!\left[\frac{1}{m}\sum_{k\,=\,1}^{m}{\bf N}({\bf a}_r,\,{\bf a}_d,\,0,\,\lambda^k)\,{\bf N}({\bf b}_r,\,{\bf b}_d,\,0,\,\lambda^k)\,{\bf N}({\bf c}_r,\,{\bf c}_d,\,0,\,\lambda^k)\,{\bf N}({\bf d}_r,\,{\bf d}_d,\,0,\,\lambda^k)\,\dots\,\right] \label{gg110}\\
&\;\;\;\;=\!\!\lim_{\,m\,\rightarrow\,\infty}\!\left[\frac{1}{m}\sum_{k\,=\,1}^{m}{\bf N}({\bf x}_r,\,{\bf x}_d,\,0,\,\lambda^k)\,{\bf N}({\bf y}_r,\,{\bf y}_d,\,0,\,\lambda^k)\,\right] \label{gg111}\\
&\;\;\;\;=\,-\,{\bf x}_r\cdot{\bf y}_r\,-\,{\bf x}_d\cdot{\bf y}_d\,-\!\!\lim_{\,m\,\rightarrow\,\infty}\left[\frac{1}{m}\sum_{k\,=\,1}^{m}{\bf N}\left({\bf x}_r\times{\bf y}_r + {\bf x}_d\times{\bf y}_d,\,{\bf x}_r\times{\bf y}_d + {\bf x}_d\times{\bf y}_r,\,0,\,\lambda\right)\right] \label{gg112} \\
&\;\;\;\;=\,-\,{\bf x}_r\cdot{\bf y}_r\,-\,{\bf x}_d\cdot{\bf y}_d\,-\!\!\lim_{\,m\,\rightarrow\,\infty}\left[\frac{1}{m}\sum_{k\,=\,1}^{m}\,\lambda^k\right]{\bf D}\left({\bf x}_r\times{\bf y}_r + {\bf x}_d\times{\bf y}_d,\,{\bf x}_r\times{\bf y}_d + {\bf x}_d\times{\bf y}_r,\,0\right) \label{gg113} \\
&\;\;\;\;=\,-\,\cos\theta_{{\bf x}{\bf y}}({\bf a},\,{\bf b},\,{\bf c},\,{\bf d},\,\dots\,)\,-\,0\,, \label{gg114}
\end{align}
because ${\lambda^k}$ is a fair coin. We can now identify this locally causal expectation with its quantum mechanical counterpart:
\begin{equation}
\langle\Psi|\;{\cal\widehat O}({\bf a},\,{\bf b},\,{\bf c},\,{\bf d},\,\dots\,)\,|\Psi\rangle\,=\,{\cal E}_{{\!}_{L.R.}}({\bf a},\,{\bf b},\,{\bf c},\,{\bf d},\,\dots\,)\,=\,-\,\cos\theta_{{\bf x}{\bf y}}({\bf a},\,{\bf b},\,{\bf c},\,{\bf d},\,\dots\,)\,.
\end{equation}
This completes the proof of the theorem for the general quantum state stated at the beginning of the subsection${\;}$\ref{theo}.

\subsection{Derivation of Tsirel'son's Bounds on the Correlation Strength}\label{T-bound}

Let us now investigate the bounds on the strengths of the local-realistic correlations (\ref{gg114}) by deriving Tsirel'son's bounds \cite{IJTP} for arbitrary quantum states \cite{disproof}. To this end, instead of (\ref{meform}) consider an alternative set of measurement results such as 
\begin{equation}
{\mathscr A}({\bf a'}\,,\,\lambda^k)\;{\mathscr B}({\bf b'}\,,\,\lambda^k)\;{\mathscr C}({\bf c'}\,,\,\lambda^k)\;{\mathscr D}({\bf d'}\,,\,\lambda^k)\;{\mathscr E}({\bf e'}\,,\,\lambda^k)\;{\mathscr F}({\bf f'}\,,\,\lambda^k)\;{\mathscr G}({\bf g'}\,,\,\lambda^k)\,\dots\,, \label{youmeform}
\end{equation}
with each pair such as ${({\mathscr C},\,{\mathscr D})}$ defined for contexts such as ${{\bf n'}^3={\bf c'}\longleftrightarrow{\bf c'}_r+{\bf c'}_d\,\varepsilon_+\,}$ and ${{\bf n'}^4={\bf d'}\longleftrightarrow{\bf d'}_r+{\bf d'}_d\,\varepsilon_+\,}$. The correlation between these results can then be derived following steps analogous to those in the previous subsection: 
\begin{align}
&\!{\cal E}_{{\!}_{L.R.}}({\bf a'},\,{\bf b'},\,{\bf c'},\,{\bf d'},\,\dots\,)\,=\!\!\lim_{\,m\,\rightarrow\,\infty}\!\left[\frac{1}{m}\!\sum_{k\,=\,1}^{m}{\mathscr A}({\bf a'}\,,\,\lambda^k)\;{\mathscr B}({\bf b'}\,,\,\lambda^k)\;{\mathscr C}({\bf c'}\,,\,\lambda^k)\;{\mathscr D}({\bf d'}\,,\,\lambda^k)\dots\right] \label{7357111ffpp} \\
&\!=\!\!\lim_{\,m\,\rightarrow\,\infty}\!\left[\frac{1}{m}\!\sum_{k\,=\,1}^{m}\!{\bf N}({\bf a'}_r,\,{\bf a'}_d,\,0,\,\lambda^k)\,{\bf N}({\bf b'}_r,\,{\bf b'}_d,\,0,\,\lambda^k)\,{\bf N}({\bf c'}_r,\,{\bf c'}_d,\,0,\,\lambda^k)\,{\bf N}({\bf d'}_r,\,{\bf d'}_d,\,0,\,\lambda^k)\dots\!\right] \label{7362111ffpp}\\
&\!=\!\!\lim_{\,m\,\rightarrow\,\infty}\!\left[\frac{1}{m}\sum_{k\,=\,1}^{m}{\bf N}({\bf x'}_r,\,{\bf x'}_d,\,0,\,\lambda^k)\,{\bf N}({\bf y'}_r,\,{\bf y'}_d,\,0,\,\lambda^k)\,\right] \label{99762111ffpp}\\
&\!=\,-\,{\bf x'}_r\cdot{\bf y'}_r\,-\,{\bf x'}_d\cdot{\bf y'}_d\,-\!\!\lim_{\,m\,\rightarrow\,\infty}\left[\frac{1}{m}\sum_{k\,=\,1}^{m}{\bf N}\!\left({\bf x'}_r\times{\bf y'}_r + {\bf x'}_d\times{\bf y'}_d,\,{\bf x'}_r\times{\bf y'}_d + {\bf x'}_d\times{\bf y'}_r,\,0,\,\lambda\right)\right] \label{ggg112} \\
&\!=\,-\,{\bf x'}_r\cdot{\bf y'}_r\,-\,{\bf x'}_d\cdot{\bf y'}_d\,-\!\!\lim_{\,m\,\rightarrow\,\infty}\left[\frac{1}{m}\sum_{k\,=\,1}^{m}\,\lambda^k\right]\!{\bf D}\!\left({\bf x'}_r\times{\bf y'}_r + {\bf x'}_d\times{\bf y'}_d,\,{\bf x'}_r\times{\bf y'}_d + {\bf x'}_d\times{\bf y'}_r,\,0\right) \label{ggg113} \\
&\!=\,-\,\cos\theta_{{\bf x'}{\bf y'}}({\bf a'},\,{\bf b'},\,{\bf c'},\,{\bf d'},\,\dots\,)\,-\,0\,. \label{ggg114}
\end{align}
In particular, in Eq.${\,}$(\ref{99762111ffpp}) we then have the relation 
\begin{equation}
{\cal E}_{{\!}_{L.R.}}({\bf a'},\,{\bf b'},\,{\bf c'},\,{\bf d'},\,\dots\,)\,=\,{\cal E}_{{\!}_{L.R.}}({\bf x'},\,{\bf y'})\,=\!\!\lim_{\,m\,\rightarrow\,\infty}\!\left[\frac{1}{m}\!\sum_{k\,=\,1}^{m}\!{\bf N}({\bf x'}_r,\,{\bf x'}_d,\,0,\,\lambda^k)\,{\bf N}({\bf y'}_r,\,{\bf y'}_d,\,0,\,\lambda^k)\right]\!.\label{ffppoo}
\end{equation}
Needless to say, we are free to choose the contexts different from the primed and unprimed ones chosen in (\ref{youmeform}) and (\ref{meform}), as well as any combinations and/or mixtures of them, such as ${({\bf a},\,{\bf b'},\,{\bf c''},\,{\bf d'''},\,{\bf e''''},\,\dots\,)}$. Consequently, we may consider the following four relations corresponding to some alternative combinations of measurement contexts so that
\begin{align}
{\cal E}_{{\!}_{L.R.}}({\bf x},\,{\bf y})\,&=\!\!\lim_{\,m\,\rightarrow\,\infty}\!\left[\frac{1}{m}\sum_{k\,=\,1}^{m}{\bf N}({\bf x}_r,\,{\bf x}_d,\,0,\,\lambda^k)\,{\bf N}({\bf y}_r,\,{\bf y}_d,\,0,\,\lambda^k)\,\right], \\ \notag \\
{\cal E}_{{\!}_{L.R.}}({\bf x},\,{\bf y'})\,&=\!\!\lim_{\,m\,\rightarrow\,\infty}\!\left[\frac{1}{m}\sum_{k\,=\,1}^{m}{\bf N}({\bf x}_r,\,{\bf x}_d,\,0,\,\lambda^k)\,{\bf N}({\bf y'}_r,\,{\bf y'}_d,\,0,\,\lambda^k)\,\right], \\ \notag \\
{\cal E}_{{\!}_{L.R.}}({\bf x'},\,{\bf y})\,&=\!\!\lim_{\,m\,\rightarrow\,\infty}\!\left[\frac{1}{m}\sum_{k\,=\,1}^{m}{\bf N}({\bf x'}_r,\,{\bf x'}_d,\,0,\,\lambda^k)\,{\bf N}({\bf y}_r,\,{\bf y}_d,\,0,\,\lambda^k)\,\right], \\ \notag \\
\text{and}\;\;\,\;\;{\cal E}_{{\!}_{L.R.}}({\bf x'},\,{\bf y'})\,&=\!\!\lim_{\,m\,\rightarrow\,\infty}\!\left[\frac{1}{m}\sum_{k\,=\,1}^{m}{\bf N}({\bf x'}_r,\,{\bf x'}_d,\,0,\,\lambda^k)\,{\bf N}({\bf y'}_r,\,{\bf y'}_d,\,0,\,\lambda^k)\,\right]. 
\end{align}
Using the above four expressions the corresponding Bell-CHSH string of expectation values \cite{IJTP}, namely the coefficient
\begin{equation}
{\cal E}_{{\!}_{L.R.}}({\bf x},\,{\bf y})\,+\,{\cal E}_{{\!}_{L.R.}}({\bf x},\,{\bf y'})\,+\,
{\cal E}_{{\!}_{L.R.}}({\bf x'},\,{\bf y})\,-\,{\cal E}_{{\!}_{L.R.}}({\bf x'},\,{\bf y'}) \label{B1-11}
\end{equation}
corresponding to this fully general case of arbitrary number of contexts and measurement results, can be written as
\begin{align}
{\cal E}_{{\!}_{L.R.}}({\bf x},\,{\bf y})\,+\,&{\cal E}_{{\!}_{L.R.}}({\bf x},\,{\bf y'})\,+\,
{\cal E}_{{\!}_{L.R.}}({\bf x'},\,{\bf y})\,-\,{\cal E}_{{\!}_{L.R.}}({\bf x'},\,{\bf y'}) \nonumber \\
&=\lim_{\,m\,\rightarrow\,\infty}\!\Bigg[\frac{1}{m}\sum_{k\,=\,1}^{m}\big\{\,{\bf N}({\bf x}_r,\,{\bf x}_d,\,0,\,\lambda^k)\,{\bf N}({\bf y}_r,\,{\bf y}_d,\,0,\,\lambda^k) \notag\\
&\;\;\;\;\;\;\;\;\;\;\;\;\;\;\;\;\;\;\;\;\;\;\;\;\;\;\;\;\;\;\;\;\;\;\;\;\;\;\;\;\;\;\;\;\;\;\;\;\;+\,{\bf N}({\bf x}_r,\,{\bf x}_d,\,0,\,\lambda^k)\,{\bf N}({\bf y'}_r,\,{\bf y'}_d,\,0,\,\lambda^k) \notag\\ \notag\\
&\;\;\;\;\;\;\;\;\;\;\;\;\;\;\;\;\;\;\;\;\;\;\;\;\;\;\;\;\;\;\;\;\;\;\;\;\;\;\;\;\;\;\;\;\;\;\;\;\;\;\;\;\;\;+\,{\bf N}({\bf x'}_r,\,{\bf x'}_d,\,0,\,\lambda^k)\,{\bf N}({\bf y}_r,\,{\bf y}_d,\,0,\,\lambda^k) \notag\\
&\;\;\;\;\;\;\;\;\;\;\;\;\;\;\;\;\;\;\;\;\;\;\;\;\;\;\;\;\;\;\;\;\;\;\;\;\;\;\;\;\;\;\;\;\;\;\;\;\;\;\;\;\;\;\;\;\;\;-\,{\bf N}({\bf x'}_r,\,{\bf x'}_d,\,0,\,\lambda^k)\,{\bf N}({\bf y'}_r,\,{\bf y'}_d,\,0,\,\lambda^k)\,\big\}\Bigg]. \label{probnonint}
\end{align}
But since ${{\bf N}({\bf x}_r,\,{\bf x}_d,\,0,\,\lambda^k)}$ and ${{\bf N}({\bf y}_r,\,{\bf y}_d,\,0,\,\lambda^k)}$ represent two independent equatorial points of an ${S^6}$ within ${S^7}$, we take them to belong to two disconnected ``sections" of the bundle ${S^5\times S^1}$ ({\it i.e.}, two disconnected ${S^5\subset S^6}$), satisfying
\begin{equation}
\left[\,{\bf N}({\bf x}_r,\,{\bf x}_d,\,0,\,\lambda^k),\;{\bf N}({\bf y}_r,\,{\bf y}_d,\,0,\,\lambda^k)\,\right]\,=\,0\,
\;\;\;\forall\;\,{\bf x}_r\;\,{\rm and}\;\,{\bf y}_d\,\in\,{\rm I\!R}^3,\label{com}
\end{equation}
which is equivalent to anticipating null outcomes along the directions ${{\bf x}_r\times{\bf y}_d}$ exclusive to both ${{\bf x}_r}$ and ${{\bf y}_d}$. If we now square the integrand of equation (\ref{probnonint}), use the above commutation relations, and use the fact that all ${{\bf N}({\bf n}_r,\,{\bf n}_d,\,0,\,\lambda^k)}$ square to ${-1}$, then the absolute value of the above Bell-CHSH string (\ref{B1-11}) leads to the following variance inequality:
\begin{equation}
|{\cal E}_{{\!}_{L.R.}}({\bf x},\,{\bf y})\,+\,{\cal E}_{{\!}_{L.R.}}({\bf x},\,{\bf y'})\,+\,{\cal E}_{{\!}_{L.R.}}({\bf x'},\,{\bf y})\,-\,{\cal E}_{{\!}_{L.R.}}({\bf x'},\,{\bf y'})| \leqslant \sqrt{\lim_{\,m\,\rightarrow\,\infty}\left[\frac{1}{m}\sum_{k\,=\,1}^{m}\,\big\{\,4\,+\,4\,{\mathscr T}_{\,{\bf x\,x'}}({\lambda}^k)\,{\mathscr T}_{\,{\bf y'\,y}}({\lambda}^k)\,\big\}\right]},\label{youever}
\end{equation}
where the classical commutators
\begin{equation}
{\mathscr T}_{\,{\bf x\,x'}}(\lambda^k):=\frac{1}{2}\left[\,{\bf N}({\bf x}_r,\,{\bf x}_d,\,0,\,\lambda^k),\;{\bf N}({\bf x'}_r,\,{\bf x'}_d,\,0,\,\lambda^k)\,\right]\,=\,-\,{\bf N}\left({\bf x}_r\times{\bf x'}_r + {\bf x}_d\times{\bf x'}_d,\,{\bf x}_r\times{\bf x'}_d + {\bf x}_d\times{\bf x'}_r,\,0,\,\lambda^k\right) \label{aa-potorsion-666}
\end{equation}
and
\begin{equation}
{\mathscr T}_{\,{\bf y'\,y}}(\lambda^k):=\frac{1}{2}\left[\,{\bf N}({\bf y'}_r,\,{\bf y'}_d,\,0,\,\lambda^k),\;{\bf N}({\bf y}_r,\,{\bf y}_d,\,0,\,\lambda^k)\right]\,=\,-\,{\bf N}\left({\bf y'}_r\times{\bf y}_r + {\bf y'}_d\times{\bf y}_d,\,{\bf y'}_r\times{\bf y}_d + {\bf y'}_d\times{\bf y}_r,\,0,\,\lambda^k\right) \label{bb-potor}
\end{equation}
are the geometric measures of the torsion within ${S^7}$ \cite{disproof,IJTP}. Thus, it is the non-vanishing torsion ${\mathscr T}$ within ${S^7}$ --- the parallelizing torsion which makes the Riemann curvature of this representation space vanish --- that is responsible for the stronger-than-linear correlations. We can see this from Eq.${\,}$(\ref{youever}) by setting ${{\mathscr T}=0}$, and in more detail as follows.

Using the above expressions for the intrinsic torsions ${{\mathscr T}_{\,{\bf x\,x'}}(\lambda^k)}$ and ${{\mathscr T}_{\,{\bf y'\,y}}(\lambda^k)}$ and defining the unnormalized vectors
\begin{equation}
{\bf u}_r:=\left({\bf x}_r\times{\bf x'}_r + {\bf x}_d\times{\bf x'}_d\right) \;\;\;\text{and}\;\;\;
{\bf u}_d:=\left({\bf x}_r\times{\bf x'}_d + {\bf x}_d\times{\bf x'}_r\right)
\end{equation}
and
\begin{equation}
{\bf v}_r:=\left({\bf y'}_r\times{\bf y}_r + {\bf y'}_d\times{\bf y}_d\right) \;\;\;\text{and}\;\;\;
{\bf v}_d:=\left({\bf y'}_r\times{\bf y}_d + {\bf y'}_d\times{\bf y}_r\right),
\end{equation}
together with ${{\bf u}\cdot{\bf v}:={\bf u}_r\cdot{\bf v}_r\,+\,{\bf u}_d\cdot{\bf v}_d\,}$ analogous to ${{\bf a}\cdot{\bf b}:={\bf a}_r\cdot{\bf b}_r\,+\,{\bf a}_d\cdot{\bf b}_d\,}$ given in Eq.${\,}$(\ref{new-dotp}), we have the product
\begin{align}
{\mathscr T}_{\,{\bf x\,x'}}(\lambda^k)\,{\mathscr T}_{\,{\bf y'\,y}}(\lambda^k)\,
&=\,-{\bf u}_r\cdot{\bf v}_r-{\bf u}_d\cdot{\bf v}_d -{\bf N}\left({\bf u}_r\times{\bf v}_r + {\bf u}_d\times{\bf v}_d,\,{\bf u}_r\times{\bf v}_d + {\bf u}_d\times{\bf v}_r,\,0,\,\lambda^k\right) \notag \\
&=\,-{\bf u}\cdot{\bf v}-{\bf N}\left({\bf u}_r\times{\bf v}_r + {\bf u}_d\times{\bf v}_d,\,{\bf u}_r\times{\bf v}_d + {\bf u}_d\times{\bf v}_r,\,0,\,\lambda^k\right).
\end{align}
As a result, we have
\begin{align}
\lim_{\,m\,\rightarrow\,\infty}\left[\frac{1}{m}\sum_{k\,=\,1}^{m}{\mathscr T}_{\,{\bf x\,x'}}(\lambda^k)\,{\mathscr T}_{\,{\bf y'\,y}}(\lambda^k)\right] &\,=-\,{\bf u}\cdot{\bf v}\,-\!\lim_{\,m\,\rightarrow\,\infty}\left[\frac{1}{m}\sum_{k\,=\,1}^{m}{\bf N}\left({\bf u}_r\times{\bf v}_r + {\bf u}_d\times{\bf v}_d,\,{\bf u}_r\times{\bf v}_d + {\bf u}_d\times{\bf v}_r,\,0,\,\lambda^k\right)\right] \notag \\
&\,=-\,{\bf u}\cdot{\bf v}\,-\!\lim_{\,m\,\rightarrow\,\infty}\left[\frac{1}{m}\sum_{k\,=\,1}^{m}\lambda^k\right]{\bf D}\left({\bf u}_r\times{\bf v}_r + {\bf u}_d\times{\bf v}_d,\,{\bf u}_r\times{\bf v}_d + {\bf u}_d\times{\bf v}_r,\,0\right) \notag \\
&\,=-\,{\bf u}\cdot{\bf v}\,-\,0\,=\,-\,{\bf u}_r\cdot{\bf v}_r\,-\,{\bf u}_d\cdot{\bf v}_d\,,
\end{align}
where ${\bf u}$ and ${\bf v}$ are unnormalized vectors. Using the constraints analogous to those expressed in Eq.${\,}$(\ref{new-dotp}), we then have
\begin{equation}
\lim_{\,m\,\rightarrow\,\infty}\left[\frac{1}{m}\sum_{k\,=\,1}^{m}{\mathscr T}_{\,{\bf x\,x'}}(\lambda^k)\,{\mathscr T}_{\,{\bf y'\,y}}(\lambda^k)\right] =\,-\,{\bf u}_r\cdot{\bf v}_r\,-\,{\bf u}_d\cdot{\bf v}_d \,=\,-\,\left({\bf x}\times{\bf x'}\right)\cdot\left({\bf y'}\times{\bf y}\right),
\end{equation}
upon using a vector identity. Consequently, substituting the above value in the variance inequality (\ref{youever}), it simplifies to
\begin{equation}
\left|\,{\cal E}_{{\!}_{L.R.}}({\bf x},\,{\bf y})\,+\,{\cal E}_{{\!}_{L.R.}}({\bf x},\,{\bf y'})\,+\,{\cal E}_{{\!}_{L.R.}}({\bf x'},\,{\bf y})\,-\,{\cal E}_{{\!}_{L.R.}}({\bf x'},\,{\bf y'})\,\right|\,\leqslant\,2\,\sqrt{\,1\,-\,\left({\bf x}\times{\bf x'}\right)\cdot\left({\bf y'}\times{\bf y}\right)\,}\,.
\end{equation}
Finally, since trigonometry dictates the geometrical bounds ${
-1\leqslant\,\left({\bf x}\times{\bf x'}\right)\cdot\left({\bf y'}\times{\bf y}\right)\,\leqslant +1}$, this inequality reduces to
\begin{equation}
\left|\,{\cal E}_{{\!}_{L.R.}}({\bf x},\,{\bf y})\,+\,{\cal E}_{{\!}_{L.R.}}({\bf x},\,{\bf y'})\,+\,{\cal E}_{{\!}_{L.R.}}({\bf x'},\,{\bf y})\,-\,{\cal E}_{{\!}_{L.R.}}({\bf x'},\,{\bf y'})\,\right|\,\leqslant\,2\sqrt{2}\,,
\label{My-CHSH}
\end{equation}
exhibiting the bounds on all possible correlations. This result can also be derived directly from the correlations (\ref{gg114}):
\begin{equation}
|\,{\cal E}_{{\!}_{L.R.}}({\bf x},\,{\bf y})\,+\,{\cal E}_{{\!}_{L.R.}}({\bf x},\,{\bf y'})\,+\,{\cal E}_{{\!}_{L.R.}}({\bf x'},\,{\bf y})\,-\,{\cal E}_{{\!}_{L.R.}}({\bf x'},\,{\bf y'})\,| \,=\,\left|\,-\,\cos\theta_{{\bf x}{\bf y}}\,
-\,\cos\theta_{{\bf x}{\bf y'}}\,-\,\cos\theta_{{\bf x'}{\bf y}}\,+\,\cos\theta_{{\bf x'}{\bf y'}}\,\right|\,\leqslant\,2\sqrt{2}\,. \label{Your-CHSH}
\end{equation}
Let us stress again that these bounds are completely general, valid for any quantum state, such as the one in Eq.${\,}$(\ref{gen-q-state-b}).

\subsection{Fragility of Strong Correlations Increases with Number of Contexts}\label{Frag}

As we saw in Eq.${\,}$(\ref{365}), in the case of two contexts the scalar part of the geometric product ${{\bf N}({\bf a}_r,\,{\bf a}_d,\,0,\,\lambda^k)\,{\bf N}({\bf b}_r,\,{\bf b}_d,\,0,\,\lambda^k)}$ is
\begin{equation}
-\,\cos\theta_{{\bf x}{\bf y}}({\bf a},\,{\bf b})\,=\,-\,{\bf a}_r\cdot{\bf b}_r\,-\,{\bf a}_d\cdot{\bf b}_d\,=\,-\,{\bf a}\cdot{\bf b}\,=\,-\,\cos\theta_{{\bf a}{\bf b}}\,.
\end{equation}
And it is this scalar part that captures the pattern of strong correlations exhibited by the singlet system. Analogously, for three contexts the scalar part of the corresponding geometric product ${{\bf N}({\bf a}_r,\,{\bf a}_d,\,0,\,\lambda^k)\,{\bf N}({\bf b}_r,\,{\bf b}_d,\,0,\,\lambda^k)\,{\bf N}({\bf c}_r,\,{\bf c}_d,\,0,\,\lambda^k)}$ works out to give
\begin{equation}
-\,\cos\theta_{{\bf x}{\bf y}}({\bf a},\,{\bf b},\,{\bf c})\,=\,{\bf a}_r\cdot\{({\bf b}_r\times{\bf c}_r)\,+\,({\bf b}_d\times{\bf c}_d)\}\,+\,{\bf a}_d\cdot\{({\bf b}_r\times{\bf c}_d)\,+\,({\bf b}_d\times{\bf c}_r)\}\,, \label{oddeven}
\end{equation}
with the geometric complexity of the scalar part now increased considerably. And for four contexts the scalar part of the geometric product
\begin{equation}
{\bf N}({\bf a}_r,\,{\bf a}_d,\,0,\,\lambda^k)\,{\bf N}({\bf b}_r,\,{\bf b}_d,\,0,\,\lambda^k)\,{\bf N}({\bf c}_r,\,{\bf c}_d,\,0,\,\lambda^k)\,{\bf N}({\bf d}_r,\,{\bf d}_d,\,0,\,\lambda^k) \notag
\end{equation}
works out to be even more intricate:
\begin{align}
-\,\cos\theta_{{\bf x}{\bf y}}({\bf a},\,{\bf b},\,{\bf c},\,{\bf d})\,&=\,({\bf a}_r\cdot{\bf b}_r)({\bf c}_r\cdot{\bf d}_r)\,+\,({\bf a}_d\cdot{\bf b}_d)({\bf c}_r\cdot{\bf d}_r)\,+\,({\bf a}_r\cdot{\bf b}_r)({\bf c}_d\cdot{\bf d}_d)\,+\,({\bf a}_d\cdot{\bf b}_d)({\bf c}_d\cdot{\bf d}_d)\notag \\
&\;\,-\left({\bf a}_r\times{\bf b}_r + {\bf a}_d\times{\bf b}_d\right)\cdot\left({\bf c}_r\times{\bf d}_r + {\bf c}_d\times{\bf d}_d\right)\,-\,\left({\bf a}_r\times{\bf b}_d + {\bf a}_d\times{\bf b}_r\right)\cdot\left({\bf c}_r\times{\bf d}_d + {\bf c}_d\times{\bf d}_r\right), \label{expectedfour}
\end{align}
because
\begin{equation}
{\bf N}({\bf a}_r,\,{\bf a}_d,\,0,\,\lambda)\,{\bf N}({\bf b}_r,\,{\bf b}_d,\,0,\,\lambda)\,=\,-\,{\bf a}_r\cdot{\bf b}_r\,-\,{\bf a}_d\cdot{\bf b}_d\,-\,{\bf N}\left({\bf a}_r\times{\bf b}_r + {\bf a}_d\times{\bf b}_d,\,{\bf a}_r\times{\bf b}_d + {\bf a}_d\times{\bf b}_r,\,0,\,\lambda\right)
\end{equation}
and
\begin{equation}
{\bf N}({\bf c}_r,\,{\bf c}_d,\,0,\,\lambda)\,{\bf N}({\bf d}_r,\,{\bf d}_d,\,0,\,\lambda) \,=\,-\,{\bf c}_r\cdot{\bf d}_r\,-\,{\bf c}_d\cdot{\bf d}_d\,-\,{\bf N}\left({\bf c}_r\times{\bf d}_r + {\bf c}_d\times{\bf d}_d,\,{\bf c}_r\times{\bf d}_d + {\bf c}_d\times{\bf d}_r,\,0,\,\lambda\right).
\end{equation}
Needless to say, this pattern of increased geometrical complexity continues with the addition of each new context. As a result, the fragility of the strong correlations also increases rapidly with the number of contexts. This is easy to see already from the above scalar part for just four contexts. It is easy to see that even a slight change, such as ${{\bf a}_r\pm\Delta{\bf a}_r}$, in only one of the four contexts ${{\bf a}_r+{\bf a}_d\,\varepsilon_+}$ would lead to a dramatic change in the pattern of the corresponding correlation.

\subsection{Reproducing the Strong Correlations Exhibited by the GHSZ States}\label{SecGHZ}

Now, as a second example of strong correlations, consider the four-particle Greenberger-Horne-Zeilinger state (or the GHSZ state \cite{GHZ}):
\begin{equation}
|\Psi_{\bf z}\rangle\,=\,\frac{1}{\sqrt{2}\,}\,\Bigl\{|{\bf z},\,+\rangle_1\otimes
|{\bf z},\,+\rangle_2\otimes|{\bf z},\,-\rangle_3\otimes
|{\bf z},\,-\rangle_4\,
-\,|{\bf z},\,-\rangle_1\otimes|{\bf z},\,-\rangle_2\otimes
|{\bf z},\,+\rangle_3\otimes|{\bf z},\,+\rangle_4\Bigr\}.\label{ghz-single}
\end{equation}
Unlike the singlet state, this entangled state is not rotationally invariant \cite{GHZ}. There is a privileged direction, and it is taken to be the ${\bf z}$-direction of the experimental setup \cite{GHZ}. The ${\bf z}$-direction thus represents the axis of anisotropy of the system. The quantum mechanical expectation value of the product of the four outcomes of the spin components in this state --- namely, the products of finding the spin of particle 1 along ${\bf a}$, the spin of particle 2 along ${\bf b}$, {\it etc.} --- is given by
\begin{equation}
{\cal E}^{\Psi_{\bf z}}_{{\!}_{Q.M.}\!}({\bf a},\,{\bf b},\,{\bf c},\,{\bf d})\,:=\,
\langle\Psi_{\bf z}|\,{\boldsymbol\sigma}\cdot{\bf a}\,\otimes\,
{\boldsymbol\sigma}\cdot{\bf b}\,\otimes\,{\boldsymbol\sigma}\cdot{\bf c}\,\otimes\,
{\boldsymbol\sigma}\cdot{\bf d}\,|\Psi_{\bf z}\rangle.\label{realobserve}
\end{equation}
This expectation value has been calculated in the Appendix F of Ref.${\,}$\cite{GHZ}. In the spherical coordinates --- with angles such as ${\theta_{\bf a}}$ and ${\phi_{\bf a}}$ representing the polar and azimuthal angles, respectively, of the direction ${\bf a}$, ${\bf b}$, {\it etc.} --- it works out to be
\begin{equation}
{\cal E}^{\Psi_{\bf z}}_{{\!}_{Q.M.}}({\bf a},\,{\bf b},\,{\bf c},\,{\bf d}) \,=\,\cos\theta_{\bf a}\,\cos\theta_{\bf b}\,\cos\theta_{\bf c}\,\cos\theta_{\bf d}\,-\,\sin\theta_{\bf a}\,
\sin\theta_{\bf b}\,\sin\theta_{\bf c}\,\sin\theta_{\bf d}\,\cos\,\left(\,\phi_{\bf a}\,+\,\phi_{\bf b}\,-\,\phi_{\bf c}\,-\,\phi_{\bf d}\,\right).\label{q-preghz}
\end{equation}

Our goal now is to reproduce this result within our locally causal framework described above (see also Chapter 6 of Ref.~\cite{disproof}). To this end, we note that the state (\ref{ghz-single}) represents, not a two-level, but a four-level quantum system\textsuperscript{\ref{Hardy}}. Each of the two pairs of the spin-${\frac{1}{2}}$ particles it represents has four alternatives available to it. These alternatives can be represented by a state-vector of the form
\begin{equation}
|\psi\rangle\,=\,\gamma_1\,|\,+\,+\,\rangle\,+\,\gamma_2\,|\,+\,-\,\rangle\,+\,\gamma_3\,|\,-\,+\,\rangle\,+\,\gamma_4\,|\,-\,-\,\rangle\,,\label{schmidt-single}
\end{equation}
where ${\gamma_1}$, ${\gamma_2}$, ${\gamma_3}$, and ${\gamma_4}$ are complex numbers satisfying ${|\,\gamma_1\,|^2+|\,\gamma_2\,|^2+|\,\gamma_3\,|^2+|\,\gamma_4\,|^2=1\,}$, which is equivalent to defining a unit 7-sphere, with ${|\,\gamma_1\,|^2}$, ${|\,\gamma_2\,|^2}$, ${|\,\gamma_3\,|^2}$, and ${|\,\gamma_4\,|^2}$ being the probabilities of actualizing the states ${|\,+\,+\,\rangle}$, ${|\,+\,-\,\rangle}$, ${|\,-\,+\,\rangle}$, and ${|\,-\,-\,\rangle}$, respectively. Therefore we may begin with four local maps of the form
\begin{align}
S^7\ni{\mathscr A}({\bf a}\,,\,\lambda^k):=\!\!&\lim_{\,\substack{{\bf s}_{r1}\,\rightarrow\;{\bf a}_r \\ {\bf s}_{d1}\,\rightarrow\;{\bf a}_d}}\!\left\{\,-\,{\bf D}({\bf a}_r,\,{\bf a}_d,\,0)\,{\bf N}({\bf s}_{r1},\,{\bf s}_{d1},\,0,\,\lambda^k)\,\right\}=\,
\begin{cases}
\,+\,1\;\;\;\;\;{\rm if} &\lambda^k\,=\,+\,1 \\
\,-\,1\;\;\;\;\;{\rm if} &\lambda^k\,=\,-\,1
\end{cases} \Bigg\} \notag \\
&\,\;\text{with}\,\;\Bigl\langle\, {\mathscr A}({\bf a}\,,\,\lambda^k) \,\Bigr\rangle\,=\,0\,, \label{onedoneggg} \\
S^7\ni{\mathscr B}({\bf b}\,,\,\lambda^k):=\!\!&\lim_{\,\substack{{\bf s}_{r2}\,\rightarrow\;{\bf b}_r \\ {\bf s}_{d2}\,\rightarrow\;{\bf b}_d}}\!\left\{\,+\,{\bf N}({\bf s}_{r2},\,{\bf s}_{d2},\,0,\,\lambda^k)\,{\bf D}({\bf b}_r,\,{\bf b}_d,\,0)\,\right\}=\,
\begin{cases}
\,-\,1\;\;\;\;\;{\rm if} &\lambda^k\,=\,+\,1 \\
\,+\,1\;\;\;\;\;{\rm if} &\lambda^k\,=\,-\,1
\end{cases} \Bigg\} \notag \\
&\,\;\text{with}\,\;\Bigl\langle\, {\mathscr B}({\bf b}\,,\,\lambda^k) \,\Bigr\rangle\,=\,0\,, \label{twodoneggg} \\
S^7\ni{\mathscr C}({\bf c}\,,\,\lambda^k)\;:=\!&\lim_{\,\substack{{\bf t}_{r1}\,\rightarrow\;{\bf c}_r \\ {\bf t}_{d1}\,\rightarrow\;{\bf c}_d}}\!\left\{\,-\,{\bf D}({\bf c}_r,\,{\bf c}_d,\,0)\,{\bf N}({\bf t}_{r1},\,{\bf t}_{d1},\,0,\,\lambda^k)\,\right\}\,=\,
\begin{cases}
\,+\,1\;\;\;\;\;{\rm if} &\lambda^k\,=\,+\,1 \\
\,-\,1\;\;\;\;\;{\rm if} &\lambda^k\,=\,-\,1
\end{cases} \Bigg\} \notag \\
&\,\;\text{with}\,\;\Bigl\langle\, {\mathscr C}({\bf c}\,,\,\lambda^k) \,\Bigr\rangle\,=\,0\,, \label{sssdoneggg}
\end{align}
and
\begin{align}
S^7\ni{\mathscr D}({\bf d}\,,\,\lambda^k):=\!\!&\lim_{\,\substack{{\bf t}_{r2}\,\rightarrow\;{\bf d}_r \\ {\bf t}_{d2}\,\rightarrow\;{\bf d}_d}}\!\left\{\,+\,{\bf N}({\bf t}_{r2},\,{\bf t}_{d2},\,0,\,\lambda^k)\,{\bf D}({\bf d}_r,\,{\bf d}_d,\,0)\,\right\}=\,
\begin{cases}
\,-\,1\;\;\;\;\;{\rm if} &\lambda^k\,=\,+\,1 \\
\,+\,1\;\;\;\;\;{\rm if} &\lambda^k\,=\,-\,1
\end{cases} \Bigg\} \notag \\
&\,\;\text{with}\,\;\Bigl\langle\, {\mathscr D}({\bf d}\,,\,\lambda^k) \,\Bigr\rangle\,=\,0\,, \label{tttdoneggg}
\end{align}
together with their geometric product
\begin{equation}
({\mathscr A}_{\bf a}\,{\mathscr B}_{\bf b}\,{\mathscr C}_{\bf c}\,{\mathscr D}_{\bf d})(\lambda^k)\,=\,{\mathscr A}({\bf a}\,,\,\lambda^k)\,{\mathscr B}({\bf b}\,,\,\lambda^k)\,{\mathscr C}({\bf c}\,,\,\lambda^k)\,{\mathscr D}({\bf d}\,,\,\lambda^k)\,=\,\pm\,1\,\in\,S^7\label{myourlocality}
\end{equation}
(cf. appendix \ref{ApB} below), and the corresponding conservation laws
\begin{equation}
{\bf N}({\bf s}_{r1},\,{\bf s}_{d1},\,0,\,\lambda^k)\,{\bf N}({\bf s}_{r2},\,{\bf s}_{d2},\,0,\,\lambda^k)\,=\,\left\{{\bf N}({\bf s}_{r},\,{\bf s}_{d},\,0,\,\lambda^k)\right\}^2\!=\,{\bf N}^2({\bf s}_{r},\,{\bf s}_{d},\,0,\,\lambda^k)\,=\,-1 \label{75699AB}
\end{equation}
and
\begin{equation}
{\bf N}({\bf t}_{r1},\,{\bf t}_{d1},\,0,\,\lambda^k)\,{\bf N}({\bf t}_{r2},\,{\bf t}_{d2},\,0,\,\lambda^k)\,=\,\left\{{\bf N}({\bf t}_{r},\,{\bf t}_{d},\,0,\,\lambda^k)\right\}^2\!=\,{\bf N}^2({\bf t}_{r},\,{\bf t}_{d},\,0,\,\lambda^k)\,=\,-1\,. \label{75699DC}
\end{equation}
As we saw above, the expectation value of the product of the outcomes ${{\mathscr A}({\bf a}\,,\,\lambda^k)}$, ${{\mathscr B}({\bf b}\,,\,\lambda^k)}$, ${{\mathscr C}({\bf c}\,,\,\lambda^k)}$ and ${{\mathscr D}({\bf d}\,,\,\lambda^k)}$ then works out to be the scalar part of the geometric product ${{\bf N}({\bf a}_r,\,{\bf a}_d,\,0,\,\lambda^k)\,{\bf N}({\bf b}_r,\,{\bf b}_d,\,0,\,\lambda^k)\,{\bf N}({\bf c}_r,\,{\bf c}_d,\,0,\,\lambda^k)\,{\bf N}({\bf d}_r,\,{\bf d}_d,\,0,\,\lambda^k)}$, as spelled out in Eq.${\,}$(\ref{expectedfour}). Using a simple vector identity this expectation value can be further simplified to take the form
\begin{align}
{\cal E}^{\rm GHZ}_{{\!}_{L.R.}}({\bf a},\,{\bf b},\,{\bf c},\,{\bf d})
\,&=\,({\bf a}_r\cdot{\bf b}_r)({\bf c}_r\cdot{\bf d}_r)\,+\,({\bf a}_d\cdot{\bf b}_d)({\bf c}_r\cdot{\bf d}_r)\,+\,({\bf a}_r\cdot{\bf b}_r)({\bf c}_d\cdot{\bf d}_d)\,+\,({\bf a}_d\cdot{\bf b}_d)({\bf c}_d\cdot{\bf d}_d)\notag \\
&-\,({\bf a}_r\cdot{\bf c}_r)({\bf b}_r\cdot{\bf d}_r)\,+\,({\bf b}_r\cdot{\bf c}_r)({\bf a}_r\cdot{\bf d}_r)\,-\,({\bf a}_r\cdot{\bf c}_d)({\bf b}_r\cdot{\bf d}_d)\,+\,({\bf b}_r\cdot{\bf c}_d)({\bf a}_r\cdot{\bf d}_d) \notag \\
&-\,({\bf a}_d\cdot{\bf c}_r)({\bf b}_d\cdot{\bf d}_r)\,+\,({\bf b}_d\cdot{\bf c}_r)({\bf a}_d\cdot{\bf d}_r)\,-\,({\bf a}_d\cdot{\bf c}_d)({\bf b}_d\cdot{\bf d}_d)\,+\,({\bf b}_d\cdot{\bf c}_d)({\bf a}_d\cdot{\bf d}_d) \notag \\
&-\,({\bf a}_r\cdot{\bf c}_r)({\bf b}_d\cdot{\bf d}_d)\,+\,({\bf b}_d\cdot{\bf c}_r)({\bf a}_r\cdot{\bf d}_d)\,-\,({\bf a}_r\cdot{\bf c}_d)({\bf b}_d\cdot{\bf d}_r)\,+\,({\bf b}_d\cdot{\bf c}_d)({\bf a}_r\cdot{\bf d}_r) \notag \\
&-\,({\bf a}_d\cdot{\bf c}_r)({\bf b}_r\cdot{\bf d}_d)\,+\,({\bf b}_r\cdot{\bf c}_r)({\bf a}_d\cdot{\bf d}_d)\,-\,({\bf a}_d\cdot{\bf c}_d)({\bf b}_r\cdot{\bf d}_r)\,+\,({\bf b}_r\cdot{\bf c}_d)({\bf a}_d\cdot{\bf d}_r).
\end{align}
Upon using the constraints in Eq.${\,}$(\ref{new-dotp}) to set the terms involving ${{\bf a}_r\cdot{\bf c}_d}$ {\it etc.} to zero, this expected value reduces to
\begin{align}
{\cal E}^{\rm GHZ}_{{\!}_{L.R.}}({\bf a},\,{\bf b},\,{\bf c},\,{\bf d}) \,&=\,({\bf a}_r\cdot{\bf b}_r)({\bf c}_r\cdot{\bf d}_r)\,+\,({\bf a}_d\cdot{\bf b}_d)({\bf c}_r\cdot{\bf d}_r)\,+\,({\bf a}_r\cdot{\bf b}_r)({\bf c}_d\cdot{\bf d}_d)\,+\,({\bf a}_d\cdot{\bf b}_d)({\bf c}_d\cdot{\bf d}_d)\notag \\
&-\,({\bf a}_r\cdot{\bf c}_r)({\bf b}_r\cdot{\bf d}_r)\,+\,({\bf b}_r\cdot{\bf c}_r)({\bf a}_r\cdot{\bf d}_r)\,-\,({\bf a}_d\cdot{\bf c}_d)({\bf b}_d\cdot{\bf d}_d)\,+\,({\bf b}_d\cdot{\bf c}_d)({\bf a}_d\cdot{\bf d}_d) \notag \\
&-\,({\bf a}_r\cdot{\bf c}_r)({\bf b}_d\cdot{\bf d}_d)\,+\,({\bf b}_d\cdot{\bf c}_d)({\bf a}_r\cdot{\bf d}_r)\,+\,({\bf b}_r\cdot{\bf c}_r)({\bf a}_d\cdot{\bf d}_d)\,-\,({\bf a}_d\cdot{\bf c}_d)({\bf b}_r\cdot{\bf d}_r).
\end{align}
Then, again using the constraints in Eq.${\,}$(\ref{new-dotp}) to identify ${{\bf a}_r\cdot{\bf b}_r}$ with ${{\bf a}_d\cdot{\bf b}_d}$, {\it etc.}, the expected value takes the form
\begin{align}
{\cal E}^{\rm GHZ}_{{\!}_{L.R.}\!}({\bf a},\,{\bf b},\,{\bf c},\,{\bf d})\,&=\,\;\;\;2({\bf a}_r\cdot{\bf b}_r)({\bf c}_r\cdot{\bf d}_r)\,+\,2({\bf a}_d\cdot{\bf b}_d)({\bf c}_d\cdot{\bf d}_d)\,-\,2({\bf a}_r\cdot{\bf c}_r)({\bf b}_r\cdot{\bf d}_r) \notag \\
&\;\;\;\,+\,2({\bf b}_r\cdot{\bf c}_r)({\bf a}_r\cdot{\bf d}_r)\,+\,2({\bf b}_d\cdot{\bf c}_d)({\bf a}_d\cdot{\bf d}_d)\,-\,2({\bf a}_d\cdot{\bf c}_d)({\bf b}_d\cdot{\bf d}_d). \label{fiexpval}
\end{align}
Next, in order to satisfy the above constraints, we relate the external measurement directions ${\bf a}$, ${\bf b}$, ${\bf c}$ and ${\bf d\in{\rm I\!R}^3}$, chosen freely by the experimenters, with the directions ${{\bf a}_r}$, ${{\bf a}_d}$, {\it etc.} within our representation space ${{S^7}\subset{\cal K}^{\lambda}}$, as follows: 
\begin{align}
{\mathscr A}(a_x,\,a_y,\,a_z,\,\lambda^k)\,=\,\pm\,1\,\in\,S^7, &\text{ to be detected by} \notag \\
&{\bf D}({\bf a}_r;\,{\bf a}_d;\,0)\,=\,{\bf D}\left(\,-\frac{a_x}{\sqrt[4]{2}},\,+\frac{a_y}{\sqrt[4]{2}},\,0;\,\;\;0,\,0,\,-\frac{a_z}{\sqrt[4]{2}};\,\;\;0\,\right), \label{defin-lll} \\
{\mathscr B}(b_x,\;b_y,\;b_z,\;\lambda^k)\,=\,\pm\,1\,\in\,S^7, &\text{ to be detected by} \notag \\
&{\bf D}({\bf b}_r;\,{\bf b}_d;\,0)\,=\,{\bf D}\left(\,+\frac{b_x}{\sqrt[4]{2}},\,+\frac{b_y}{\sqrt[4]{2}},\,0;\,\;\;0,\,0,\,+\frac{b_z}{\sqrt[4]{2}};\,\;\;0\,\right), \\
{\mathscr C}(c_x,\;c_y,\;c_z,\;\lambda^k)\,=\,\pm\,1\,\in\,S^7, &\text{ to be detected by} \notag \\
&{\bf D}({\bf c}_r;\,{\bf c}_d;\,0)\;=\;{\bf D}\left(\,+\frac{c_x}{\sqrt[4]{2}},\,+\frac{c_y}{\sqrt[4]{2}},\,0;\,\;\;0,\,0,\,+\frac{c_z}{\sqrt[4]{2}};\,\;\;0\,\right), 
\end{align}
and
\begin{align}
{\mathscr D}(d_x,\,d_y,\,d_z,\,\lambda^k)\,=\,\pm\,1\,\in\,S^7, &\text{ to be detected by} \notag \\
&{\bf D}({\bf d}_r;\,{\bf d}_d;\,0)\,=\,{\bf D}\left(\,+\frac{d_x}{\sqrt[4]{2}},\,-\frac{d_y}{\sqrt[4]{2}},\,0;\,\;\;0,\,0,\,-\frac{d_z}{\sqrt[4]{2}};\,\;\;0\,\right).\label{defin-nnn}
\end{align}
Here the ${4^{\rm th}}$ roots of ${2}$ in the denominators of ${\bf D}$ [instead of ${\sqrt{2}}$ as in Eq.~(\ref{71-root})] arise because the product of four factors, ${{\bf N}({\bf a}_r,\,{\bf a}_d,\,0,\,\lambda^k)\,{\bf N}({\bf b}_r,\,{\bf b}_d,\,0,\,\lambda^k)\,{\bf N}({\bf c}_r,\,{\bf c}_d,\,0,\,\lambda^k)\,{\bf N}({\bf d}_r,\,{\bf d}_d,\,0,\,\lambda^k)}$, instead of two, ${{\bf N}({\bf a}_r,\,{\bf a}_d,\,0,\,\lambda^k)\,{\bf N}({\bf b}_r,\,{\bf b}_d,\,0,\,\lambda^k)}$, is involved in the calculation (\ref{gg110}) of the correlation, while maintaining the unity of the radius of ${S^7}$. Note also that components of only external vectors are involved in the definitions of the four detectors. And they do not mix with each other, so that Bell's condition of local causality, or parameter independence \cite{Bell-1964}, is strictly respected throughout. Substituting these coordinate values into the remaining vectors in the expected value (\ref{fiexpval}) then reduces that value to 
\begin{align}
{\cal E}^{\rm GHZ}_{{\!}_{L.R.}\!}({\bf a},\,{\bf b},\,{\bf c},\,{\bf d})\,=\,+\,{a}_{z}\,{b}_{z}\,{c}_{z}\,{d}_{z}
&\,-\,{a}_{y}\,{b}_{y}\,{c}_{y}\,{d}_{y}
\,-\,{a}_{x}\,{b}_{y}\,{c}_{x}\,{d}_{y}
\,-\,{a}_{y}\,{b}_{x}\,{c}_{y}\,{d}_{x}
\,-\,{a}_{x}\,{b}_{x}\,{c}_{x}\,{d}_{x} \notag \\
&\,+\,{a}_{x}\,{b}_{x}\,{c}_{y}\,{d}_{y}
\,-\,{a}_{x}\,{b}_{y}\,{c}_{y}\,{d}_{x}
\,-\,{a}_{y}\,{b}_{x}\,{c}_{x}\,{d}_{y}
\,+\,{a}_{y}\,{b}_{y}\,{c}_{x}\,{d}_{x}\,.
\end{align}
In the spherical coordinates -- with angles ${\theta_{\bf a}}$ and ${\phi_{\bf a}}$ representing respectively the polar and azimuthal angles of the direction ${\bf a}$, {\it etc.}, for all four measurement directions -- this expression of the expected value can be further simplified to
\begin{equation}
{\cal E}^{\rm GHZ}_{{\!}_{L.R.}}({\bf a},\,{\bf b},\,{\bf c},\,{\bf d}) \,=\,\cos\theta_{\bf a}\,\cos\theta_{\bf b}\,\cos\theta_{\bf c}\,\cos\theta_{\bf d}\,-\,\sin\theta_{\bf a}\,\sin\theta_{\bf b}\,\sin\theta_{\bf c}\,\sin\theta_{\bf d}\,\cos\left(\,\phi_{\bf a}\,+\,\phi_{\bf b}\,-\,\phi_{\bf c}\,-\,\phi_{\bf d}\,\right). \label{203}
\end{equation}
This is exactly the quantum mechanical prediction (\ref{q-preghz}) for the four-particle GHZ state (\ref{ghz-single}). We have derived this prediction, however, as purely geometric effects within our locally causal framework. The GHZ correlations thus simply exhibit the classical, deterministic, local, and realistic correlations among four points of our representation space ${S^7}$.

\section{Bell's Theorem, its Experimental Tests, and the GHSZ Variant}

\subsection{Bell-Test Experiments: From Inceptions to Loophole-Free Advances}\label{Bell-test}

Contrary to what we have demonstrated above, it is widely believed that the so-called Bell-test experiments --- from their initial conceptions summarized in the classic review paper by Clauser and Shimony \cite{Clauser} to their state-of-the-art ``loophole-free" variants \cite{Aspect} --- undermine any prospects of a locally causal understanding of quantum correlations. It is important to appreciate, however, that all such experiments simply confirm the predictions of quantum mechanics. They neither contradict the quantum mechanical predictions, nor go beyond them in any sense. Moreover, since in the subsection \ref{twosect} above we have reproduced all of the quantum mechanical predictions for the singlet state {\it exactly}, the Bell-test experiments \cite{Aspect} do not contradict the predictions of our model either. Rather, they simply corroborate them.      

More precisely, in the analysis of all such experiments one averages over ``coincidence counts" to calculate expectation values in the form 
\begin{align}
{\cal E}({\bf a},\,{\bf b})\,&=\lim_{\,n\,\gg\,1}\left[\frac{1}{n}\sum_{k\,=\,1}^{n}\,{\mathscr A}({\bf a},\,{\lambda}^k)\;{\mathscr B}({\bf b},\,{\lambda}^k)\right] \notag \\
\,&\equiv\,\frac{\Big[C_{++}({\bf a},\,{\bf b})\,+\,C_{--}({\bf a},\,{\bf b})\,-\,C_{+-}({\bf a},\,{\bf b})\,-\,C_{-+}({\bf a},\,{\bf b})\Big]}{\Big[C_{++}({\bf a},\,{\bf b})\,+\,C_{--}({\bf a},\,{\bf b})\,+\,C_{+-}({\bf a},\,{\bf b})\,+\,C_{-+}({\bf a},\,{\bf b})\Big]} \notag \\
&=\,-\cos\theta_{{\bf a}{\bf b}}\,, \label{a.1}
\end{align}
where ${C_{+-}({\bf a},\,{\bf b})}$ etc. represent the number of simultaneous occurrences of detections ${+1}$ along ${\bf a}$ and ${-1}$ along ${\bf b}$, etc. In addition, they observe individual results ${{\mathscr A}({\bf a},\,{\lambda}^k)}$ and ${{\mathscr B}({\bf b},\,{\lambda}^k)}$ at each remote station to find that on average 
\begin{equation}
\Bigl\langle\;{\mathscr A}({\bf a},\,{\lambda}^k)\,\Bigr\rangle\,=\,0\;\;\;\text{and}\;\;\;\Bigl\langle\;{\mathscr B}({\bf b},\,{\lambda}^k)\,\Bigr\rangle\,=\,0. \label{a.2}
\end{equation}
Finally, they observe that Bell-CHSH inequalities \cite{CHSH} with the absolute bound of 2 are exceeded by a factor of ${\sqrt{2}}$: 
\begin{equation}
-\,2\sqrt{2}\,\leqslant\,{\cal E}({\bf a},\,{\bf b})\,+\,{\cal E}({\bf a},\,{\bf b'})\,+\,{\cal E}({\bf a'},\,{\bf b})\,-\,{\cal E}({\bf a'},\,{\bf b'})\leqslant +\,2\sqrt{2}\,. \label{a.3}
\end{equation}
The relations (\ref{a.1}), (\ref{a.2}), and (\ref{a.3}) are precisely the predictions of quantum mechanics for the singlet state (\ref{single}). In practice, however, it is often difficult to perform such experiments with a pair of spin-${\frac{1}{2}}$ particles considered in Fig.~\ref{fig6}. For this reason the usual preference for preforming the Bell-test experiments is to measure correlations in a pair of linearly polarized photons instead of in a pair of spin-${\frac{1}{2}}$ particles. But the predictions (\ref{a.1}), (\ref{a.2}), and (\ref{a.3}) of quantum mechanics do not change for either choice, apart from a factor of 2 in the angular-dependence of the correlations (cf. Eqs.~(1) and (33) of Ref.~\cite{restoring}). In this paper, however, we are not concerned about the practical difficulties in performing the experiments, and therefore the original reformulation of the EPR argument by Bohm in terms of a pair of spin-${\frac{1}{2}}$ particles, such as the electron-positron pair depicted in Fig.~\ref{fig6}, is sufficient for our purposes. What is more important to appreciate is the fact that our ${S^7}$ model predicts precisely the relations (\ref{a.1}), (\ref{a.2}), and (\ref{a.3}) for the entangled state (\ref{single}), as can be verified from our predictions (\ref{365}), (\ref{onedone}), (\ref{twodone}), and (\ref{My-CHSH}).

Thus the crucial difference between the predictions of our ${S^7}$ model and those of quantum mechanics is {\it not} in the observational content, but in the interpretation of the latter in terms of non-locality\footnote{It is however possible to distinguish between the two interpretations in a macroscopic experiment. Such a macroscopic experiment has been proposed in Ref.~\cite{IJTP}. If realized, it will explore whether or not Bell inequalities are violated for a manifestly local, classical system, without involving either quantum entanglement or quantum superposition.}. And this interpretation depends entirely on the argument put forward by Bell and his followers \cite{Bell-1964,Clauser}. This argument, however, is fatally flawed, as we now demonstrate.

\subsection{Surprising Oversight in the Derivation of the Bell-CHSH Inequalities}\label{Bell-flaw}

From the outset let us stress that Bell's so-called theorem is by no means a ``theorem" in the sense that word is used by mathematicians but rather a word-statement, which claims that {\it no physical theory which is realistic as well as local in the strict senses espoused by} Einstein \cite{EPR} {\it and later formulated by} Bell\textsuperscript{\ref{Bellfoot}} \cite{Bell-1964} {\it can reproduce all of the statistical predictions of quantum theory} \cite{Stanford}. This word-statement is based on ``violations" of certain mathematical inequalities, which are derived by considering four {\it incompatible} EPR-Bohm type experiments, and without using a single concept from quantum theory. While the bounds thus derived on the inequalities are exceeded by the predictions of quantum theory and ``violated" in actual experiments, their derivation happens to be marred by a serious conceptual oversight.

To appreciate this, consider the standard EPR type spin-${\frac{1}{2}}$ experiment, as proposed by Bohm and later used by Bell to prove his theorem. Alice is free to choose a detector direction ${\bf a}$ or ${\bf a'}$ and Bob is free to choose a detector direction ${\bf b}$ or ${\bf b'}$ to detect spins of the fermions they receive from a common source, at a space-like distance from each other. The objects of interest then are the bounds on the sum of possible averages put together in the manner of CHSH \cite{CHSH},
\begin{equation}
{\cal E}({\bf a},\,{\bf b})\,+\,{\cal E}({\bf a},\,{\bf b'})\,+\,{\cal E}({\bf a'},\,{\bf b})\,-\,{\cal E}({\bf a'},\,{\bf b'})\,, \label{B1-11-2}
\end{equation}
with each average defined as
\begin{equation}
{\cal E}({\bf a},\,{\bf b})\,=\lim_{\,n\,\gg\,1}\left[\frac{1}{n}\sum_{k\,=\,1}^{n}\,
{\mathscr A}({\bf a},\,{\lambda}^k)\;{\mathscr B}({\bf b},\,{\lambda}^k)\right]\,\equiv\,\Bigl\langle\,{\mathscr A}_{k}({\bf a})\,{\mathscr B}_{k}({\bf b})\,\Bigr\rangle\,,\label{exppeu-2}
\end{equation}
where ${\mathscr A({\bf a},\,{\lambda}^k)\equiv {\mathscr A}_{k}({\bf a})=\pm1}$ and ${\mathscr B({\bf b},\,{\lambda}^k)\equiv {\mathscr B}_{k}({\bf b})=\pm1}$ are the respective measurement results of Alice and Bob. Now, since ${{\mathscr A}_{k}({\bf a})=\pm1}$ and ${{\mathscr B}_{k}({\bf b})=\pm1}$, the average of their product is ${-1\leqslant\Bigl\langle\,{\mathscr A}_{k}({\bf a})\,{\mathscr B}_{k}({\bf b})\,\Bigr\rangle\leqslant +1}$. As a result, we can immediately read off the upper and lower bounds on the string of the four averages considered above in (\ref{B1-11-2}):
\begin{equation}
-\,4\,\leqslant\,\Bigl\langle\,{\mathscr A}_{k}({\bf a})\,{\mathscr B}_{k}({\bf b})\,\Bigr\rangle\,+\, \Bigl\langle\,{\mathscr A}_{k}({\bf a})\,{\mathscr B}_{k}({\bf b'})\,\Bigr\rangle\,+\,\Bigl\langle\,{\mathscr A}_{k}({\bf a'})\,{\mathscr B}_{k}({\bf b})\,\Bigr\rangle\,-\, \Bigl\langle\,{\mathscr A}_{k}({\bf a'})\,{\mathscr B}_{k}({\bf b'})\,\Bigr\rangle\,\leqslant\,+\,4\,. \label{3}
\end{equation}

This should have been Bell's final conclusion. However, by continuing, Bell overlooked something that is physically unjustifiable. He replaced the above sum of four separate averages of real numbers with the following single average:
\begin{equation}
{\cal E}({\bf a},\,{\bf b})\,+\,{\cal E}({\bf a},\,{\bf b'})\,+\,{\cal E}({\bf a'},\,{\bf b})\,-\,{\cal E}({\bf a'},\,{\bf b'}) \,\longrightarrow\,\Bigl\langle\,{\mathscr A}_{k}({\bf a})\,{\mathscr B}_{k}({\bf b})\,+\,{\mathscr A}_{k}({\bf a})\,{\mathscr B}_{k}({\bf b'})\,+\,{\mathscr A}_{k}({\bf a'})\,{\mathscr B}_{k}({\bf b})\,-\,{\mathscr A}_{k}({\bf a'})\,{\mathscr B}_{k}({\bf b'})\,\Bigr\rangle\,. \label{rep}
\end{equation}
As innocuous as this step may seem mathematically, it is in fact an illegitimate step physically, because what is being averaged on its RHS are {\it unobservable} and {\it unphysical} quantities. But it allows us to reduce the sum of four averages to 
\begin{equation}
\Bigl\langle\,{\mathscr A}_{k}({\bf a})\,\big\{\,{\mathscr B}_{k}({\bf b})+{\mathscr B}_{k}({\bf b'})\,\big\}\,+\,{\mathscr A}_{k}({\bf a'})\,\big\{\,{\mathscr B}_{k}({\bf b})-{\mathscr B}_{k}({\bf b'})\,\big\}\,\Bigr\rangle\,.\label{absurd}
\end{equation}
And since ${{\mathscr B}_{k}({\bf b})=\pm1}$, if ${|{\mathscr B}_{k}({\bf b})+{\mathscr B}_{k}({\bf b'})|=2}$, then ${|{\mathscr B}_{k}({\bf b})-{\mathscr B}_{k}({\bf b'})|=0}$, and vice versa \cite{IJTP}. Consequently, using ${{\mathscr A}_{k}({\bf a})=\pm1}$, it is easy to conclude that the absolute value of the above average cannot exceed 2, just as Bell concluded\footnote{\label{BooleIn}A similar inequality was first considered by Boole in 1862, but without the interpretation attributed to it by Bell (cf. Ref.~\cite{Boole}).}: 
\begin{equation}
-\,2\,\leqslant\,\Bigl\langle\,{\mathscr A}_{k}({\bf a})\,{\mathscr B}_{k}({\bf b})\,+\,
{\mathscr A}_{k}({\bf a})\,{\mathscr B}_{k}({\bf b'})\,+\,{\mathscr A}_{k}({\bf a'})\,{\mathscr B}_{k}({\bf b})\,-\,{\mathscr A}_{k}({\bf a'})\,{\mathscr B}_{k}({\bf b'})\,\Bigr\rangle\,\leqslant\,+\,2\,.\label{5}
\end{equation}

Let us now try to understand why the replacement in (\ref{rep}) above is illegitimate\footnote{In the derivation of the absolute bounds on the Bell-CHSH correlator, such as those in Eq.~(\ref{5}) above, one usually employs factorized probabilities of observing binary measurement results rather than the actual measurement results we have used in our derivation. But employing probabilities in that manner only manages to obfuscate the conceptual flaw in Bell's argument we intend to bring out here.}. To begin with, Einstein's (or even Bell's own) notion of local-realism does not, by itself, demand this replacement. Since this notion is captured already in the very definition\textsuperscript{\ref{Bellfoot}} \cite{Bell-1964} of the functions ${{\mathscr A}({\bf a},\,{\lambda}^k)}$, the LHS of (\ref{rep}) satisfies the demand of local-realism perfectly well. Nor can a possible statistical independence of the four separate averages on the LHS of (\ref{rep}) justify their replacement with the single average on its RHS, {\it at the expense of what is physically possible in the actual experiments}. To be sure, mathematically there is nothing wrong with a replacement of four separate averages with a single average. Indeed, every school child knows that the sum of averages is equal to the average of the sum. But this rule of thumb is not valid in the above case, because ${({\bf a},\,{\bf b})}$, ${({\bf a},\,{\bf b'})}$, ${({\bf a'},\,{\bf b})}$, and ${({\bf a'},\,{\bf b'})}$ are {\it mutually exclusive pairs of measurement directions}, corresponding to four {\it incompatible} experiments. Each pair can be used by Alice and Bob for a given experiment, for all runs ${1}$ to ${n}$, but no two of the four pairs can be used by them simultaneously. This is because Alice and Bob do not have the ability to make measurements along counterfactually possible pairs of directions such as ${({\bf a},\,{\bf b})}$ and ${({\bf a},\,{\bf b'})}$ simultaneously. Alice, for example, can make measurements along ${\bf a}$ or ${\bf a'}$, but not along ${\bf a}$ {\it and} ${\bf a'}$ at the same time.

But this fact is rather devastating for Bell's argument, because it means that his replacement (\ref{rep}) is illegitimate. Consider, for example, a specific run of the EPR-Bohm type experiment and the corresponding quantity being averaged in (\ref{rep}):
\begin{equation}
{\mathscr A}_{k}({\bf a})\,{\mathscr B}_{k}({\bf b})\,+\,
{\mathscr A}_{k}({\bf a})\,{\mathscr B}_{k}({\bf b'})\,+\,{\mathscr A}_{k}({\bf a'})\,{\mathscr B}_{k}({\bf b})\,-\,
{\mathscr A}_{k}({\bf a'})\,{\mathscr B}_{k}({\bf b'})\,. \label{riy}
\end{equation}
Here the index ${k=1}$ now represents a specific run of the experiment. But since Alice and Bob have only two particles at their disposal for each run, only one of the four terms of the above sum is physically meaningful. In other words, the above quantity is physically meaningless, because Alice, for example, cannot align her detector along ${\bf a}$ and ${\bf a'}$ at the same time. And likewise, Bob cannot align his detector along ${\bf b}$ and ${\bf b'}$ at the same time. What is more, this will be true for all possible runs of the experiment, or equivalently for all possible pairs of particles. Which implies that all of the quantities listed below, as they appear in the average (\ref{5}), are unobservable, and hence physically meaningless:
\begin{align}
&{\mathscr A}_{1}({\bf a})\,{\mathscr B}_{1}({\bf b})\,+\,
{\mathscr A}_{1}({\bf a})\,{\mathscr B}_{1}({\bf b'})\,+\,{\mathscr A}_{1}({\bf a'})\,{\mathscr B}_{1}({\bf b})\,-\,
{\mathscr A}_{1}({\bf a'})\,{\mathscr B}_{1}({\bf b'})\,, \nonumber \\
&{\mathscr A}_{2}({\bf a})\,{\mathscr B}_{2}({\bf b})\,+\,
{\mathscr A}_{2}({\bf a})\,{\mathscr B}_{2}({\bf b'})\,+\,{\mathscr A}_{2}({\bf a'})\,{\mathscr B}_{2}({\bf b})\,-\,
{\mathscr A}_{2}({\bf a'})\,{\mathscr B}_{2}({\bf b'})\,,  \nonumber \\
&{\mathscr A}_{3}({\bf a})\,{\mathscr B}_{3}({\bf b})\,+\,
{\mathscr A}_{3}({\bf a})\,{\mathscr B}_{3}({\bf b'})\,+\,{\mathscr A}_{3}({\bf a'})\,{\mathscr B}_{3}({\bf b})\,-\,
{\mathscr A}_{3}({\bf a'})\,{\mathscr B}_{3}({\bf b'})\,,  \nonumber \\
&{\mathscr A}_{4}({\bf a})\,{\mathscr B}_{4}({\bf b})\,+\,
{\mathscr A}_{4}({\bf a})\,{\mathscr B}_{4}({\bf b'})\,+\,{\mathscr A}_{4}({\bf a'})\,{\mathscr B}_{4}({\bf b})\,-\,
{\mathscr A}_{4}({\bf a'})\,{\mathscr B}_{4}({\bf b'})\,,  \nonumber \\
&\;\;\;\;\boldsymbol{\cdot}\nonumber \\
&\;\;\;\;\boldsymbol{\cdot}\nonumber \\
&\;\;\;\;\boldsymbol{\cdot}\nonumber \\
&\!{\mathscr A}_{n}({\bf a})\,{\mathscr B}_{n}({\bf b})\,+\,
{\mathscr A}_{n}({\bf a})\,{\mathscr B}_{n}({\bf b'})\,+\,{\mathscr A}_{n}({\bf a'})\,{\mathscr B}_{n}({\bf b})\,-\,
{\mathscr A}_{n}({\bf a'})\,{\mathscr B}_{n}({\bf b'})\,.\nonumber
\end{align}
But since each of the quantities above is physically meaningless, their average appearing on the RHS of (\ref{rep}), namely
\begin{equation}
\Bigl\langle\,{\mathscr A}_{k}({\bf a})\,{\mathscr B}_{k}({\bf b})\,+\,
{\mathscr A}_{k}({\bf a})\,{\mathscr B}_{k}({\bf b'})\,+\,{\mathscr A}_{k}({\bf a'})\,{\mathscr B}_{k}({\bf b})\,-\,
{\mathscr A}_{k}({\bf a'})\,{\mathscr B}_{k}({\bf b'})\,\Bigr\rangle\,,
\end{equation}
is also physically meaningless\footnote{\label{homely}The possible space-like separated events being averaged in (\ref{absurd}) cannot possibly occur in any possible world, classical or quantum. To appreciate this elementary fact, consider the following homely analogy: Imagine a couple, say Jack and Jill, who decide to separate while in Kansas City, and travel to the West and East Coasts respectively. Jack decides to travel to Los Angeles, while Jill can't make up her mind and might travel to either New York or Miami. So while Jack reaches Los Angeles, Jill might reach either New York or Miami. Thus there are two possible destinations for the couple. Either Jack reaches Los Angeles and Jill reaches New York, or Jack reaches Los Angeles and Jill reaches Miami. Now suppose that, upon reaching New York, Jill decides to buy either apple juice or orange juice. And likewise, upon reaching Miami, Jill decides to buy either apple juice or orange juice. Consequently, there are following four counterfactually possible events that can realistically occur, at least in our familiar world: (1) While Jack reaches Los Angeles and buys apple juice, Jill reaches New York and buys apple juice; Or, (2) while Jack reaches Los Angeles and buys apple juice, Jill reaches New York and buys orange juice; Or, (3) while Jack reaches Los Angeles and buys apple juice, Jill reaches Miami and buys apple juice; Or, (4) while Jack reaches Los Angeles and buys apple juice, Jill reaches Miami and buys orange juice. So far so good. But what is being averaged in (\ref{absurd}) are impossible events of the following kind: (5) While Jack reaches Los Angeles and buys apple juice, Jill reaches New York and buys apple juice {\it and} Jill reaches Miami and buys orange juice {\it at exactly the same time!} Needless to say, no such events can possibly occur in any possible world, even counterfactually. In particular, Einstein's conception of local realism by no means demands such absurd or impossible events in any possible world \cite{EPR}. It is therefore not at all surprising why the unphysical bounds of ${\pm2}$ on the CHSH sum of expectation values obtained by averaging over the absurd events like (\ref{riy}) are not respected in the actual experiments \cite{Aspect}.} \cite{Bell-1964,Stanford,Clauser}. That is to say, no physical experiment can ever be performed --- {\it even in principle} --- that can meaningfully allow to measure or evaluate the above average, since none of the above list of quantities could have experimentally observable values. Therefore the innocuous looking replacement (\ref{rep}) made by Bell is, in fact, illegal.

On the other hand, it is important to note that each of the averages appearing on the LHS of replacement (\ref{rep}),
\begin{align}
{\cal E}({\bf a},\,{\bf b})\,&=\lim_{\,n\,\gg\,1}\left[\frac{1}{n}\sum_{k\,=\,1}^{n}\,
{\mathscr A}({\bf a},\,{\lambda}^k)\;{\mathscr B}({\bf b},\,{\lambda}^k)\right]\,\equiv\,\Bigl\langle\,{\mathscr A}_{k}({\bf a})\,{\mathscr B}_{k}({\bf b})\,\Bigr\rangle\,, \\
{\cal E}({\bf a},\,{\bf b'})\,&=\lim_{\,n\,\gg\,1}\left[\frac{1}{n}\sum_{k\,=\,1}^{n}\,
{\mathscr A}({\bf a},\,{\lambda}^k)\;{\mathscr B}({\bf b'},\,{\lambda}^k)\right]\,\equiv\,\Bigl\langle\,{\mathscr A}_{k}({\bf a})\,{\mathscr B}_{k}({\bf b'})\,\Bigr\rangle\,, \\
{\cal E}({\bf a'},\,{\bf b})\,&=\lim_{\,n\,\gg\,1}\left[\frac{1}{n}\sum_{k\,=\,1}^{n}\,
{\mathscr A}({\bf a'},\,{\lambda}^k)\;{\mathscr B}({\bf b},\,{\lambda}^k)\right]\,\equiv\,\Bigl\langle\,{\mathscr A}_{k}({\bf a'})\,{\mathscr B}_{k}({\bf b})\,\Bigr\rangle\,, \\
{\text{and}}\;\;\;{\cal E}({\bf a'},\,{\bf b'})\,&=\lim_{\,n\,\gg\,1}\left[\frac{1}{n}\sum_{k\,=\,1}^{n}\,
{\mathscr A}({\bf a'},\,{\lambda}^k)\;{\mathscr B}({\bf b'},\,{\lambda}^k)\right]\,\equiv\,\Bigl\langle\,{\mathscr A}_{k}({\bf a'})\,{\mathscr B}_{k}({\bf b'})\,\Bigr\rangle\,,
\end{align} 
is a perfectly well defined and observable physical quantity. Therefore the bounds (\ref{3}) on their sum are harmless. These bounds of ${\{-4,\,+4\}}$, however, have never been violated in any experiment. Indeed, nothing can violate them. 

In summary, Bell and his followers derive the upper bound of 2 on the CHSH string of averages by an illegal move. In the middle of their derivation they unjustifiably replace an observable, and hence physically meaningful quantity,  
\begin{equation}
\Bigl\langle\,{\mathscr A}_{k}({\bf a})\,{\mathscr B}_{k}({\bf b})\,\Bigr\rangle\,+\, \Bigl\langle\,{\mathscr A}_{k}({\bf a})\,{\mathscr B}_{k}({\bf b'})\,\Bigr\rangle\,+\,\Bigl\langle\,{\mathscr A}_{k}({\bf a'})\,{\mathscr B}_{k}({\bf b})\,\Bigr\rangle\,-\, \Bigl\langle\,{\mathscr A}_{k}({\bf a'})\,{\mathscr B}_{k}({\bf b'})\,\Bigr\rangle\,,
\end{equation}
with an experimentally unobservable, and hence physically entirely meaningless quantity (regardless of the method): 
\begin{equation}
\Bigl\langle\,{\mathscr A}_{k}({\bf a})\,{\mathscr B}_{k}({\bf b})\,+\,
{\mathscr A}_{k}({\bf a})\,{\mathscr B}_{k}({\bf b'})\,+\,{\mathscr A}_{k}({\bf a'})\,{\mathscr B}_{k}({\bf b})\,-\,
{\mathscr A}_{k}({\bf a'})\,{\mathscr B}_{k}({\bf b'})\,\Bigr\rangle\,.
\end{equation}
If they do not make this illegitimate replacement, then the absolute upper bound on the CHSH string of averages is 4, not 2. And the absolute upper bound of 4 has never been exceeded --- and can never exceed --- in any experiment \cite{IJTP}.

One may suspect that the above conclusion is perhaps an artifact of the discrete version, (\ref{exppeu-2}), of the expectation values ${\,{\cal E}({\bf a},\,{\bf b})}$. Perhaps it can be ameliorated if we considered the CHSH sum (\ref{B1-11-2}) in the following continuous form:
\begin{equation}
\int_{\Lambda}{\mathscr A}({\bf a},\,\lambda)\,{\mathscr B}({\bf b},\,\lambda)\,d\rho(\lambda)\,+\int_{\Lambda}{\mathscr A}({\bf a},\,\lambda)\,{\mathscr B}({\bf b'},\,\lambda)\,d\rho(\lambda)\,+\int_{\Lambda}{\mathscr A}({\bf a'},\,\lambda)\,{\mathscr B}({\bf b},\,\lambda)\,d\rho(\lambda)\,-\int_{\Lambda}{\mathscr A}({\bf a'},\,\lambda)\,{\mathscr B}({\bf b'},\,\lambda)\,d\rho(\lambda)\,, \label{d15}
\end{equation}
where ${\Lambda}$ is the space of all hidden variables ${\lambda}$ and ${\rho(\lambda)}$ is the probability measure of ${\lambda}$ \cite{Bell-1964,GHZ}. Written in this form, it is now easy to see that the above CHSH sum of expectation values is both mathematically and physically identical to
\begin{equation}
\int_{\Lambda}\;\Big[\;{\mathscr A}({\bf a},\,\lambda)\,\big\{\,{\mathscr B}({\bf b},\,\lambda)\,+\,{\mathscr B}({\bf b'},\,\lambda)\,\big\}\,+\,{\mathscr A}({\bf a'},\,\lambda)\,\big\{\,{\mathscr B}({\bf b},\,\lambda)\,-\,{\mathscr B}({\bf b'},\,\lambda)\,\big\}\Big]\;\,d\rho(\lambda)\,. \label{d16}
\end{equation}
But since the above two integral expressions are identical to each other, we can use the second expression without loss of generality to prove that the criterion of reality used by Bell is unreasonably restrictive compared to that of EPR.

To begin with, expression (\ref{d16}) involves an integration over fictitious quantities\textsuperscript{\ref{homely}} such as ${{\mathscr A}({\bf a},\,\lambda)\left\{{\mathscr B}({\bf b},\,\lambda) + {\mathscr B}({\bf b'},\,\lambda)\right\}}$ and ${{\mathscr A}({\bf a'},\,\lambda)\left\{{\mathscr B}({\bf b},\,\lambda) - {\mathscr B}({\bf b'},\,\lambda)\right\}}$. These quantities are not parts of the space of all possible measurement outcomes such as ${{\mathscr A}({\bf a},\,\lambda)}$, ${{\mathscr A}({\bf a'},\,\lambda)}$, ${{\mathscr B}({\bf b},\,\lambda)}$, ${{\mathscr B}({\bf b'},\,\lambda)}$, {\it etc.}; because that space --- although evidently closed under multiplication --- is {\it not} closed under addition. Since each function ${{\mathscr B}({\bf b},\,\lambda)}$ is by definition either ${+1}$ or ${-1}$, their sum such as ${{\mathscr B}({\bf b},\,\lambda)+{\mathscr B}({\bf b'},\,\lambda)}$ can only take values from the set ${\{-2,\,0,\,+2\}}$, and therefore it is not a part of the unit 2-sphere representing the space of all possible measurement results. Consequently, the quantities ${{\mathscr A}({\bf a},\,\lambda)\left\{{\mathscr B}({\bf b},\,\lambda) + {\mathscr B}({\bf b'},\,\lambda)\right\}}$ and ${{\mathscr A}({\bf a'},\,\lambda)\left\{{\mathscr B}({\bf b},\,\lambda) - {\mathscr B}({\bf b'},\,\lambda)\right\}}$ appearing in the integrand of (\ref{d16}) do not themselves exist, despite the fact that ${{\mathscr A}({\bf a},\,\lambda)}$, ${{\mathscr A}({\bf a'},\,\lambda)}$, ${{\mathscr B}({\bf b},\,\lambda)}$ and ${{\mathscr B}({\bf b'},\,\lambda)}$ exist, at least counterfactually, in accordance with the hypothesis of local realism. This is analogous to the fact that the set ${{\cal O}:=\{1, 2, 3, 4, 5, 6\}}$ of all possible outcomes of a die throw is not closed under addition. For example, the sum ${3 + 6}$ is not a part of the set ${\cal O}$.

But there is also a much more serious physical problem with Bell's version of reality. As noted above, the quantities ${{\mathscr A}({\bf a},\,\lambda)\left\{{\mathscr B}({\bf b},\,\lambda) + {\mathscr B}({\bf b'},\,\lambda)\right\}}$ and ${{\mathscr A}({\bf a'},\,\lambda)\left\{{\mathscr B}({\bf b},\,\lambda) - {\mathscr B}({\bf b'},\,\lambda)\right\}}$ are not physically meaningful in {\it any} possible physical world, classical or quantum. That is because ${{\mathscr B}({\bf b},\,\lambda)}$ and ${{\mathscr B}({\bf b'},\,\lambda)}$ can coexist with ${{\mathscr A}({\bf a},\,\lambda)}$ only counterfactually, since ${\bf b}$ and ${\bf b'}$ are mutually exclusive directions. If ${{\mathscr B}({\bf b},\,\lambda)}$ coexists with ${{\mathscr A}({\bf a},\,\lambda)}$, then ${{\mathscr B}({\bf b'},\,\lambda)}$ cannot coexist with ${{\mathscr A}({\bf a},\,\lambda)}$, and vice versa. But in the proof of his theorem Bell assumes that both ${{\mathscr B}({\bf b},\,\lambda)}$ and ${{\mathscr B}({\bf b'},\,\lambda)}$ can coexist with ${{\mathscr A}({\bf a},\,\lambda)}$ simultaneously. That is analogous to being in New York and Miami at exactly the same time\textsuperscript{\ref{homely}}. But no reasonable criterion of reality can justify such an unphysical demand. The EPR criterion of reality most certainly does not demand any such thing. 

In conclusion, since the two integrands of (\ref{d16}) are physically meaningless, the stringent bounds of ${\pm2}$ on the expression (\ref{d15}) are also physically meaningless \cite{local}. They are mathematical curiosities, without any relevance for the question of local realism.

\begin{corollary}\label{C41}
{\sl It is not possible to be in two places at once.}
\end{corollary}

It is instructive to consider the converse of the above argument. Consider the following hypothesis\textsuperscript{\ref{homely}}: {\it It is possible --- at least momentarily --- to be in two places at once --- for example, in New York and Miami --- at exactly the same time.}

From this hypothesis it follows that in a world in which it is possible to be in two places at once, it would be possible for Bob to detect a component of spin along two mutually exclusive directions, say ${\bf b}$ {\it and} ${\,{\bf b'}}$, at exactly the same time as Alice detects a component of spin along the direction ${\bf a}$, or ${\bf a'}$. If we denote the measurement functions of Alice and Bob by ${{\mathscr A}({\bf a},\,\lambda)}$ and ${{\mathscr B}({\bf b},\,\lambda)}$, respectively, then we can posit that in such a world it would be possible for the measurement event like ${{\mathscr A}({\bf a},\,\lambda)}$ observed by Alice to coexist with both the measurement events ${{\mathscr B}({\bf b},\,\lambda)}$ and ${{\mathscr B}({\bf b'},\,\lambda)}$ that are otherwise only counterfactually observable by Bob, where ${\lambda}$ is the initial state of the singlet system. Therefore, hypothetically, we can represent such a simultaneous event observed by Alice and Bob by a random variable 
\begin{equation}
X({\bf a},\,{\bf b},\,{\bf b'},\,\lambda)\,:=\,{\mathscr A}({\bf a},\,\lambda)\left\{{\mathscr B}({\bf b},\,\lambda)\,+\,{\mathscr B}({\bf b'},\,\lambda)\right\}\,=\,+\,2,\;\,{\rm or}\;\,0,\;\,{\rm or}\;-2\,,
\end{equation}
notwithstanding the fact that there are in fact only two localized particles available to Alice and Bob for each run of their EPR-Bohm type experiment. It is also worth stressing here that in our familiar macroscopic world (after all the vectors ${\bf a}$ and ${\bf b}$ represent macroscopic directions) such a bizarre spacetime event is never observed, because the measurement directions ${\bf a}$ and ${\bf b}$ freely chosen by Alice and Bob are mutually exclusive macroscopic measurement directions in physical space.

Likewise, nothing prevents Alice and Bob in such a bizarre world to simultaneously observe an event represented by
\begin{equation}
Y({\bf a'},\,{\bf b},\,{\bf b'},\,\lambda)\,:=\,{\mathscr A}({\bf a'},\,\lambda)\left\{{\mathscr B}({\bf b},\,\lambda)\,-\,{\mathscr B}({\bf b'},\,\lambda)\right\}\,=\,+\,2,\;\,{\rm or}\;\,0,\;\,{\rm or}\;-2\,.
\end{equation}
And of course nothing prevents Alice and Bob in such a bizarre world to simultaneously observe the sum of the above two events as a single event ({\it i.e.}, four simultaneous clicks of their four detectors), represented by the random variable 
\begin{equation}
Z({\bf a},\,{\bf a'},\,{\bf b},\,{\bf b'},\,\lambda)\,:=\,X({\bf a},\,{\bf b},\,{\bf b'},\,\lambda)\,+\,Y({\bf a'},\,{\bf b},\,{\bf b'},\,\lambda)\,=\,+\,2\;\;{\rm or}\;-2\,. \label{anse}
\end{equation}

Consider now a large number of such initial states ${\lambda}$ and corresponding simultaneous events like ${Z({\bf a},\,{\bf a'},\,{\bf b},\,{\bf b'},\,\lambda)}$. We can then calculate the expected value of such an event occurring in this bizarre world, by means of the integral
\begin{equation}
\int_{\Lambda}\;Z({\bf a},\,{\bf a'},\,{\bf b},\,{\bf b'},\,\lambda)\,   \;\,d\rho(\lambda) \,=\,\int_{\Lambda}\;\Big[\;{\mathscr A}({\bf a},\,\lambda)\,\big\{\,{\mathscr B}({\bf b},\,\lambda)\,+\,{\mathscr B}({\bf b'},\,\lambda)\,\big\}\,+\,{\mathscr A}({\bf a'},\,\lambda)\,\big\{\,{\mathscr B}({\bf b},\,\lambda)\,-\,{\mathscr B}({\bf b'},\,\lambda)\,\big\}\Big]\;\,d\rho(\lambda)\,, \label{expevale}
\end{equation}
where ${\Lambda}$ is the space of all hidden variables ${\lambda}$ and ${\rho(\lambda)}$ is the corresponding normalized probability measure of ${\lambda\in\Lambda}$.

Note that we are assuming nothing about the hidden variables ${\lambda}$. They can be as non-local as we do not like. They can be functions of ${\mathscr A}$ and ${\mathscr B}$, as well as of ${\bf a}$ and ${\bf b}$. In which case we would be dealing with a highly non-local model:
\begin{equation}
\Lambda\ni\lambda=f\left({\bf a},\,{\bf a'},\,{\bf b},\,{\bf b'},\,{\mathscr A},\,{\mathscr B}\right).
\end{equation}

Next we ask: What are the upper and lower bounds on the expected value (\ref{expevale})? The answer is given by (\ref{anse}). Since ${Z({\bf a},\,{\bf a'},\,{\bf b},\,{\bf b'},\,\lambda)}$ can only take two values, ${-2}$ and ${+2}$, the bounds on its integration over ${\rho(\lambda)}$ are necessarily 
\begin{equation}
-\,2\,\leqslant \int_{\Lambda}\;\Big[\;{\mathscr A}({\bf a},\,\lambda)\,{\mathscr B}({\bf b},\,\lambda)\,+\,{\mathscr A}({\bf a},\,\lambda)\,{\mathscr B}({\bf b'},\,\lambda) \,+\,{\mathscr A}({\bf a'},\,\lambda)\,{\mathscr B}({\bf b},\,\lambda)\,-\,{\mathscr A}({\bf a'},\,\lambda)\,{\mathscr B}({\bf b'},\,\lambda)\,\Big]\;\,d\rho(\lambda)\;\leqslant +\,2\,. \label{onetofour}
\end{equation}
But using the addition property of anti-derivatives this expected value can be written as a sum of four expected values,
\begin{equation}
\int_{\Lambda}{\mathscr A}({\bf a},\,\lambda)\,{\mathscr B}({\bf b},\,\lambda)\,d\rho(\lambda)\,+\int_{\Lambda}{\mathscr A}({\bf a},\,\lambda)\,{\mathscr B}({\bf b'},\,\lambda)\,d\rho(\lambda) \,+\int_{\Lambda}{\mathscr A}({\bf a'},\,\lambda)\,{\mathscr B}({\bf b},\,\lambda)\,d\rho(\lambda)\,-\int_{\Lambda}{\mathscr A}({\bf a'},\,\lambda)\,{\mathscr B}({\bf b'},\,\lambda)\,d\rho(\lambda)\,,
\end{equation}
{\it despite our allowing of ${\,\lambda\left({\bf a},\,{\bf a'},\,{\bf b},\,{\bf b'},\,{\mathscr A},\,{\mathscr B}\right)}$ to be non-local}. As a result (\ref{onetofour}) can be written in a familiar form as
\begin{equation}
-\,2\leqslant \,{\cal E}({\bf a},\,{\bf b})\,+\,{\cal E}({\bf a},\,{\bf b'})\,+\,{\cal E}({\bf a'},\,{\bf b})\,-\,{\cal E}({\bf a'},\,{\bf b'})\leqslant +\,2\,. \label{bouchsh}
\end{equation}
Note that the {\it only} hypothesis used to derive these stringent bounds of ${\pm\,2}$ is the one stated above: {\it It is possible -- at least momentarily -- to be in two places at once}. Locality was never assumed; nor was the realism of EPR compromised.

Now we perform the experiments and find that our results exceed the bounds of ${\pm\,2}$ we found in (\ref{bouchsh}) theoretically:  
\begin{equation}
-\,2\sqrt{2}\,\leqslant\,{\cal E}({\bf a},\,{\bf b})\,+\,{\cal E}({\bf a},\,{\bf b'})\,+\,{\cal E}({\bf a'},\,{\bf b})\,-\,{\cal E}({\bf a'},\,{\bf b'})\leqslant +\,2\sqrt{2}\,.
\end{equation}
Consequently, we conclude that the hypothesis we started out with must be false: We do not actually live in a bizarre world in which it is possible -- even momentarily -- to be in New York and Miami at exactly the same time. This is what Bell proved. He proved that we do not live in such a bizarre world. But EPR never demanded, nor hoped that we do.

To summarize our Corollary, Bell inequalities are usually derived by assuming locality and realism, and therefore violations of the Bell-CHSH inequality are usually taken to imply violations of either locality or realism, or both. But we have derived the Bell-CHSH inequality above by assuming only that Bob can measure along the directions ${\bf b}$ and ${\bf b'}$ simultaneously while Alice measures along either ${\bf a}$ or ${\bf a'}$, and likewise Alice can measure along the directions ${\bf a}$ and ${\bf a'}$ simultaneously while Bob measures along either ${\bf b}$ or ${\bf b'}$, {\it without assuming locality}. The violations of the Bell-CHSH inequality therefore simply confirm the impossibility of measuring along ${\bf b}$ and ${\bf b'}$ (or along ${\bf a}$ and ${\bf a'}\,$) simultaneously.

\subsection{The GHZ Variant of Bell's Theorem without Involving Inequalities}\label{GHZ-flaw}

Apart from Bell's argument discussed above there is also an argument, originally proposed by Greenberger, Horne, and Zeilinger (GHZ), that purports to prove the impossibility of any local-realistic understanding of quantum correlations. A remarkable feature of their argument is that, unlike Bell's argument, it does not involve either inequalities or statistics \cite{GHZ}. Instead, they consider the quantum mechanical expectation value (\ref{q-preghz}) for a restricted case in which the measurement settings ${\bf a}$, ${\bf b}$, ${\bf c}$, and ${\bf d}$ are confined to the x-y plane. In that case the expectation value (\ref{q-preghz}) for the state (\ref{ghz-single}) simplifies to
\begin{equation}
\left.{\cal E}^{\rm GHZ}_{{\!}_{Q.M.}}({\bf a},\,{\bf b},\,{\bf c},\,{\bf d})\right|_{\text{x-y}}
=\;-\,\cos\left(\,\phi_{\bf a}\,+\,\phi_{\bf b}\,-\,\phi_{\bf c}\,-\,\phi_{\bf d}\,\right). \label{Maud-q}
\end{equation}
Then, for ${\phi_{\bf a}\,+\,\phi_{\bf b}\,-\,\phi_{\bf c}\,-\,\phi_{\bf d}=0}$, the above expectation value reduces to ${-1}$ for all runs, and thus even for a single run of the experiment. Similarly, for ${\phi_{\bf a}\,+\,\phi_{\bf b}\,-\,\phi_{\bf c}\,-\,\phi_{\bf d}=\pi}$ the above expectation value reduces to ${+1}$ for all runs, and thus even for a single run of the experiment. This is quite similar to the condition ${{\cal E}^{\rm EPR}_{{\!}_{Q.M.}}({\bf a},\,{\bf b})= -1}$ or ${+1}$ for the 2-particle state (\ref{single}) for the specific settings ${{\bf a} = {\bf b}}$ and ${{\bf a} = -{\bf b}}$, respectively, for all runs, and thus even for a single run \cite{GHZ}. These are the conditions of perfect anti-correlations and perfect correlations predicted by quantum mechanics. The claim of Greenberger, Horne, and Zeilinger is that the corresponding conditions ${{\mathscr A}{\mathscr B}{\mathscr C}{\mathscr D}=-1}$ and ${{\mathscr A}{\mathscr B}{\mathscr C}{\mathscr D}=+1}$ for respective settings are impossible to reproduce for a single run of the experiment within any locally causal theory. In what follows we disprove this claim, first analytically, and then by a complete event-by-event numerical simulation.  

\subsubsection{Analytical Disproof of the GHZ Argument}\label{GHZ-disproof}

What we wish to prove is ${{\mathscr A}{\mathscr B}{\mathscr C}{\mathscr D}=-1}$ for any given run of the experiment, for ${\bf a}$, ${\bf b}$, ${\bf c}$, and ${\bf d}$ confined to the x-y plane with ${\phi_{\bf a}\,+\,\phi_{\bf b}\,-\,\phi_{\bf c}\,-\,\phi_{\bf d}=0}$ (cf. Eqs.${\,}$(8) and (11a) of the GHSZ paper \cite{GHZ}). Since the case ${{\mathscr A}{\mathscr B}{\mathscr C}{\mathscr D}=+1}$ for ${\phi_{\bf a}\,+\,\phi_{\bf b}\,-\,\phi_{\bf c}\,-\,\phi_{\bf d}=\pi}$ follows similarly, it will suffice to prove only the ${{\mathscr A}{\mathscr B}{\mathscr C}{\mathscr D}=-1}$ case. We start with our equations (\ref{onedoneggg}) to (\ref{tttdoneggg}), which define the functions ${{\mathscr A}=\pm1}$, ${{\mathscr B}=\pm1}$, ${{\mathscr C}=\pm1}$, and ${{\mathscr D}=\pm1}$. Now, as derived in subsection \ref{SecGHZ}, the expectation value (\ref{203}) follows from the very construction of these functions as a geometrical identity within our model. Therefore we can use this geometrical identity to demonstrate that ${{\mathscr A}{\mathscr B}{\mathscr C}{\mathscr D}=-1}$ for the chosen settings, for which it reduces to
\begin{align}
\left.{\cal E}^{\rm GHZ}_{{\!}_{L.R.}}({\bf a},\,{\bf b},\,{\bf c},\,{\bf d})\right|_{\text{x-y}}&=\!\lim_{\,m\,\rightarrow\,\infty}\!\left[\frac{1}{m}\sum_{k\,=\,1}^{m}\,{\mathscr A}({\bf a}\,,\,\lambda^k)\;{\mathscr B}({\bf b}\,,\,\lambda^k)\;{\mathscr C}({\bf c}\,,\,\lambda^k)\;{\mathscr D}({\bf d}\,,\,\lambda^k)\,\right] \notag \\
&=\,-\,\cos\left(\,\phi_{\bf a}\,+\,\phi_{\bf b}\,-\,\phi_{\bf c}\,-\,\phi_{\bf d}\,\right). \label{Maud}
\end{align}
In fact, for the chosen settings this identity reduces simply to ${{\cal E}^{\rm GHZ}_{{\!}_{L.R.}}({\bf a},\,{\bf b},\,{\bf c},\,{\bf d})=\langle\,{\mathscr A}{\mathscr B}{\mathscr C}{\mathscr D}\,\rangle=-1}$ (see Eq.${\,}$(10a) of GHSZ for details). This tells us that the average of the number ${{\mathscr A}{\mathscr B}{\mathscr C}{\mathscr D}}$ is a constant, and it is equal to ${-1}$. But that is mathematically possible only if ${{\mathscr A}{\mathscr B}{\mathscr C}{\mathscr D}=-1}$ for all runs, for the chosen settings. But if ${{\mathscr A}{\mathscr B}{\mathscr C}{\mathscr D}=-1}$ for all runs, then ${{\mathscr A}{\mathscr B}{\mathscr C}{\mathscr D}=-1}$ holds also for any given run. Therefore ${{\mathscr A}{\mathscr B}{\mathscr C}{\mathscr D}=-1}$ for any run, for the chosen settings. QED.

\section{Numerical Simulations of EPR-Bohm and GHZ Correlations}\label{Numerical}

While our analytical result (\ref{203}) for the general 4-particle GHZ correlations stands on its own, we have nevertheless verified the specific correlations (\ref{Maud}) in an event-by-event numerical simulation of our ${S^7}$ model using the settings prescribed by GHSZ discussed above, but without the restrictions they imposed on the azimuthal or ${\phi_{\bf n}}$-angles \cite{GHZ}. The code for this simulation is reproduced below in the subsection \ref{numghz}, and the graph generated by it is depicted in Fig.~\ref{fig-7}. It is evident from this graph that the predictions of the ${S^7}$ model match exactly with the predictions of quantum mechanics ({\it i.e.}, with the negative cosine curve), despite the ${S^7}$ model being manifestly local-realistic. In particular, it is evident from the graph that the product ${{\mathscr A}{\mathscr B}{\mathscr C}{\mathscr D}}$ takes both positive and negative values for the settings chosen by GHSZ, contrary to their impossibility claim. What is more, the graph has been generated by 200,000 runs of the simulated 4-particle experiment. Thus statistically our simulation is far more robust compared to the mere 250 or so events observed in the ``loophole-free" experiments \cite{Aspect}. Since some familiarity with the languages of Geometric Algebra \cite{Clifford} and the GAViewer program \cite{GAView} is prerequisite for understanding our simulation, we first present a simpler simulation of the 2-particle singlet correlations (\ref{365}), which, as we discussed in previous sections [cf. subsubsection \ref{twosect}], are also predicted by our ${S^7}$ model:

\subsection{Numerical Simulation of the 2-Particle EPR-Bohm Correlations}

\begin{figure}
\hrule
\scalebox{0.7}{
\begin{pspicture}(4.5,-1.0)(5.0,5.9)
\psset{xunit=0.5mm,yunit=4cm}
\psaxes[axesstyle=frame,linewidth=0.01mm,tickstyle=full,ticksize=0pt,dx=90\psxunit,Dx=180,dy=1
\psyunit,Dy=+2,Oy=-1](0,0)(180,1.0)
\psline[linewidth=0.2mm,arrowinset=0.3,arrowsize=2pt 3,arrowlength=2]{->}(0,0.5)(190,0.5)
\psline[linewidth=0.2mm]{-}(45,0)(45,1)
\psline[linewidth=0.2mm]{-}(90,0)(90,1)
\psline[linewidth=0.2mm]{-}(135,0)(135,1)
\psline[linewidth=0.2mm,arrowinset=0.3,arrowsize=2pt 3,arrowlength=2]{->}(0,0)(0,1.2)
\psline[linewidth=0.35mm,linestyle=dashed,linecolor=gray]{-}(0,0)(90,1)
\psline[linewidth=0.35mm,linestyle=dashed,linecolor=gray]{-}(90,1)(180,0)
\put(2.1,-0.38){${90}$}
\put(6.5,-0.38){${270}$}
\put(-0.63,3.92){${+}$}
\put(-0.9,5.0){{\large ${{\cal E}^{\rm EPR}_{{\!}_{L.R.}}}$}${({\bf a},\,{\bf b})}$}
\put(-0.38,1.93){${0}$}
\put(9.65,1.85){\large ${(\phi_{\bf b}\!-\phi_{\bf a})}$}
\psplot[linewidth=0.35mm,linecolor=black]{0.0}{180}{x dup cos exch cos mul 1.0 mul neg 1 add}
\end{pspicture}}
\hrule
\caption{Plot of an event-by-event numerical simulation of the EPR-Bohm correlations predicted by the ${S^7}$ model. The setting vectors ${\bf a}$ and ${\bf b}$ in this simulation were confined to the x-y plane as is usually done in the Bell-test experiments \cite{Aspect}.}
\vspace{0.3cm}
\hrule
\label{fig-6}
\end{figure}
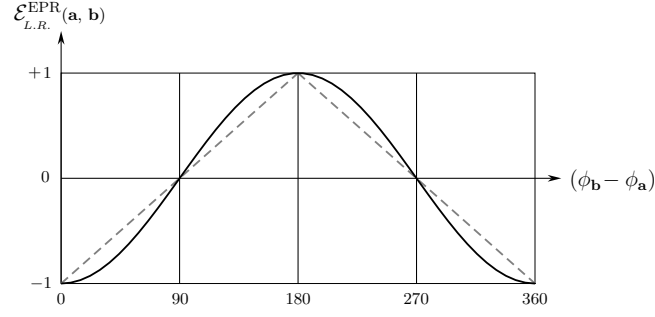

The following code (as well as the one in subsection \ref{numghz}) was written in collaboration with Carl F. Diether III.

\begin{lstlisting}[escapechar=\%]
//Adaptation of A-J. Wonnink's code in GAViewer for the %${S^7}$% model of the
//2-particle EPR-Bohm correlations%~\cite{Wonnink}%:

function getRandomLambda() 
{
   if( rand()>0.5) {return 1;} else {return -1;}
}
function getRandomUnitVector()     
                //http://mathworld.wolfram.com/SpherePointPicking.html
{
   v=randGaussStd()*e1 + randGaussStd()*e2 + 0.00*e3;
                               //vectors are confined to the x-y plane
                               // as done in the Bell-test experiments
   return normalize(v);
}
   batch test()
{
   set_window_title(%``%Test of the S^7 Model for the 2-particle EPR-Bohm
   correlations");
   
   default_model(p3ga);              //choice of the model in GAViewer 
   m=200000;                         //number of iterations, or trials
   I=e1^e2^e3;                       //the fundamental trivector of GA
   s=0;
   t=0;
   u=0;
   for(nn=0;nn<m;nn=nn+1)            //performs the experiment m times
   {
      a_r=getRandomUnitVector()/(sqrt(2)); //vector a_r defined in %(\ref{71-root})%
      a_d=normalize(a_r.(e1*e2))/(sqrt(2)); //a_d is orthogonal to a_r
      D_a=((I a_r) + (a_d e0));          //bivector D_a defined in %(\ref{det})%
      b_r=getRandomUnitVector()/(sqrt(2));
      b_d=normalize(b_r.(e1*e2))/(sqrt(2)); //b_d is orthogonal to b_r
      D_b=((I b_r) + (b_d e0));
      lambda=getRandomLambda();   //lambda = %${\pm1}$% is the hidden variable 
      A=(-D_a).(lambda*D_a);      //the manifestly local function %(\ref{onedone})%
      B=(lambda*D_b).(D_b);       //the manifestly local function %(\ref{twodone})%
      NA=A/-D_a;
      NB=B/D_b;    //implements the twist %(\ref{75699})% in the Hopf bundle of%~${S^3}$%
      q=0;
      if(lambda==1) {q=(NA NB);} else {q=(NB NA);}
                         //calculates the correlations between A and B
      s=s+q;
      print(corrs=scalar(q), %``%f");
                                //outputs correlations shown in Fig.%~~\ref{fig-6}%
      phi_a=atan2(scalar(D_a/(e3^e1)), scalar(D_a/(e2^e3)))*180/pi;
                                               //angle between x and a
      phi_b=atan2(scalar(D_b/(e3^e1)), scalar(D_b/(e2^e3)))*180/pi;
                                               //angle between x and b
      angle=abs(phi_b - phi_a);
      print(angle);         //outputs azimuthal angles between a and b
      t=t+A;
      u=u+B;
   }
      mean=s/m;
      print(mean, %``%f");   //shows the vanishing of the non-scalar part
      aveA=t/m;
      print(aveA, %``%f");   //verifies that individual average < A > = 0
      aveB=u/m;
      print(aveB, %``%f");   //verifies that individual average < B > = 0
      prompt();
}
\end{lstlisting}

The graph generated by this simulation is shown in Fig.~\ref{fig-6}. It is evident from it that the predictions of ${S^7}$ model match exactly with those of quantum mechanics ({\it i.e.}, with the negative cosine curve), despite the model being local-realistic.

\subsection{Numerical Simulation of the 4-Particle GHZ Correlations}\label{numghz}

It is now straightforward to generalize the above code to simulate the 4-particle GHZ correlations (\ref{Maud}) as follows:

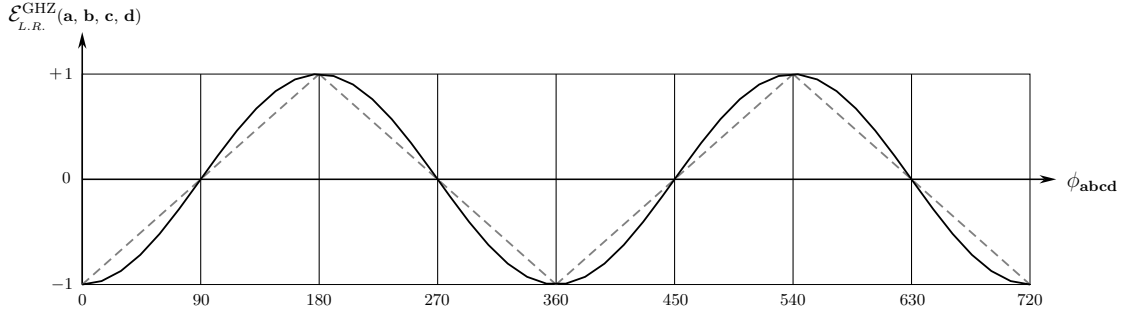
\begin{figure}
\hrule
\scalebox{0.7}{
\begin{pspicture}(13.5,-1.0)(5.0,5.9)
\psset{xunit=0.5mm,yunit=4cm}
\psaxes[axesstyle=frame,linewidth=0.01mm,tickstyle=full,ticksize=0pt,dx=90\psxunit,Dx=180,dy=1\psyunit,Dy=+2,Oy=-1](0,0)(360,1.0)
\psline[linewidth=0.3mm,arrowinset=0.3,arrowsize=2pt 3,arrowlength=2]{->}(0,0.5)(370,0.5)
\psline[linewidth=0.2mm]{-}(45,0)(45,1)
\psline[linewidth=0.2mm]{-}(90,0)(90,1)
\psline[linewidth=0.2mm]{-}(135,0)(135,1)
\psline[linewidth=0.2mm]{-}(180,0)(180,1)
\psline[linewidth=0.2mm]{-}(225,0)(225,1)
\psline[linewidth=0.2mm]{-}(270,0)(270,1)
\psline[linewidth=0.2mm]{-}(315,0)(315,1)
\psline[linewidth=0.3mm,arrowinset=0.3,arrowsize=2pt 3,arrowlength=2]{->}(0,0)(0,1.2)
\psline[linewidth=0.35mm,linestyle=dashed,linecolor=gray]{-}(0,0)(90,1)
\psline[linewidth=0.35mm,linestyle=dashed,linecolor=gray]{-}(90,1)(180,0)
\psline[linewidth=0.35mm,linestyle=dashed,linecolor=gray]{-}(180,0)(270,1)
\psline[linewidth=0.35mm,linestyle=dashed,linecolor=gray]{-}(270,1)(360,0)
\put(2.1,-0.38){${90}$}
\put(6.5,-0.38){${270}$}
\put(11.03,-0.38){${450}$}
\put(15.53,-0.38){${630}$}
\put(-0.63,3.92){${+}$}
\put(-1.4,5.0){{\large ${{\cal E}^{\rm GHZ}_{{\!}_{L.R.}}}$}${({\bf a},\,{\bf b},\,{\bf c},\,{\bf d})}$}
\put(-0.38,1.93){${0}$}
\put(18.7,1.85){\large ${\phi_{{\bf a}{\bf b}{\bf c}{\bf d}}}$}
\psplot[linewidth=0.35mm,linecolor=black]{0.0}{360}{x dup cos exch cos mul 1.0 mul neg 1 add}
\end{pspicture}}
\hrule
\caption{Plot of an event-by-event numerical simulation of the 4-particle GHZ correlations predicted by the ${S^7}$ model. The settings in this simulation were confined to the x-y plane, with the azimuthal angle ${\phi_{{\bf a}{\bf b}{\bf c}{\bf d}}=\phi_{\bf a}+\phi_{\bf b}-\phi_{\bf c}-\phi_{\bf d}}$.}
\vspace{0.3cm}
\label{fig-7}
\hrule
\end{figure}

\begin{lstlisting}[escapechar=\%]
//Adaptation of A-J. Wonnink's code in GAViewer for the %${S^7}$% model of the
//4-particle GHSZ correlations%~\cite{Wonnink}%:

function getRandomLambda() 
{
   if( rand()>0.5) {return 1;} else {return -1;}
}
function getRandomUnitVector()            
               //http://mathworld.wolfram.com/SpherePointPicking.html
{
v=randGaussStd()*e1 + randGaussStd()*e2 + 0.00*e3;
                                //vectors are restricted to x-y plane
                                //as in Eq. (9) of the GHSZ paper %\cite{GHZ}%
   return normalize(v);
}
   batch test()
{
   set_window_title(%``%Test of the S^7 Model for the 4-particle GHSZ
   correlations");
   
   default_model(p3ga);             //choice of the model in GAViewer
   m=200000;                        //number of iterations, or trials
   I=e1^e2^e3;                      //the fundamental trivector of GA
   s=0;
   t=0;
   u=0;
   x=0;
   y=0;
   for(nn=0;nn<m;nn=nn+1)           //performs the experiment m times
   {
     a_r=getRandomUnitVector()/(sqrt(2)); //vector a_r defined in %(\ref{71-root})%
     a_d=normalize(a_r.(e1*e2))/(sqrt(2)); //a_d is orthogonal to a_r
     D_a=((I a_r) + (a_d e0));          //bivector D_a defined in %(\ref{det})%
     b_r=getRandomUnitVector()/(sqrt(2));
     b_d=normalize(b_r.(e1*e2))/(sqrt(2)); //b_d is orthogonal to b_r
     D_b=((I b_r) + (b_d e0));          //bivector D_b defined in %(\ref{det})%
     c_r=getRandomUnitVector()/(sqrt(2));
     c_d=normalize(c_r.(e1*e2))/(sqrt(2)); //c_d is orthogonal to c_r
     D_c=((I c_r) + (c_d e0));          //bivector D_c defined in %(\ref{det})%
     d_r=getRandomUnitVector()/(sqrt(2));
     d_d=normalize(d_r.(e1*e2))/(sqrt(2)); //d_d is orthogonal to d_r
     D_d=((I d_r) + (d_d e0));          //bivector D_d defined in %(\ref{det})%
     lambda=getRandomLambda();   //lambda = %${\pm1}$% is the hidden variable
     A=(-D_a).(lambda*D_a);      //the manifestly local function %(\ref{onedoneggg})%
     B=(lambda*D_b).(D_b);       //the manifestly local function %(\ref{twodoneggg})%
     NA=A/-D_a;
     NB=B/D_b;    //implements the twist %(\ref{75699AB})% in the Hopf bundle of%~${S^3}$%
     C=(-D_c).(lambda*D_c);      //the manifestly local function %(\ref{sssdoneggg})%
     D=(lambda*D_d).(D_d);       //the manifestly local function %(\ref{tttdoneggg})%
     NC=C/-D_c;
     ND=D/D_d;    //implements the twist %(\ref{75699DC})% in the Hopf bundle of%~${S^3}$%
     q=0;
     if(lambda==1) {q=(NA NB NC ND);} else {q=(ND NC NB NA);}
                       //calculates correlations among A, B, C, and D
     s=s+q;
     print(corrs=scalar(q), %``%f");
                          //outputs correlations among A, B, C, and D
     phi_a=atan2(scalar(-D_a/(e3^e1)), scalar(D_a/(e2^e3)))*180/pi;
                                              //angle between x and a
     phi_b=atan2(scalar(D_b/(e3^e1)), scalar(D_b/(e2^e3)))*180/pi;
                                              //angle between x and b
     phi_c=atan2(scalar(D_c/(e3^e1)), scalar(D_c/(e2^e3)))*180/pi;
                                              //angle between x and c
     phi_d=atan2(scalar(D_d/(e3^e1)), scalar(-D_d/(e2^e3)))*180/pi;
                                              //angle between x and d
     angle=abs(phi_a + phi_b - phi_c - phi_d);
                           //GHZ combination of four azimuthal angles 
     print(angle); //outputs the combination of four azimuthal angles
     t=t+A;
     u=u+B;
     x=x+C;
     y=y+D;     
  }
     mean=s/m;
     print(mean, %``%f");   //shows the vanishing of the non-scalar part
     aveA=t/m;
     print(aveA, %``%f");   //verifies that individual average < A > = 0
     aveB=u/m;
     print(aveB, %``%f");   //verifies that individual average < B > = 0
     aveC=x/m;
     print(aveC, %``%f");   //verifies that individual average < C > = 0
     aveD=y/m;
     print(aveD, %``%f");   //verifies that individual average < D > = 0
     prompt();
}
\end{lstlisting}

In addition to the above simulations with all settings confined to the x-y plane, we have also built three-dimensional simulations for both the 2-particle EPR-Bohm state and the 4-particle Greenberger-Horne-Zeilinger state using arbitrary settings, not confined to the x-y plane \cite{3D-Sim}.

\section{Concluding Remarks}\label{conc}

Any experiment in physics can be reduced to a series of elementary questions with possible ``yes" or ``no" answers. These answers in turn may be observed as ``clicks" of event-detectors, as is usually done in the EPR-Bohm type correlation experiments \cite{von}. When we compare such answers -- possibly recorded by remotely located observers -- we find that they are correlated in a remarkably disciplined manner, with the strength of the correlations exceeding the expectations based on Bell's theorem \cite{Bell-1964,GHZ,Stanford}. The natural question then is: Why are these answers correlated in such a disciplined manner when in quantum mechanics there appears to be no predetermined cause dictating the correlations? In this paper we have shown that the discipline and strength exhibited in the correlation experiments are natural consequences of the fact that the three-dimensional physical space in which all experiments are conducted respects the symmetries of a Clifford-algebraic 7-sphere, which arises from an associative interplay of the graded Euclidean primitives, such as points, lines, planes and volumes. These primitives provide the basis for the conformal\textsuperscript{\ref{Conformal}} geometry of the physical space (\ref{3-sphere}), namely that of a quaternionic 3-sphere, ${S^3}$, embedded in an eight-dimensional Clifford-algebraic manifold, ${{\cal K}^{\lambda}}$. They allow us to understand the origins and strengths of {\it all} quantum correlations locally, as aspects of the geometry of the compactified physical space ${S^3}$, with ${S^7\subset{\cal K}^{\lambda}}$ being its algebraic representation space. Thus every quantum correlation can be understood as a correlation among a set of points of this ${S^7}$. We have demonstrated this by proving a comprehensive theorem about the geometric origins of the correlations predicted by arbitrary quantum states:
\begin{equation}
{\cal E}_{{\!}_{L.R.}\!}({\bf a},\,{\bf b},\,{\bf c},\,{\bf d},\,\dots\,)\,=\!\!\lim_{\,m\,\rightarrow\,\infty}\!\left[\frac{1}{m}\sum_{k\,=\,1}^{m}\,{\mathscr A}({\bf a}\,,\,\lambda^k)\;{\mathscr B}({\bf b}\,,\,\lambda^k)\;{\mathscr C}({\bf c}\,,\,\lambda^k)\;{\mathscr D}({\bf d}\,,\,\lambda^k)\;\dots\,\right] \,=\,-\cos\theta_{{\bf x}{\bf y}}({\bf a},\,{\bf b},\,{\bf c},\,{\bf d},\,\dots\,).
\end{equation}
We have also proved within our framework that the strengths of these correlations are bounded by Tsirel'son's bounds:
\begin{align}
\left|\,{\cal E}_{{\!}_{L.R.}}({\bf x},\,{\bf y})\,+\,{\cal E}_{{\!}_{L.R.}}({\bf x},\,{\bf y'})\,+\,{\cal E}_{{\!}_{L.R.}}({\bf x'},\,{\bf y})\,-\,{\cal E}_{{\!}_{L.R.}}({\bf x'},\,{\bf y'})\,\right|\,&\leqslant\,2\,\sqrt{\,1\,-\,\left({\bf x}\times{\bf x'}\right)\cdot\left({\bf y'}\times{\bf y}\right)\,} \notag \\
&\leqslant\,2\sqrt{2}\,.
\end{align}
We have then explicitly reproduced the strong correlations predicted by the EPR-Bohm state within our framework,
\begin{align}
{\cal E}^{\rm EPR}_{{\!}_{L.R.}\!}({\bf a},\,{\bf b})\,&=\!\lim_{\,m\,\rightarrow\,\infty}\left[\frac{1}{m}\sum_{k\,=\,1}^{m}\,{\mathscr A}({\bf a}\,,\,\lambda^k)\;{\mathscr B}({\bf b}\,,\,\lambda^k)\right] =\,-\,\cos\theta_{{\bf a}{\bf b}}\,, \notag \\
&\;\,\text{together with}\;\;\,{\cal E}^{\rm EPR}_{{\!}_{L.R.}\!}({\bf n})\,=\!\lim_{\,m\,\rightarrow\,\infty}\left[\frac{1}{m}\sum_{k\,=\,1}^{m}\,{\mathscr N}({\bf n}\,,\,\lambda^k)\right] =\,0\,, \label{203-1}
\end{align}
as well as explicitly reproduced the strong correlations predicted by the 4-particle Greenberger-Horne-Zeilinger state:
\begin{align}
{\cal E}^{\rm GHZ}_{{\!}_{L.R.}}({\bf a},\,{\bf b},\,{\bf c},\,{\bf d})\,&=\!\!\lim_{\,m\,\rightarrow\,\infty}\!\left[\frac{1}{m}\sum_{k\,=\,1}^{m}\,{\mathscr A}({\bf a}\,,\,\lambda^k)\;{\mathscr B}({\bf b}\,,\,\lambda^k)\;{\mathscr C}({\bf c}\,,\,\lambda^k)\;{\mathscr D}({\bf d}\,,\,\lambda^k)\right] \notag \\
&=\,\cos\theta_{\bf a}\,\cos\theta_{\bf b}\,\cos\theta_{\bf c}\,\cos\theta_{\bf d}\,-\,\sin\theta_{\bf a}\,\sin\theta_{\bf b}\,\sin\theta_{\bf c}\,\sin\theta_{\bf d}\,\cos\left(\,\phi_{\bf a}\,+\,\phi_{\bf b}\,-\,\phi_{\bf c}\,-\,\phi_{\bf d}\,\right), \notag \\
&\;\;\;\;\;\,\text{together with}\;\;\,{\cal E}^{\rm GHZ}_{{\!}_{L.R.}\!}({\bf n})\,=\!\lim_{\,m\,\rightarrow\,\infty}\left[\frac{1}{m}\sum_{k\,=\,1}^{m}\,{\mathscr N}({\bf n}\,,\,\lambda^k)\right] =\,0\,. \label{203-2}
\end{align}
We have also presented two event-by-event numerical simulations of the correlations (\ref{203-1}) and (\ref{203-2}) predicted within our local-realistic framework. The comprehensive theorem we have proved dictates that --- at least in principle --- it is always possible to locally reproduce the strong correlations predicted by any arbitrary quantum state. The {\it raison d'\^etre} for the strength of the correlations turns out to be the non-trivial twist in the Hopf bundle of ${S^3}$ \cite{disproof,IJTP}, or in its algebraic representation space ${S^7}$. Given the fact that we started out our analysis with the most primitive elements of the physical space in the spirit of Euclid's elements for geometry, our demonstration suggests that the quantum correlations observed in Nature are best viewed as consequences of spacetime, rather than spacetime as an emergent property of quantum entanglement.

\appendix

\counterwithin{equation}{section}

\section{Local Causality and the Normed Division Algebras}\label{ApA}

In section \ref{222} we started out by reviewing the algebraic properties of the physical space, which was crystallized in the definition (\ref{3-sphere}) and turned out to be a quaternionic 3-sphere. We then constructed the algebraic representation space (\ref{spring}) of this physical space, which turned out to be an octonion-like 7-sphere. In the subsequent sections we saw the crucial role played by these two spheres in understanding the existence, origins, and strengths of all quantum correlations. But what is so special about the 3 and 7 dimensions? Why is the vector cross product definable only in 3 and 7 dimensions and no other? Why are ${\mathbb R}$, ${\mathbb C}$, ${\mathbb H}$, and ${\mathbb O}$ the only possible normed division algebras? Why are only the 3- and 7-dimensional spheres non-trivially parallelizable out of infinitely many possible spheres? Why is it possible to derive all quantum mechanical correlations as local-realistic correlations among the points of only the 3- and 7-spheres?

The answers to all of these questions are intimately connected to the notion of factorizability introduced by Bell within the context of his theorem \cite{Bell-1964}. Mathematicians have long been asking: When is a product of two squares itself a square: ${x^2\,y^2 = z^2\,}$? If the number ${z}$ is factorizable, then it can be written as a product of two other numbers, ${z = x\,y}$, and then the above equality is seen to hold for the numbers ${x}$, ${y}$, and ${z}$. For ordinary numbers this is easy to check. The number 8 can be factorized into a product of 2 and 4, and we then have ${64 = 8^2 = (2 \times 4)^2 = 2^2\times 4^2 = 64}$. But what about sums of squares? A more profound equality holds, in fact, for a sum of two squares times a sum of two squares as a third sum of two squares:
\begin{equation}
(x_1^2 + x_2^2)\,(y_1^2 + y_2^2) \,=\, (x^{}_1y^{}_1 - x^{}_2y^{}_2)^2 + (x^{}_1y^{}_2 + x^{}_2y^{}_1)^2\,=\,z_1^2 + z_2^2.
\end{equation}
There is also an identity like this for the sums of four squares. It was first discovered by Euler, and then rediscovered and popularized by Hamilton in the ${19^{th}}$ century through his work on quaternions. It is also known that Graves and Cayley independently discovered a similar identity for the sums of eight squares. This naturally leads to the question of whether the product of two sums of squares of ${n}$ different numbers can be a sum of ${n}$ different squares? In other words, does the following equality hold in general for any ${n}$?
\begin{equation}
(x_1^2 + x_2^2 + \dots + x_n^2)\,(y_1^2 + y_2^2 + \dots + y_n^2) \,=\, z_1^2 + z_2^2 + \dots + z_n^2.\label{8.39}
\end{equation}
It turns out that this equality holds only for ${n}$ = 1, 2, 4, and 8. This was proved by Hurwitz in 1898 \cite{Hurwitz-rrr}. It reveals a deep and surprising fact about the world we live in. Much of what we see around us, from elementary particles to distant galaxies, is an inevitable consequence of this simple mathematical fact. The world is the way it is because the above equality holds only for ${n = \text{1}}$, 2, 4, and 8. For example, the above identity is equivalent to the existence of a division algebra of dimension ${n}$ over the field ${\mathbb R}$ of real numbers. Indeed, if we define vectors ${{\bf x}=(x_1,\dots, x_n)}$, ${{\bf y}=(y_1,\dots, y_n)}$, and ${{\bf z}={\bf x}*{\bf y}}$ in ${{\mathbb R}^n}$ such that ${z_i}$'s are functions of ${x_j}$'s and ${y_k}$'s determined by (\ref{8.39}), then we have
\begin{equation}
||{\bf x}||\,||{\bf y}|| \,=\, ||{\bf x}*{\bf y}||\,.
\end{equation}
Thus the division algebras ${\mathbb R}$ (real), ${\mathbb C}$ (complex), ${\mathbb H}$ (quaternion), and ${\mathbb O}$ (octonion) we use in much of our science are intimately related to the dimensions ${n}$ = 1, 2, 4, and 8. Moreover, from the equation of a unit sphere in ${m}$ dimensions,
\begin{equation}
x_0^2 + x_1^2 + x_2^2 + \dots + x_m^2 \,=\,1\,,
\end{equation}
it is easy to see that the four parallelizable spheres ${S^0}$, ${S^1}$, ${S^3}$, and ${S^7}$ correspond to ${n}$ = 1, 2, 4, and 8, which are the dimensions of the respective embedding spaces of these four spheres. What is not so easy to see, however, is the fact that there is a deep connection between Hurwitz's theorem and the quantum correlations (cf. Chapter 7 of Ref.~\cite{disproof}).

As we saw, quantum correlations are inevitable consequences of the non-trivial geometry and topology of ${S^7}$, which in turn is the largest parallelizable sphere permitted by Hurwitz's theorem. In the language of Hopf fibrations, ${S^7}$ is locally (but not globally) equal to the product ${S^4\times S^3}$, and thus is a Hopf bundle made up of 4-sphere worth of 3-spheres with a non-trivial twist in the bundle. Similarly, ${S^3}$ is locally (but not globally) equal to the product ${S^2\times S^1}$, and thus is a Hopf bundle made up of 2-sphere worth of 1-spheres with a non-trivial twist in the bundle. Thus the innocent looking algebraic equality (\ref{8.39}) has far reaching consequences, not only for the edifice of mathematics, but also for that of quantum physics. In fact ${S^7}$ turns out to be both necessary and sufficient for understanding the strong correlations locally. That may seem surprising, but this necessity stems from the profound relationship between the normed division algebras and the parallelizability of the four spheres we noted above. Quantum correlations thus exist and exhibit the remarkable strengths they do because the equality (\ref{8.39}) holds only for ${n}$ = 1, 2, 4, and 8.

\section{Proof of Bell's Condition of Factorizability within ${S^7}$}\label{ApB}

In a deterministic hidden-variable framework local causality can be specified by Bell's condition of factorizability, as in Eq.~(3.4) of Ref.~\cite{Clauser}. For the GHZS measurement results within 7-sphere this condition is given by Eq.~(\ref{myourlocality}):
\begin{equation}\tag{192}
({\mathscr A}_{\bf a}\,{\mathscr B}_{\bf b}\,{\mathscr C}_{\bf c}\,{\mathscr D}_{\bf d})(\lambda^k)\,=\,{\mathscr A}({\bf a}\,,\,\lambda^k)\,{\mathscr B}({\bf b}\,,\,\lambda^k)\,{\mathscr C}({\bf c}\,,\,\lambda^k)\,{\mathscr D}({\bf d}\,,\,\lambda^k)\,=\,\pm\,1\,\in\,S^7.
\end{equation}
To prove it, we begin with four maps of the following form, which --- without the limits --- are generic elements of ${S^7}$:
\begin{align}
S^7\ni{\mathscr A}({\bf a}\,,\,\lambda^k):=\!\!\lim_{\,\substack{{\bf s}_{r1}\,\rightarrow\;{\bf a}_r \\ {\bf s}_{d1}\,\rightarrow\;{\bf a}_d}}\!\left\{\,-\,{\bf D}({\bf a}_r,\,{\bf a}_d,\,0)\,{\bf N}({\bf s}_{r1},\,{\bf s}_{d1},\,0,\,\lambda^k)\,\right\}&=\,
\begin{cases}
\,+\,1\;\;\;\;\;{\rm if} &\lambda^k\,=\,+\,1 \\
\,-\,1\;\;\;\;\;{\rm if} &\lambda^k\,=\,-\,1
\end{cases} \Bigg\}\,, \label{aponedoneggg} \\
S^7\ni{\mathscr B}({\bf b}\,,\,\lambda^k):=\!\!\lim_{\,\substack{{\bf s}_{r2}\,\rightarrow\;{\bf b}_r \\ {\bf s}_{d2}\,\rightarrow\;{\bf b}_d}}\!\left\{\,+\,{\bf N}({\bf s}_{r2},\,{\bf s}_{d2},\,0,\,\lambda^k)\,{\bf D}({\bf b}_r,\,{\bf b}_d,\,0)\,\right\}&=\,
\begin{cases}
\,-\,1\;\;\;\;\;{\rm if} &\lambda^k\,=\,+\,1 \\
\,+\,1\;\;\;\;\;{\rm if} &\lambda^k\,=\,-\,1
\end{cases} \Bigg\}\,, \label{aptwodoneggg} \\
S^7\ni{\mathscr C}({\bf c}\,,\,\lambda^k)\;:=\lim_{\,\substack{{\bf t}_{r1}\,\rightarrow\;{\bf c}_r \\ {\bf t}_{d1}\,\rightarrow\;{\bf c}_d}}\!\left\{\,-\,{\bf D}({\bf c}_r,\,{\bf c}_d,\,0)\,{\bf N}({\bf t}_{r1},\,{\bf t}_{d1},\,0,\,\lambda^k)\,\right\}&=\,
\begin{cases}
\,+\,1\;\;\;\;\;{\rm if} &\lambda^k\,=\,+\,1 \\
\,-\,1\;\;\;\;\;{\rm if} &\lambda^k\,=\,-\,1
\end{cases} \Bigg\}\,, \\
\text{and}\;\;\;\;\;\;\;\;\;\;\;\;\;\;\;\;\;\;\;\;\;\;\;\;\;\;\;\;\;\;\;\;\;\;\;\;\;\;\;\;\;\;\;\;\;\;\;\;\;\;\;\;\;\;\;\;\;\;\;\;\;\;\;\;\;\;\;\;\;\;\;\;\;\;\;\;\;\;\;\;\;\;\;\;\;\;\;\;\;\;\; &\notag \\
S^7\ni{\mathscr D}({\bf d}\,,\,\lambda^k):=\!\!\lim_{\,\substack{{\bf t}_{r2}\,\rightarrow\;{\bf d}_r \\ {\bf t}_{d2}\,\rightarrow\;{\bf d}_d}}\!\left\{\,+\,{\bf N}({\bf t}_{r2},\,{\bf t}_{d2},\,0,\,\lambda^k)\,{\bf D}({\bf d}_r,\,{\bf d}_d,\,0)\,\right\}&=\,
\begin{cases}
\,-\,1\;\;\;\;\;{\rm if} &\lambda^k\,=\,+\,1 \\
\,+\,1\;\;\;\;\;{\rm if} &\lambda^k\,=\,-\,1
\end{cases} \Bigg\}\,.
\end{align}
But since ${S^7}$ remains closed under multiplication, the geometric product of these four maps (or, for that matter, the geometric product of any number of such maps) --- because of the ``product of limits equal to limits of product" rule --- is necessarily a fifth map of the form
\begin{align}
S^7\ni{\mathscr X}({\bf x}\,,\,\lambda^k):=\!\!\lim_{\,\substack{{\bf u}_{r}\,\rightarrow\;{\bf x}_r \\ {\bf u}_{d}\,\rightarrow\;{\bf x}_d}}\!\left\{\,-\,{\bf D}({\bf x}_r,\,{\bf x}_d,\,0)\,{\bf N}({\bf u}_{r},\,{\bf u}_{d},\,0,\,\lambda^k)\,\right\}&=\,
\begin{cases}
\,+\,1\;\;\;\;\;{\rm if} &\lambda^k\,=\,+\,1 \\
\,-\,1\;\;\;\;\;{\rm if} &\lambda^k\,=\,-\,1
\end{cases} \Bigg\}\,.
\end{align}
Consequently, as a consequence of the non-trivial twists (\ref{75699AB}) and (\ref{75699DC}) in the Hopf bundle of ${S^7}$, we necessarily have
\begin{equation}
({\mathscr A}_{\bf a}\,{\mathscr B}_{\bf b}\,{\mathscr C}_{\bf c}\,{\mathscr D}_{\bf d})(\lambda^k)\,=\,{\mathscr A}({\bf a}\,,\,\lambda^k)\,{\mathscr B}({\bf b}\,,\,\lambda^k)\,{\mathscr C}({\bf c}\,,\,\lambda^k)\,{\mathscr D}({\bf d}\,,\,\lambda^k)\,=\,{\mathscr X}({\bf x}\,,\,\lambda^k)\,=\,\pm\,1\,\in\,S^7.
\end{equation}

\section*{Acknowledgements}
The author thanks Carl F. Diether III for his generous help with the simulations of the 2-particle EPR-Bohm and 4-particle GHSZ correlations predicted by the ${S^7}$ model, by adapting the original GAViewer code of Albert Jan Wonnink for the simulations of the earlier models based on ${S^3}$.

\end{document}